\pdfoutput=1
\documentclass[aps,prd,amsmath,amssymb,superscriptaddress,altaffillsymbol, preprintnumbers,preprint,nofootinbib,a4paper]{revtex4}

\usepackage{amsthm}
\usepackage{amsmath}
\usepackage{graphicx}
\usepackage{color}
\usepackage{subfigure}
\usepackage{multirow}
\usepackage{ulem}

\newcommand{\beq}{\begin{eqnarray}}
\newcommand{\eeq}{\end{eqnarray}}

\newcommand{\bmp}{\noindent\begin{minipage}{16cm}}
\newcommand{\emp}{\end{minipage}\vskip 7mm} 

\newcommand{\GeV}{\mbox{ ${\mathrm{GeV}}$}}

\usepackage{dcolumn}
\usepackage{bm}
\usepackage{slashed}

\theoremstyle{definition}

\theoremstyle{plain}

\newcommand{\be}{\begin{equation}}
\newcommand{\ee}{\end{equation}}

\newcommand{\GHC}{G_{\mathrm{HC}}}

\newcommand{\NHC}{N_{\mathrm{HC}}}

\newcommand{\Rcoset}{SU(5)/SO(5)}
\newcommand{\PRcoset}{SU(4)/Sp(4)}
\newcommand{\Ccoset}{SU(4)\times SU(4)'/SU(4)_D}
\newcommand{\Rcolor}{SU(6)/SO(6)}
\newcommand{\PRcolor}{SU(6)/Sp(6)}
\newcommand{\Ccolor}{SU(3)\times SU(3)'/SU(3)_D}

\newcommand{\Fu}{{\mathbf{F}}}
\newcommand{\An}{{\mathbf{A}}}
\newcommand{\Sy}{{\mathbf{S}}}
\newcommand{\Ad}{{\mathbf{Ad}}}
\newcommand{\Sp}{{\mathbf{Spin}}}

\usepackage{epsfig}

\usepackage[ margin=5pt, font=normalsize,labelfont=bf,justification=raggedright]{caption}

\usepackage{hyperref}
\definecolor{rossoCP3}{cmyk}{0,.88,.77,.40}
\definecolor{verdeCP3}{rgb}{0.09765625, 0.57421875, 0.1015625}
\definecolor{bluCP3}{rgb}{0, 0.23, 0.67}
\hypersetup{colorlinks, bookmarksopen, bookmarksnumbered,citecolor=verdeCP3, linkcolor=bluCP3, pdfstartview=FitH, urlcolor=rossoCP3}
\preprint{CTPU-16-24, LU TP 16-56, LYCEN 2016-08}

\begin{document}
\title{\Large Di-boson signatures as Standard Candles\\
 for Partial Compositeness}

\author{Alexander Belyaev}

\affiliation{School of Physics \& Astronomy, University of Southampton, UK;}
\affiliation{Particle Physics Department, Rutherford Appleton Laboratory, Chilton, Didcot, Oxon OX11 0QX, UK;}

\author{Giacomo Cacciapaglia}
\author{Haiying Cai}
\affiliation{\mbox{Univ Lyon, Universit\'e Lyon 1, CNRS/IN2P3, IPNL, F-69622, Villeurbanne, France;}}

\author{Gabriele Ferretti}
\affiliation{Department of Physics, Chalmers University of Technology, Fysikg\r{a}rden, 41296 G\"oteborg, Sweden;}

\author{Thomas Flacke}
\affiliation{Center for Theoretical Physics of the Universe, Institute for Basic Science (IBS), Daejeon, 34051, Korea;}
\affiliation{Department of Physics, Korea University, Seoul 136-713, Korea;}

\author{Alberto Parolini}
\affiliation{Department of Physics, Korea University, Seoul 136-713, Korea;}

\author{Hugo Serodio}
\affiliation{Department of Physics, Korea University, Seoul 136-713, Korea;}
\affiliation{Department of Astronomy and Theoretical Physics, Lund University, SE-223 62 Lund, Sweden.}


\begin{abstract}
Composite Higgs Models are often constructed including fermionic top partners with a mass around the TeV scale, with the top partners playing the role of stabilizing the Higgs potential and enforcing partial compositeness for the top quark. A class of models of this kind can be formulated in terms of fermionic strongly coupled gauge theories. A common feature they all share is the presence of specific additional scalar resonances, namely two neutral singlets and a colored octet, described by a simple effective Lagrangian. We study the phenomenology of these scalars, both in a model independent and model dependent way, including the bounds from all the available searches in the relevant channels with di-boson and di-top final states. We develop a generic framework which can be used to constrain any model containing pseudo-scalar singlets or octets. Using it, we  find that such signatures 
provide strong bounds on the compositeness scale complementary to the traditional EWPT and Higgs couplings deviations. In many cases a relatively light scalar can be on the verge of discovery as a first sign of new physics.
 \\[.1cm]

 \end{abstract}

\maketitle
\tableofcontents


\section{Introduction}

The LHC has  entered a phase
with exceptional  potential for discovering new physics, and new data is being collected at an unprecedented rate during the Run--II that started last year. Not surprisingly, this fact has led to a flurry of model-building activity, with the intent of charting the landscape Beyond the Standard Model (BSM) and proposing new discovery channels.

Among the various BSM proposals, the idea that the Higgs sector of the Standard Model (SM) is dynamically generated by a confining strong dynamics is playing an important role and is being continuously tested experimentally.
In particular, the models discussed in this work are four dimensional gauge theories combining the concept of the Higgs as a pseudo-Nambu-Goldstone boson (pNGB) \cite{Kaplan:1983fs} with that of partial compositeness \cite{Kaplan:1991dc}, where the top quark mass arises by a linear coupling with a spin-1/2 ``top-partner''. Therefore, the main requirement on the underlying theory is to provide a viable Higgs sector together with the appropriate colored fermionic bound states.
The construction of these models has been discussed elsewhere \cite{Ferretti:2013kya,Ferretti:2016upr}, and some specific examples were given in \cite{Barnard:2013zea,Ferretti:2014qta,Vecchi:2015fma}.
With the exception of  \cite{Vecchi:2015fma}, all models contain at least two species of underlying fermions belonging to different irreducible representations (irreps) under the confining hypercolor (HC) gauge group. This observation will play a crucial role in the rest of this paper.
The chiral perturbation theory for these models has been recently presented in Ref.~\cite{DeGrand:2016pgq}. The coupling to tops has been addressed in~\cite{Golterman:2015zwa}.

The phenomenology of Composite Higgs models has been already extensively studied, with particular focus on the minimal symmetry breaking pattern $SO(5)/SO(4)$ that leads to only a Higgs boson in the pNGB spectrum (see \cite{Bellazzini:2014yua,Panico:2015jxa} for recent reviews). Because of the lack of additional light scalars, collider searches have focused on colored top partners, 
together with other indirect constraints on SM quantities. The current bounds on the masses of top partners range around 700-900 GeV \cite{ATLAS:2013ima,TheATLAScollaboration:2013sha,Chatrchyan:2013uxa,Chatrchyan:2013wfa,Aad:2014efa}. However, it is very challenging to obtain the minimal scenario starting from a four dimensional fermionic theory: attempts present in the literature are either relying on supersymmetry \cite{Caracciolo:2012je} or on effective four-fermion interactions {\it \`a la} Nambu-Jona Lasinio (NJL) \cite{vonGersdorff:2015fta}.

In the class of models we consider, based on a confining gauged HC and with only fermionic matter fields~\footnote{The possibility of top partners arising as bound states of a fermion and a scalar has been recently proposed in~\cite{Sannino:2016sfx}.}, the symmetry breaking patterns are determined by the representations of the underlying fermions~\cite{Peskin:1980gc,Preskill:1980mz}, giving rise to non-minimal cosets with additional pNGBs.
Thus, the main message we want to convey is that the first evidence of this class of models of partial compositeness may come from the discovery of the additional pNGBs rather than from the direct observation of the top partners.

The phenomenological relevance of pNGBs in composite models is not new \cite{Ellis:1980hz,Kaplan:1983sm,Dugan:1984hq,Casalbuoni:1998fs,Arbey:2015exa}.
What we aim at, on the other hand, is to directly link their presence to the mechanism of partial compositeness.
To do so, instead of looking at the details of each model \cite{Ferretti:2016upr}, we focus on two types of scalars that are universally present in all models: singlet pseudo-scalars associated to global U(1) symmetries \cite{Belyaev:2015hgo}, and a color octet arising from the presence of colored underlying fermions.

The presence of two types of fermions in the underlying theory guarantees that there is always a combination of the two U(1)'s which is non anomalous with respect to the hypercolor group. Thus, contrary to the anomalous axial current in QCD, the associated pseudo-scalar will be light. Inspired by large-$N_c$ estimates in QCD, we will also keep the anomalous U(1) scalar in the spectrum because, depending on the model, it may also be light. These two states will be denoted $a$ and $\eta'$ in the mass eigenstate basis (as non-trivial mixing is present).

We will briefly review the salient points of these constructions, however the focus of the paper is to investigate their phenomenology, derive all constraints from up-to-date searches,
point to the promising  signatures and their correlations, and make concrete suggestions for their
further exploration at the LHC. In particular, we will focus on the two singlets and on the color octet, commonly present in all models. Their couplings to the SM gauge bosons are generated via the Wess-Zumino-Witten \cite{Wess:1971yu,Witten:1983tw} anomalous term, and are thus computable in terms of the properties of the underlying theory. 

Additional couplings to tops can also appear: we prove that the singlets always couple to tops via partial compositeness, while this coupling may be absent for the octet, and we present an estimate of the couplings to tops (and other SM fermions) proportional to their mass. The calculability of the phenomenologically relevant couplings makes these three pseudoscalars standard candles for fundamental models of partial compositeness, that will shine in particular via di-boson searches at the LHC. In fact, the observation of resonances in di-boson channels would allow to extract information about the WZW couplings, which are directly related to the properties of the underlying theory. 

The scalar singlet production via gluon fusion and its subsequent decay to a pair of gauge bosons, both mediated by the WZW interactions, leads to clean signatures at the LHC. Such final states are intensely searched for at the LHC, and recently the emergence of excesses in both di-boson and di-photon, now less prominent or entirely disappeared, was the source of big excitement and inspiration in the particle physics community. A pseudo-scalar decaying via WZW interactions can easily accommodate such signatures \cite{Cacciapaglia:2015nga}, and the case of the models under investigation has been already pointed out by a subset of the authors \cite{Cai:2015bss,Belyaev:2015hgo}.

The paper is organized as follows: in Section \ref{sec:models} we briefly present the content of the models under consideration and their salient dynamical properties. In Section \ref{sec:PNGBtheory} we discuss the pNGBs of relevance for this work. We present their chiral lagrangian, their couplings and their masses. Section \ref{sec:pheno} discusses their phenomenology and presents up-to-date (post ICHEP2016) bounds on their couplings in a model-independent way. 
We focus on the most updated constraints deriving from di-boson searches, di-top resonances and other relevant channels (like pair production in the case of the color octet).
Section \ref{sec:specific pheno} confronts these bounds with the models presented in Section \ref{sec:models}. We explore two specific models and extract a combined lower bound on the decay constant of the pNGBs. Being associated to the Higgs sector, the value of such scale is a direct measure of the fine tuning involved in these models. As a result of this study we set up the strategy and
create the framework for a generic exploration of the models with di-boson, di-jet and top-quarks final state signatures.
We summarize our findings and conclude in Section \ref{sec:conclusions}.

\section{Classification of the models}\label{sec:models}

The models we consider are gauge theories based on a simple HC group $\GHC$ characterized by having \emph{two} distinct irreps of fermions, denoted by $\psi$ and $\chi$~\footnote{We always work with Weyl fermions, unless otherwise specified, and consider only vector-like theories. A complex irrep and its conjugate is counted as one, see~\cite{Ferretti:2016upr} for details.}. In addition to hypercolor, the $\psi$ fermions carry electroweak (EW) quantum numbers chosen in order to generate a composite pNGB Higgs, while the $\chi$ carry QCD color and hypercharge.

The phenomenological reasons for this choice are two-fold. On the one hand, these models easily accommodate the presence of potential top partners obtained from HC neutral bound states of three fermions. To this end a non zero hypercharge $Y_\chi$ has to be consistently assigned to the fermions $\chi$. On the other hand, separating color (carried by the $\chi$'s) from the Higgs sector avoids problems with spontaneous color breaking and the presence of light colored pNGBs. Even more relevant for this work, the presence of two distinguished irreps allows for the existence of a light pNGB associated to a $U(1)$ axial symmetry non-anomalous under HC.

The main constraints on the  models under consideration which define their group stucture are:
implementation of a composite Higgs mechanism compatible with custodial symmetry, the existence of top partners and the protection of the $Z \to b_L \bar b_L$ branching ratio.
The last requirement eliminates some possible solutions that were present in the lists~\cite{Ferretti:2013kya}~\cite{Belyaev:2015hgo} with $\psi$ in a complex irrep and top partners in the $({\mathbf{2}}, {\mathbf{1}})$ of $SU(2)_L\times SU(2)_R$~\cite{Agashe:2006at}.

The Higgs mass is generated by the explicit breaking of the global symmetry of the strong sector. Typical sources of breaking are the coupling to the EW bosons and to the heavy quarks as well as possible bare masses for the hyperquarks. The measured Higgs mass is then used as an input to give one relation between these couplings, the low-energy coefficients of the strongly coupled theory (in principle computable on the lattice) and $f_\psi$.  A similar relation follows from fixing the Higgs vev. That it is possible to fix the Higgs mass and vev to their physical values (at the cost of some fine-tuning) is shown in various previous works:  e.g.~\cite{Ma:2015gra} and~\cite{Ferretti:2016upr} for the cosets of interest in this paper.

Within the  constraints above there are  three minimal cosets $\Rcoset$, $\PRcoset$ and $\Ccoset$ in the case of $N_\psi=5$ real, $N_\psi=4$ pseudo-real or $N_\psi=4$ complex (plus its conjugate) irreps of $\GHC$ respectively coming from condensation of the fermions $\psi$.
If this were the only condensate forming, there would be no more pNGBs in the spectrum. In particular, in this case the axial $U(1)_\psi$ rotating all $\psi$ by the same phase would be spontaneously broken but also explicitly broken by a $U(1)_\psi\GHC^2$ Adler-Bell-Jackiw (ABJ) anomaly and thus its would-be Goldstone boson would acquire a large mass.

In the present class of models, however, the $\chi$ also condense, giving rise to new colored pNGBs associated to the cosets $\Rcolor$, $\PRcolor$ and $\Ccolor$ for $N_\chi=6$ real, $N_\chi=6$ pseudo-real or $N_\chi=3$ complex (plus its conjugate) irreps of $\GHC$ respectively. Now there is also an additional axial $U(1)_\chi$ spontaneously broken and it is possible to construct an ABJ anomaly free linear combination $U(1)_a$ by choosing the charges $q_{\psi,\chi}$ to obey
\beq \label{eq:qpsiqchi}
q_\psi N_\psi T(\psi) +  q_\chi N_\chi T(\chi) = 0\,,
\eeq
where $T$ denotes the Dynkin index of the irrep and for complex irreps we must count both the complex and its conjugate, i.e.  count the index twice. The pNGB $\tilde{a}$ associated to this symmetry is naturally lighter than the typical confinement scale, while the remaining orthogonal state $\tilde{\eta}'$ acquires a larger mass. We denote these states with a tilde because they do not yet correspond to mass eigenstates, as will be discussed in the following section.

Among the remaining states, a color octet $\pi_8$ stands out as an unavoidable prediction, independent on the type of irreps in the model. For the case $\Ccolor$ this turns out to be the only one, for $\Rcolor$ and $\PRcolor$ there is an additional color sextet and triplet respectively. The full list of pNGBs is given in Table~\ref{pNGBcontent}.

The relative model independence of these three pseudo-scalars (the $a$, $\eta'$ and $\pi_8$) and the fact that they have a direct coupling to gluons via the WZW anomaly, implying a larger cross section as compared to e.g. the pNGBs in the electro-weak sector, are the reasons why we focus on them in this work. They are indeed ``standard candles'' that will allow to falsify these models with the minimal number of additional assumptions.

We conclude this section by presenting in Table~\ref{allmodels} the complete list of models  that are the focus of this work.

\begin{table}
\begin{center}
{\footnotesize
  \begin{tabular}{|c|c|c|c|c|c|c|c|}
    \hline
    $\GHC$ & $ \psi$ & $ \chi$ & Restrictions & $\,-q_\chi/q_\psi$ & $\,Y_\chi\,$ & Non Conformal & Model Name\\
    \hline
      \multicolumn{1}{c}{} &    \multicolumn{1}{c}{Real} & \multicolumn{1}{c}{Real} & \multicolumn{3}{c}{SU(5)/SO(5) $\times$ SU(6)/SO(6)} \\
      \hline
    $SO(\NHC)$ & $5\times \Sy_2$ & $6\times \Fu$ & $\NHC \geq 55$ & $\frac{5(\NHC+2)}{6}$ &$1/3$ & /& \\
    \hline
    $SO(\NHC)$ & $5\times \Ad$ & $6\times \Fu$ & $\NHC \geq 15$ &  $\frac{5(\NHC-2)}{6}$ & $1/3$ & /& \\
    \hline
    $SO(\NHC)$ & $5\times \Fu$ & $6\times \Sp$ & $\NHC=7,9$ & $\frac{5}{6}\,,\,\frac{5}{12}$ & $1/3$ & $\NHC=7,9$ & M1, M2\\
    \hline
    $SO(\NHC)$ & $5\times \Sp$ & $6\times \Fu$ & $\NHC = 7,9 $  & $\frac{5}{6}\,,\,\frac{5}{3}$ & $2/3$ & $\NHC = 7,9 $ & M3, M4\\
    \hline
     \multicolumn{1}{c}{} &    \multicolumn{1}{c}{Real} & \multicolumn{1}{c}{Pseudo-Real} & \multicolumn{3}{c}{SU(5)/SO(5) $\times$ SU(6)/Sp(6)} \\
     \hline
    $Sp(2 \NHC)$ & $5\times \Ad$ & $6\times \Fu$ & $2 \NHC \geq 12$ & $\frac{5(\NHC+1)}{3}$ & $1/3$ & / &\\
    \hline
    $Sp(2 \NHC)$ & $5\times \An_2$ & $6\times \Fu$ & $2 \NHC \geq 4$  & $\frac{5(\NHC-1)}{3}$ &$1/3$ &$2 \NHC = 4$& M5\\
    \hline
    $SO(\NHC)$ & $5\times \Fu$ & $6\times \Sp$ & $\NHC=11,13$  & $\frac{5}{24}\,,\,\frac{5}{48}$ & $1/3$ & /&\\
     \hline
     \multicolumn{1}{c}{} &    \multicolumn{1}{c}{Real} & \multicolumn{1}{c}{Complex} & \multicolumn{3}{c}{ SU(5)/SO(5) $\times$ SU(3)$^2$/SU(3)} \\
     \hline
    $SU(\NHC)$ & $5\times \An_2$ & $3\times (\Fu, \overline{\Fu})$ & $\NHC = 4$   & $\frac{5}{3}$ & $1/3$ & $\NHC = 4$&M6\\
    \hline
    $SO(\NHC)$ & $5\times \Fu$ & $3\times (\Sp, \overline{\Sp})$ & $\NHC = 10,14$  & $\frac{5}{12}\,,\,\frac{5}{48}$ & $1/3$ &$\NHC = 10$&M7\\
    \hline
     \multicolumn{1}{c}{} &    \multicolumn{1}{c}{Pseudo-Real} & \multicolumn{1}{c}{Real} & \multicolumn{3}{c}{ SU(4)/Sp(4) $\times$ SU(6)/SO(6)} \\
     \hline
    $Sp(2 \NHC)$ & $4\times \Fu$ & $6\times \An_2$ & $2 \NHC \leq 36$  & $\frac{1}{3(\NHC-1)}$ & $2/3$ & $2\NHC = 4$&M8\\
    \hline
    $SO(\NHC)$ & $4\times \Sp$ & $6\times \Fu$ & $\NHC = 11,13$  & $\frac{8}{3}\,,\,\frac{16}{3}$ & $2/3$ & $\NHC = 11$&M9\\
    \hline
     \multicolumn{1}{c}{} &    \multicolumn{1}{c}{Complex} & \multicolumn{1}{c}{Real} & \multicolumn{3}{c}{ SU(4)$^2$/SU(4) $\times$ SU(6)/SO(6)} \\
     \hline
    $SO(\NHC)$ & $4\times (\Sp,  \overline{\Sp})$ & $6\times \Fu$ & $\NHC = 10$  & $\frac{8}{3}$ & $2/3$ & $\NHC = 10$&M10\\
    \hline
    $SU(\NHC)$ & $4\times(\Fu,  \overline{\Fu})$ & $6\times \An_2$ & $\NHC = 4$  & $\frac{2}{3}$ & $2/3$ & $\NHC = 4$&M11\\
    \hline
     \multicolumn{1}{c}{} &    \multicolumn{1}{c}{Complex} & \multicolumn{1}{c}{Complex} & \multicolumn{3}{c}{ SU(4)$^2$/SU(4) $\times$ SU(3)$^2$/SU(3)} \\
     \hline
    $SU(\NHC)$ & $4\times (\Fu, \overline{\Fu})$ & $3\times (\An_2, \overline{\An}_2)$ & $\NHC \geq 5$ & $\frac{4}{3(\NHC-2)}$ & $2/3$  & $\NHC = 5$&M12\\
    \hline
    $SU(\NHC)$ & $4\times (\Fu, \overline{\Fu})$ & $3\times (\Sy_2, \overline{\Sy}_2)$ & $\NHC \geq 5$  & $\frac{4}{3(\NHC+2)}$ & $2/3$ & /&\\
    \hline
    $SU(\NHC)$ & $4\times (\An_2, \overline{\An}_2)$ & $3\times (\Fu, \overline{\Fu})$ & $\NHC = 5$  & $4$ & $2/3$ & /&\\
    \hline
  \end{tabular}}
\end{center}
  \caption{Models of interest in this paper. ``Restrictions'' denotes requirements such as asymptotic freedom and compatibility with the reality properties of the irrep. ``Non Conformal'' indicates the sub-range for which the model is likely outside of the conformal region: a ``/'' indicates that there are no solutions, i.e. all models are likely conformal. $-q_\chi/q_\psi$ is the ratio of charges of the fermions under the non anomalous $U(1)$ combination. $\Fu, \An_2, \Sy_2, \Ad \mbox{ and } \Sp$ denote the fundamental, two-index antisymmetric,  two-index symmetric, adjoint and spinorial irreps respectively. A bar denotes the conjugate irrep.} 
  \label{allmodels}
\end{table}

We split the table according to the reality properties of the irreps, from which the pNGB coset can be read-off. We also indicate the range of hypercolors for which the theory is likely to be \emph{outside} of the conformal region~\footnote{It is notoriously difficult to exactly characterize the conformal region of non-supersymmetric gauge theories outside of the perturbative regime. There are however some heuristic arguments and, luckily, most of the models in Table~\ref{allmodels} are rather clear-cut cases~\cite{Ferretti:2016upr}. There has also been intensive work on the lattice, reviewed in~\cite{DeGrand:2015zxa}, with some more recent related contributions in~\cite{Fodor:2016zil,Hasenfratz:2016gut}, but unfortunately a universal consensus has not yet been reached.}. In fact, the mechanism of partial compositeness relies on the fact that the theory is conformal in the UV, so that a large anomalous dimension for the operator corresponding to the fermionic bound state can, in principle, be generated. This large anomalous dimension would allow to decouple the scale of flavor symmetry breaking and the EW scale. The compositeness scale $\Lambda$ then is identified with the scale where conformal invariance is broken explicitly.

One possible philosophy is to view the compositeness scale $\Lambda$ as the scale in which some hyperfermions acquire a mass and the theory exits the conformal window due to the reduced number of fermionic matter. This mechanism has recently been tested on the Lattice for a multi-flavor QCD-like theory~\cite{Brower:2015owo,Hasenfratz:2016gut}. With this interpretation, the promising models are those which are \emph{not} conformal and yet contain enough light fermions to allow for the construction of a custodial coset for EW symmetry breaking as well as top-partners. These models can then be simply brought into the conformal window by adding additional fermions of mass $\approx \Lambda$, possibly in the same irreps already used. Another possible philosophy is to rely on the top couplings responsible for partial compositeness: the operator responsible for the linear mixing grows in the IR due to the large anomalous dimensions, thus it breaks the conformal invariance when its coefficient becomes relevant. We will however rely on the former scenario.

\begin{table}
\begin{center}
{\footnotesize
\begin{tabular}{|c|c|}
\hline
Electro-weak coset  & $SU(2)_L \times U(1)_Y$\\
\hline
$\Rcoset$ & ${\mathbf 3}_{\pm 1}  + {\mathbf 3}_0  + {\mathbf 2}_{\pm 1/2}  + {\mathbf 1}_0 $\\
$\PRcoset$ & $ {\mathbf 2}_{\pm 1/2}  + {\mathbf 1}_0 $ \\
$\Ccoset$ & $ {\mathbf 3}_0 + {\mathbf 2}_{\pm 1/2} +  {\mathbf 2'}_{\pm 1/2} + {\mathbf 1}_{\pm
1}  + {\mathbf 1}_0  + {\mathbf 1'}_0 $ \\
\hline\hline
Color coset  & $SU(3)_c \times U(1)_Y$\\
\hline
$\Rcolor$ & ${\mathbf 8}_{0} +{\mathbf 6}_{(-2/3\mbox{ or } 4/3)} + \bar {\mathbf 6}_{(2/3 \mbox{ or } -4/3)}$\\
$\PRcolor$ & ${\mathbf 8}_{0} +{\mathbf 3}_{2/3} + \bar {\mathbf 3}_{-2/3}$\\
$\Ccolor$ & ${\mathbf 8}_{0}$\\
\hline
\end{tabular}}
\end{center}
\caption{The SM quantum numbers of the pNGBs appearing in the models of Table~I in addition to the ubiquitous $a$ and $\eta'$. The Electro-weak coset arises from the condensation of $\psi$ while the Color one from $\chi$. The sextets can have two possible charge assignments depending on whether the top-partners are realized as ``$\chi\psi\chi$'' or ``$\psi\chi\psi$''.}
  \label{pNGBcontent}
\end{table}

We would like to end this section by commenting on the possible symmetry breaking patterns for these theories. First of all, all models in Table~\ref{allmodels} are ``vector-like'' in the sense that a gauge invariant mass term can be added for every fermion. This implies, by the Vafa-Witten argument~\cite{Vafa:1983tf}, that the HC group remains unbroken and thus a $\langle \psi\chi\rangle$ condensate never forms.

As far as the condensation of each separate pair $\langle \psi\psi\rangle$ and $\langle \chi\chi\rangle$ goes, there is also the logical possibility of the presence of massless composite fermions in the IR matching the 't~Hooft anomaly \cite{tHooft:1980xb} of the chiral global symmetry and thus preempting the need for symmetry breaking. This possibility has been suggested as the reason behind the lightness of top partners in~\cite{Katz:2003sn,Cacciapaglia:2015vrx}. By invoking the persistent mass condition, however, we find this last scenario unlikely. In all the models classified as non-conformal in Table~\ref{allmodels}, the only possible hypercolor invariant fermionic bound states composed of at most three elementary fields must contain at least one $\psi$ and one $\chi$ fermion. Giving a common mass to one type of fermions (e.g. $\psi$) renders all the fermionic bound states massive. However, the other type of fermion (e.g. $\chi$) is still massless and with non vanishing `t~Hooft anomaly. Since such an anomaly cannot be canceled by the composite states, the corresponding symmetry must be spontaneously broken. Reversing the role of the fermions we reach the same conclusion for the other coset.
We point out that this argument is not rigorous. Its main weaknesses are the possible existence of phase transitions~\cite{Preskill:1981sr}, invalidating the massless limit, as well as the fact that we are ignoring bound states composed by five or more fundamental fermions, which can sometimes be formed using only one fermion species. We find it however sufficiently convincing to assume that both condensates form, a necessary condition for the existence of the pNGBs considered in this work.

\section{Properties of the $U(1)$ singlets and the octet \label{sec:PNGBtheory}}

\subsection{Chiral Lagrangian}

In this section we discuss in detail how an effective Lagrangian formalism can be used to describe the properties of the singlets associated to the global $U(1)$'s.
A chiral perturbation theory for the class of models of interest has been recently presented in Ref.~\cite{DeGrand:2016pgq}, including the singlet -- referred as $a$ in our work -- associated with the non-anomalous U(1). Here, we want to be more general and keep both states $a$ and $\eta'$ in the low energy Lagrangian, as the mass generated for the anomalous current may be not very large. 

As the model contains two fermion condensates, the chiral Lagrangian can be described in terms of two copies of the pNGB matrix $\Sigma_r$ and two singlets $\Phi_r$, where $r = \psi, \chi$. The $\Sigma_r$'s contain the pNGBs from the non-abelian cosets, while $\Phi_r$'s contain the singlets. Furthermore, we want to choose the normalization of the decay constants $f_r$'s in such a way that the mass of the $W$ (and $Z$) bosons can be written as
\beq
m_W = \frac{g}{2} f_\psi \sin \theta\,,
\eeq
where $\theta$ is an angle describing the misalignment of the vacuum \cite{Kaplan:1983fs} (thus, $\sin \theta = 1$ represents the ``Technicolor'' limit of the theory, where $f_\psi = v_{\rm SM} = 246$~GeV). In this way, we can define the ratio
\beq
\epsilon = \frac{v_{\rm SM}^2}{f_\psi^2} = \sin^2 \theta
\eeq
as a measure of the fine tuning needed in the alignment of the vacuum.
The presence of the parameter $\epsilon$ characterizes the main advantage of models of this type compared to earlier Technicolor models. The $S$-parameter has an additional suppression by a factor $\epsilon$ circumventing EW precision tests albeit at the price of some fine-tuning.

This notation has the additional advantage that the Higgs couplings to the vector bosons are the same for all cosets and are, in fact, the same as those of the minimal coset $SO(5)/SO(4)$ (for which EW precision tests and Higgs couplings generically require $\epsilon\lesssim0.1$~\cite{Bellazzini:2014yua,Panico:2015jxa}, or equivalently $f_\psi \gtrsim 800$~GeV). However, this forces us to normalize the chiral lagrangian differently depending on the nature of the $\psi$ irrep. To allow us to write a common expression for all cases, we introduce the quantity
\beq
c_5 = \left\{ \begin{array}{l} \sqrt{2}\quad \mbox{for $\psi$ real}\,, \\
1 \quad \mbox{elsewhere}\,; \end{array} \right.
\eeq
in terms of which
\beq
\Sigma_r = e^{i 2 \sqrt{2} c_5 \pi^a_r T^a_r/f_r} \cdot \Sigma_{0,r}\,, \quad \Phi_r = e^{i c_5 a_r/f_{a_r}}\,,
\eeq
where $T^a_r$ are the non-abelian generators in the fundamental irrep normalized so that $\mbox{Tr} [T_r^a T_r^b] = \delta^{ab}/2$, $f_r$ and $f_{a_r}$ are the decay constants for the non abelian pions and the singlets respectively. The matrix $ \Sigma_{0,r}$ is the gauge-preserving vacuum\footnote{In this approach, the EW symmetry breaking arises from the pNGB corresponding to the Brout-Englert-Higgs doublet developing a vacuum expectation value. This effect can also be seen as a misalignment of the vacuum with respect to the gauged generators. We chose the former approach for simplicity.}.

Following this convention, the lowest order chiral Lagrangian can be written as:
\beq
 \mathcal{L}_{\chi pt} = \sum_{r=\psi, \chi} \frac{f_r^2}{8 c_5^2}\ \mbox{Tr} [(D_\mu \Sigma_r)^\dagger (D^\mu \Sigma_r)]+ \frac{f_{a_r}^2}{2 c_5^2}\ (\partial_\mu \Phi_r)^\dagger (\partial^\mu \Phi_r)\,.
\eeq
Notice that we chose the same normalization (driven by the nature of the $\psi$ irrep) for both cosets, in order to simplify the notation for the abelian pNGBs later.

A few comments are in order at this stage: for the singlets, the lowest order operator simply gives a kinetic term which does not depend on $f_{a_r}$. However, the couplings of $a_r$ will always be generated by the couplings of the $U(1)$ currents to the underlying fermions, which depend on an arbitrary parameter, i.e. the charge $Q_r$ of the fermions under the global $U(1)$. This consideration justifies why the decay constants $f_r$ and $f_{a_r}$ are, in principle, unrelated. In the following, we fix the decay constants by choosing $Q_r = 1$ for $r=\psi, \chi$.
A stronger relation between the decay constants of the singlets and the non-abelian pions in each sector can only be drawn assuming that both are dominantly made of di-fermion states. In QCD, this situation is achieved in the large-$N_c$ limit~\cite{Veneziano:1979ec}, following from Zweig's rule, where the singlet associated to the anomalous U(1) is also expected to become light. All mesons can therefore be described by a single meson matrix $\Phi_r^2 \Sigma_r$ (the $\Phi^2_r$ comes from the fact that the condensate has charge 2). The chiral Lagrangian, then, looks like
\beq
\mathcal{L}_{\chi pt} = \sum_{r=\psi, \chi} \frac{f_r^2}{8 c_5^2}\ \mbox{Tr} [(D_\mu \Phi_r^2 \Sigma_r)^\dagger (D^\mu \Phi_r^2 \Sigma_r)]\,,
\eeq
which is consistent with the above formulation for $f_{a_r} = \sqrt{N_r} f_r$, $N_r$ being the dimension of the flavour matrix $\Sigma_r$ ($N_\psi = 4$ for SU(4)/Sp(4) and SU(4)$\times$SU(4)/SU(4), $N_\psi = 5$ for SU(5)/SO(5), $N_\chi = 6$ for SU(6)/Sp(6) and SU(6)/SO(6), and $N_\chi = 3$ for SU(3)$\times$SU(3)/SU(3)).
In the following, we will be interested in cases like the large-$N_c$ limit of QCD where both singlets can be light, so that we introduce the parameters
\beq
\xi_r = N_r \frac{f_r^2}{f_{a_r}^2}\,,
\eeq
which should be equal to 1 in the large-$N_c$ limit. Note that corrections to this relation will be generated by loop corrections in the chiral Lagrangian~\cite{Bijnens:1988kx,Kaiser:1998ds}.

Out of the 2 singlets we introduced, only one remains a pNGB because it is associated to the anomaly-free combination of U(1)'s. If $q_\psi$ and $q_\chi$ are the charges associated to the anomaly-free current, defined in Eq.(\ref{eq:qpsiqchi}), the pNGB gauge eigenstates, $\tilde{a}$ and the anomalous $\tilde{\eta}'$, can be defined as
\beq
\tilde{a} = \frac{q_\psi f_{a_\psi} a_\psi + q_\chi f_{a_\chi} a_\chi}{\sqrt{q_\psi^2 f_{a_\psi}^2 + q_\chi^2 f_{a_\chi}^2}}\,, \quad \tilde{\eta}' = \frac{q_\psi f_{a_\psi} a_\chi - q_\chi f_{a_\chi} a_\psi}{\sqrt{q_\psi^2 f_{a_\psi}^2 + q_\chi^2 f_{a_\chi}^2}}\,.
\eeq
For later convenience, we  define a single dimensionless parameter describing this basis, i.e. an angle $\zeta$:
\beq
\tan \zeta = \frac{q_\chi f_{a_\chi}}{q_\psi f_{a_\psi}}\,. 
\eeq
Note that all physical observables will only depend on ratios of the two charges $q_r$ .
The values of $q_\chi/q_\psi$ for the various models are listed in Table~\ref{allmodels}, always leading to values of $\tan \zeta < 0$ (for which we define the angle in the rage $-\pi/2 < \zeta < 0$ in the rest of the paper).

\subsection{Couplings within the strong sector} \label{sec:3_couplings}

The couplings of the singlets can only be generated by terms explicitly breaking the global symmetries. The partial gauging of the non-abelian global symmetries cannot do the job, as the gauged generators are not charged under the U(1)'s. (For recent lattice results on the case of charged pNGBs see~\cite{DeGrand:2016htl}. Even more recently, a detailed analysis of the reach of the LHC in the search for the doubly charged pNGB present in the $SU(5)/SO(5)$ models has been given in~\cite{Englert:2016ktc}.)
If a mass term for the underlying fermions is added, however, it necessarily carries the U(1) charge of the specific fermion.
Following~\cite{DeGrand:2016pgq}, we add the fermion masses in the Lagrangian as follows:
\begin{multline} \label{eq:Lmass}
\mathcal{L}_{m} = \sum_{r=\psi, \chi} \frac{f_r^2}{8 c_5^2}\ \Phi_r^2 \mbox{Tr} [X_r^\dagger \Sigma_r] + h.c. = \sum_{r=\psi,\chi}  \frac{f_r^2}{4 c_5^2}\ \left[ \cos \left( 2 c_5 \frac{a_r}{f_{a_r}}\right) \mbox{Re} \mbox{Tr} [X_r^\dagger \Sigma_r] \right. \\
\left. - \sin \left( 2 c_5 \frac{a_r}{f_{a_r}}\right) \mbox{Im} \mbox{Tr} [X_r^\dagger \Sigma_r] \right]\,.
\end{multline}
The spurions $X_r$ are related to the fermion masses linearly
\beq
X_r = 2 B_r m_r \qquad r = \psi, \chi\,,
\eeq
where $B_r$ is a dimensional constant (that can, in principle, be calculated on the Lattice).
Note that, without loss of generality, $m_r$ is a real matrix in the non-abelian flavour space of the fermion specie $r$.
From the above expressions, we can read off the masses of the singlets and non-abelian pions:
\beq \label{eq:GMOR}
\left( m_{\pi_r}^2\right)^{ab} = 4 B_r\ \mbox{Tr} [T^a_r T^b_r \Sigma_{0,r} m_r]\,, \quad
m_{a_r}^2 = 2 \frac{f_r^2}{f_{a_r}^2} B_r\ \mbox{Tr} [\Sigma_{0,r} m_r]\,.
\eeq
In the limit where the condensates are aligned with the mass matrices $m_r = \mu_r \Sigma_{0,r}^\dagger$, which corresponds to the EW preserving vacuum and where $\mu_r$ is a common mass for all underlying fermions, the masses simplify to (all the non abelian pions having the same mass)
\beq
m_{\pi_r}^2 = 2 B_r \mu_r\,, \quad m_{a_r}^2 = 2 N_r \frac{f_r^2}{f_{a_r}^2} B_r \mu_r = \xi_r\ m_{\pi_r}^2\,,
\eeq
where $N_r$ is the dimension of the matrix $\Sigma_r$. We recover the result that in the large-$N_c$ limit, the masses of all mesons are equal as $\xi_r = 1$.

We also note that Eq.~(\ref{eq:Lmass}) contains linear couplings of the singlets to the non-abelian pions:
\beq \label{eq:Lmix}
\mathcal{L}_m \supset -  \frac{f_r^2}{2c_5 f_{a_r}} a_r \mbox{Im} \mbox{Tr} [\Sigma_r X^\dagger_r]\,,
\eeq
which potentially include mass mixing terms between the singlet and the non-abelian pions. In the limit where both vacuum and mass matrices are aligned with the EW preserving direction, the expression simplifies to
\beq
\mathcal{L}_m \supset - B_r \mu_r\ \frac{f_r^2}{f_{a_r} c_5} a_r \mbox{Im} \mbox{Tr} [e^{i 2 \sqrt{2}c_5 \frac{\pi_r^a}{f_r} T^a_r}] = \frac{\sqrt{2} c_5^2 m_{\pi_r}^2}{3 f_r f_{a_r}}\ a_r \sum_{abc} d^{abc} \pi_r^a \pi_r^b \pi_r^c + \dots\,,
\eeq
where the dots include terms with more fields and $d^{abc} = 2 \mbox{Tr} [T_r^a \{T_r^b, T_r^c\} ]$ is a fully-symmetric tensor.
The presence of mixing with or couplings to other non-abelian pions depends on the coset. In the EW sector, 3 possible cosets are allowed. For the coset SU(4)/Sp(4), we found that no mixing and no coupling is possible as the trace $\mbox{Tr} [\Sigma_\psi X^\dagger_\psi]$ is real. In the SU(4)$\times$SU(4)/SU(4) case, at leading order in $v/f_\psi$ no mixing is generated however a coupling to the triplets and to the second doublet is generated, allowing 2-body decays into these additional pions. This coupling can potentially affect the phenomenology of the singlet, if the additional pions are light enough. In the SU(5)/SO(5) case, we found that a mass mixing with all neutral pseudo-scalar is generated by the Higgs VEV at leading order. More details on such couplings can be found in the Appendix~\ref{app:coups}. Finally, in the color sector generated by the $\chi\chi$ condensate, we found that a coupling to 3 colored pions is present in the SU(6)/Sp(6) and SU(6)/SO(6) cases.

\subsection{Couplings to SM fermions}

The link of the strong dynamics to SM fermions is another source of explicit breaking of the global symmetries that may induce direct couplings of the singlets to fermions~\cite{Golterman:2015zwa}. To generate a mass for the top, the class of models we want to investigate implements partial compositeness, where the top mass is proportional to two linear mixings of the elementary fermions to composite states~\footnote{We use Dirac spinors in this subsection.}:
\begin{align}
\mathcal{L}_{mix}&\supseteq y_L\ \bar{q}_L \Psi_{q_L}+y_R\ \bar{\Psi}_{t_R} t_R+h.c.\,,
\end{align}
where $\Psi_{q_L/t_R}$ are fermionic composite operators that have the same quantum numbers as the left-handed and right-handed tops respectively, and which contain the top partners at low energy. As such operators are made of 3 fermions, they carry charge under the two U(1)'s: the couplings of the pions can then be recovered by assigning a charge to the pre-Yukawas $y_{L/R}$ that matches the one of the composite operators. Without loss of generality,
each spurion can be associated with a combination of pion matrices
\beq
y_{L} \to \Phi_\psi^{n_{L\psi}} \Phi_\chi^{n_{L\chi}} y_L\,,
\eeq
and similarly for $y_R$.
As $m_{\rm top} \sim y_L y_R$, the singlets decouple from the top quark as long as the charges of the two pre-Yukawas are opposite~\cite{Belyaev:2015hgo,Gripaios:2016mmi}. However, this situation can never be realized in the class of models under consideration. If both pre-Yukawas involve the same operator in terms of fundamental states, then the charges are the same as the U(1)'s are axial. The charge assignments depend on the structure of the fermionic bound states: if the fermion is of type ``$\psi \psi \chi$'', then the possible charges of the spurions and of the top mass are~\footnote{The various assignments correspond to the following operators: (2,1) for $\psi \psi \chi$, (-2,1) for $\bar{\psi} \bar{\psi} \chi$ and (0,-1) for $\bar{\psi} \psi \bar{\chi}$. Here we only focus on left-handed operators, which can be made of 3 left-handed fermions, or 2 right-handed and 1 left-handed one.}
\beq
 &y_{L(R)}\rightarrow (n_{L(R)\psi},n_{L(R)\chi}) = (\pm2,1)\,, \; (0,-1)\,  & \nonumber \\
& \Downarrow &  \\
 &m_{\rm top}\rightarrow
(n_{L\psi}+n_{R\psi},n_{L\chi}+n_{R\chi})\equiv
(n_\psi,n_\chi) =
(\pm 4,2)\,, \;\; (0,\pm 2)\,, \;\; (\pm 2,0)\,. & \nonumber
\eeq
For ``$\psi \chi \chi$'', it suffices to exchange the two charges.
We see that in no case the charge of the top mass can be zero for both singlets.
The couplings of the singlets to tops can therefore be written as
\beq\label{eq: top coupling in flavor basis}
\mathcal{L}_{top} = m_{\rm top} \Phi_\psi^{n_\psi} \Phi_\chi^{n_\chi}\ \bar{t}_L t_R + h.c. = m_{\rm top}\ \bar{t} t + i c_5 \left( n_\psi \frac{a_\psi}{f_{a_\psi}} + n_\chi \frac{a_\chi}{f_{a_\chi}}\right) m_{\rm top}\ \bar{t} \gamma^5 t + \dots
\eeq
Changing basis to $\tilde{a}$ and $\tilde{\eta}'$, the couplings read
\beq
i c_5 \frac{m_{\rm top}}{\sqrt{q_\psi^2 f_{a_\psi}^2 + q_\chi^2 f_{a_\chi}^2}} \left( (n_\psi q_\psi + n_\chi q_\chi)\ \tilde{a} + \left( n_\chi q_\psi \frac{f_{a_\psi}}{f_{a_\chi}} - n_\psi q_\chi \frac{f_{a_\chi}}{f_{a_\psi}} \right) \tilde{\eta}' \right) \bar{t} \gamma^5 t\,,
\eeq
where we recognize that the couplings of the pNGB $\tilde{a}$ is proportional to the charge under the non-anomalous U(1). Note, however, that the reasoning above is only valid if the operators $\Psi_{q_L}$ and $\Psi_{t_R}$, that mix to the top, have definite charges, i.e. they correspond to a well defined combination of hyperfermions. In general, as different operators transform in the same way under the global symmetries, mixing among operators is possible.

In this class of composite Higgs models, the matter content of the confining sector cannot accommodate enough partners to realize partial compositeness for all fermions: the Yukawa couplings of the light fermions must therefore come from a different operator. A simple possibility \cite{Eichten:1979ah} is to introduce couplings of SM bilinears $\bar{f}f$ with the strong sector:
\begin{align}
\frac{y_{bil}}{\Lambda_F^{2}}\ \bar{f}f \bar{\psi}\psi\,,
\end{align}
where $y_{bil}\sim m_f$ and the flavour scale $\Lambda_F$ can be much higher than the condensation scale. While these operators are generically irrelevant, they can be large enough to reproduce light quark masses, and suppressed enough to evade flavour bounds~\cite{Matsedonskyi:2014iha,Cacciapaglia:2015dsa,Ferretti:2016upr}. Another possibility would be that the masses of light fermions are generated at higher scale, possibly via partial compositeness~\cite{Panico:2016ull}.  The U(1) symmetries can be formally restored promoting $y_{bil}$ to be a spurion only charged under U(1)$_\psi$, and this implies a low energy coupling proportional to
\beq
m_f \Phi_\psi^2\ \bar{f}_L f_R + h.c. = m_f\ \bar{f} f + 2 i c_5  \frac{m_f}{f_{a_\psi}}\ a_\psi \ \bar{f} \gamma^5 f + \dots
\eeq
This coupling has the same form as the one we derived for the top, but with fixed charges $n_\psi = 2$ and $n_\chi = 0$.

\subsection{Masses and Mixing of the Singlets}

The masses for the singlets are generated by the masses of the underlying fermions, $m_\psi$ and $m_\chi$, and the instanton related to the anomalous current. Even though couplings to tops and light fermions exist, they do not lead to corrections to the mass of the singlets. One way to see this is that all loops of fermions will be proportional to the absolute value of the spurions in order to write an operator which is gauge invariant. Thus, the dependence on the singlet pions, which comes in via exponentials, vanishes.

The mass matrix for the singlets, therefore, can be written from
\beq \label{eq:mass1}
- \mathcal{L}_{\rm mass} = \frac{1}{2} m_{a_\chi}^2 a_\chi^2 + \frac{1}{2} m_{a_\psi}^2 a_\psi^2 + \frac{1}{2} M^2_A (\cos \zeta a_\chi - \sin \zeta a_\psi)^2
\eeq
where $M_A^2$ is the mass generated by instanton effects, proportional to the topological susceptibility of the hypercolor group, for the singlet $\tilde{\eta}'$ associated with the anomalous combination of $U(1)$'s. For now, we will consider it as a free parameter, even though the topological mass is, in principle, calculable once the underlying dynamics is specified.

In the following, we want to entertain the case where the topological mass may be small, as it happens in large-$N_c$ QCD~\cite{Witten:1979vv,Veneziano:1979ec}. In fact, in many of the models we consider the number of colors is large and/or the representation of the underlying fermions is large. Another physical consideration allows us to simplify the mixing structure: the mass of the pNGBs in the EW sector, due to the condensation of the $\psi$'s, also contributes to the mass of the SM-like Higgs boson. Thus, its value is constrained to be small in order to minimize the fine tuning in the Higgs mass. While the details depend on the specific model, some general considerations are in order. The mass term can be used to stabilize the Higgs potential against the contribution of the top loops and obtain a small misalignment in the vacuum~\cite{Galloway:2010bp,Cacciapaglia:2014uja}. In such cases, one would expect $m_{\pi_\psi} \sim f_\psi$. Alternatively, if the top partners are light enough, their contribution to the Higgs potential is also enough to stabilize it and give the correct value of the Higgs mass~\cite{Matsedonskyi:2012ym,Marzocca:2012zn,Pomarol:2012qf}. In this case, therefore, one would require that the contribution of the fermion mass were small, i.e. $m_{\pi_\psi} \ll f_\psi$. This situation contrasts with the coset generated by $\chi$: here, colored pNGBs are expected and the strong constraints from searches at the LHC require their masses to be close to the TeV scale~\cite{Cacciapaglia:2015eqa}. It is thus natural to expect that $m_{\pi_\psi} \ll m_{\pi_\chi}$.\footnote{The pNGB masses are related to the hyperquark masses $m_\psi$ and $m_\chi$ by the usual Gell-Mann Oakes-Renner relation, see Eq.~(\ref{eq:GMOR}). The hierarchy $m_{\pi_\psi} \ll m_{\pi_\chi}$ can be obtained by choosing $m_{\psi} \ll m_{\chi}$.} In the following, we will work under this assumption and, for simplicity, neglect the contribution of $m_{a_\psi}$~\footnote{To restore the dependence on $m_{a_\psi}$ it is sufficient to replace $m_{a_\chi}^2 \to m_{a_\chi}^2 - m_{a_\psi}^2$, $m_a^2 \to m_a^2 - m_{a_\psi}^2$ and $m_{\eta'}^2 \to m_{\eta'}^2 - m_{a_\psi}^2$ in all the formulas in this section, as long as $m_{a_\psi} < m_{a_\chi}$.}.

We will first diagonalize the mass matrix from Eq.~(\ref{eq:mass1}), after setting $m_{a_\psi} = 0$.
We define the mass eigenstate as
\beq
\left( \begin{array}{c} a \\ \eta' \end{array} \right) = \left( \begin{array}{cc}
\cos \alpha & \sin \alpha \\
- \sin \alpha & \cos \alpha
\end{array} \right) \left( \begin{array}{c} a_\psi \\ a_\chi \end{array} \right)
\eeq
with
\beq\label{eq:eigenm}
m^2_{a/\eta'} = \frac{1}{2} \left( M_A^2 + m_{a_\chi}^2  \mp \sqrt{M_A^4 + m_{a_\chi}^4 + 2 M_A^2\ m_{a_\chi}^2 \cos 2 \zeta}\right)\,.
\eeq
The mixing angle can be expressed in terms of the mass eigenvalues and the parameter $\zeta$ as
\beq
\tan \alpha = \tan \zeta \left( 1 - \frac{m_{\eta'}^2 + m_a^2 - \sqrt{(m_{\eta'}^2 - m_a^2)^2 - 4 m_{\eta'}^2\ m_a^2 \tan^{-2} \zeta}}{2m_{\eta'}^2}\right)\,. \label{eq:alphazeta}
\eeq
Note that for $m_a \ll m_{\eta'}$ ($m_{a_\chi} \ll M_A$), then $\alpha \sim \zeta$ and the mass eigenstates coincide with the pNGB and the anomalous combination, as expected.

The mass matrix depends on 3 independent parameters: 2 masses and the angle $\zeta$. It is convenient to trade the two masses for the mass eigenvalues which have a more direct physical meaning.
Thus, we can define a ``physical basis'' thanks to the following relations:
\beq
2 m_{a_\chi}^2&=& m_{\eta'}^2 + m_a^2 - \sqrt{(m_{\eta'}^2 - m_a^2)^2 - 4 m_{\eta'}^2\ m_a^2 \tan^{-2} \zeta}\,, \label{eq:machi}\\
2 M_A^2 &=& m_{\eta'}^2 + m_a^2 + \sqrt{(m_{\eta'}^2 - m_a^2)^2 - 4 m_{\eta'}^2\ m_a^2 \tan^{-2} \zeta}\,.
\eeq
However, there are constraints on the value of the physical masses. From the positivity of the argument of the square root in the above formulas, we can derive a lower bound on the mass difference:
\beq\label{eq: mass splitting}
m_{\eta'}^2 - m_a^2 \geqslant \frac{2 \cos \zeta}{1-\cos \zeta} m_a^2\,.
\eeq
From the equation above we can see that the two masses can be equal only  in the limiting cases $\zeta = \pm \pi/2$ and $\zeta = 0$, when the two U(1) pNGBs decouple: in the former, $a_\chi$ is identified with the non-anomalous U(1), while in the latter it is $a_\psi$. Note that the apparent divergence for $\zeta = 0$ is removed by the fact that $m_a = 0$ in that limit.
The value of the lighter mass is also a monotonically increasing function of $M_A$, thus it reaches the maximum value for $M_A \to \infty$:
\beq
0 < m_a^2 < m_{a_\chi}^2 \sin^2 \zeta\,.
\eeq
The above constraint has significant physical implications as, for models with low values of $\zeta$, it implies that the mass of the lightest singlet has to be much lighter than the condensation scale $f_\chi$, as $m_{a_\chi}$ cannot be much larger than $f_\chi$ without spoiling the validity of the chiral Lagrangian expansion.
It is also interesting to notice that the mixing angle $\alpha$ is bounded between:
\beq
|\tan \frac{\zeta}{2} | < |\tan \alpha| < |\tan \zeta|\,.
\eeq
The lower bound corresponds to the minimal splitting between the two mass eigenvalues, while $\alpha = \zeta$ is achieved in the decoupling of $\eta'$.

As already mentioned the topological mass term is in principle calculable in a given underlying theory. We can then extract a simple correlation between the mass mixing angle $\alpha$ and the mass of the lightest singlet $m_a$ for fixed topological mass $M_A$
\beq
\tan \alpha=\tan\zeta\left(1-\frac{m_a^2}{M_A^2\sin^2\zeta}\right)\,.
\eeq
From Eq.~\eqref{eq:eigenm} we can extract the allowed range for each mass eigenvalue when $m_{a_\chi}\leq M_A$, i.e. (cosine is taken to be positive)
\begin{equation}
0\leq m_a^2\leq M_A^2(1-\cos\zeta)\,,\quad M_A^2\leq m_{\eta^\prime}^2\leq M_A^2(1+\cos\zeta)\,.
\end{equation}
For $m_a\ll M_A$ we get the upper bound in Eq.~\eqref{eq:eigenm}, while for $m_a$ maximal we can saturate the lower bound. Such an additional constraint can be significant when the topological mass is not larger than a few TeV.

\subsection{Non-abelian pions: the octet}

Among the many non-abelian pions present in these models, there is a common player that appears necessarily in all cosets: a color octet from the $\chi \chi$ condensation. Independently on the representation of $\chi$ under the confining HC, the octet $\pi_8$ can be identified as a bound state of $\langle \chi_1 \chi_2 \rangle$, where $\chi_{1,2}$ are the fermions transforming like a QCD color triplet and anti-triplet respectively. Due to its ubiquitous presence, and the fairly large production cross sections one may expect at the LHC, in the following we will consider its phenomenology and possible connections with the properties of the singlets.

As a first connection, we note that its mass can be expressed in terms of the $\chi$-mass as
\beq
m_{\pi_8}^2 = m_{\pi_\chi}^2 + C_g\frac{3}{4}g_s^2\ f_\chi^2 = \frac{1}{\xi_{\chi}} m_{a_\chi}^2 + C_g\frac{3}{4}g_s^2\ f_\chi^2 \,,
\eeq
where the second term comes from loop corrections from QCD, and $C_g>0$ is an unknown order one number (the loop contribution is cut off at a scale $\Lambda\sim4\pi f_\chi$). This provides a link between the mass of the octet and the masses in the singlet sector: in fact, $m_{a_\chi}$ is related to the singlet masses by Eq.~(\ref{eq:machi}). We also recall that $\xi_\chi \sim 1$, as expected in the large $N_c$-limit in QCD.
In the limit of $m_{a}\ll m_{\eta'}$, where the lighter singlet reaches its maximal mass $m_a \sim m_{a_\chi} \sin \zeta$, we obtain
\beq
m_{\pi_8}^2 \sim \frac{m_a^2}{\xi_\chi \sin^2 \zeta} + C_g\frac{3}{4}g_s^2\ f_\chi^2\,.
\eeq
The relation above shows that typically we would expect the octet to be heavier than the light singlet pNGB, even if the color corrections were small.

The octet has also the possibility to couple to tops: like in the case of singlets, the presence or not of this coupling depends on the representation of the composite top partners under the global symmetries. As the octet pNGB is associated to the bound state $\langle \chi \chi \rangle$, which is also charged under the U(1)$_\chi$, it is straightforward to find a correlation between the effective charges of the top mass and the presence of a coupling with the octet. If the top mass has a effective charge $\pm 2$ under U(1)$_\chi$, as indicated in the previous section, then the effective operator generating the mass of the top needs to be ``dressed'' by the appropriate pNGB matrix $\Phi_\chi^2 \Sigma_\chi$. If the charge is $\pm 4$, then two matrices are needed: this can also be understood in terms of indices of the global symmetry that cannot be contracted in an invariant way (but need the breaking generated by the consensate). On the other hand, if the charge is zero, then it is not needed to couple $\Sigma_\chi$ to the top mass term, and a coupling to the octet is not necessarily present. One can thus find a nice correlation between the charges determining the coupling of the singlets to the tops, and the presence of an octet coupling.
If present, the coupling will have the form:
\beq \label{eq:8top}
m_t  \bar{t_L} (\Sigma_\chi)^{n_\chi/2} t_R + h.c. \sim m_t \bar{t} t + i \frac{n_\chi}{\sqrt{2}} c_5 \frac{m_t}{f_\chi}\ \pi_8^a\ \bar{t} \gamma^5 \lambda^a t + \dots
\eeq
where $\lambda^a$ are the Gell-Mann matrices, and we have omitted the other pNGB and singlets.
For the light quarks, if their masses are generated by 4-fermion interactions then no couplings to the octet pNGB are generated.

It should also be remarked that, contrary to the case of the singlet, the presence of top couplings will also generate corrections to the masses of the octet. Those contributions are more model dependent, as they crucially depend on the representations of the top partners, and are typically of the same order as the QCD corrections but expected to be negative: we refer the reader to~\cite{Cacciapaglia:2015eqa} for an example.

\subsection{Wess-Zumino-Witten terms}

The couplings of the singlets to the SM gauge bosons, generated by the WZW term, can be computed in a similar way as in QCD~\cite{Bijnens:1988kx}. Following the normalization adopted in this work, the couplings can be written as
\beq\label{eq:Lanom_singlet}
\mathcal{L}_{\rm WZW} \supset \frac{\alpha_A}{8 \pi} c_5 \frac{C_A^r}{f_{a_r}} \delta^{ab}\ a_r\ \varepsilon^{\mu \nu \alpha \beta} A^a_{\mu \nu} A^b_{\alpha \beta}\,,
\eeq
where
\beq
& C_A^r \delta^{ab} = 2 d_r \mbox{Tr} [S^a S^b]\,, \quad \mbox{for complex reps}\,, & \nonumber \\
& C_A^r \delta^{ab} = d_r \mbox{Tr} [S^a S^b]\,, \quad \mbox{for real/pseudo-real reps}\,, &
\eeq
and $d_r$ is the dimension of the rep $r$ of HC, and $S^{a,b}$ in the trace correspond to the gauged generators with gauge coupling $\alpha_A = g_A^2/(4\pi)$. The normalization of the gauged generators depends on the global group the gauge interactions are embedded in, so that their trace is not the same as for the generators of the flavor group. Specifically, we note that, in the cases of interest
\beq
& \mbox{Tr} [S^a S^b] = \delta^{ab} \,, \quad \mbox{for SU(5) ($\psi$) and SU(6) ($\chi$)}\,;  & \nonumber \\
& \mbox{Tr} [S^a S^b] = \frac{1}{2} \delta^{ab} \,, \quad \mbox{for all other cases}\,.  & \nonumber
\eeq
 For completeness and comparison, the WZW term for the non-abelian pions is
\beq \label{eq:WZWna}
\mathcal{L}_{\rm WZW} \supset \frac{\sqrt{\alpha_{A^b} \alpha_{A^c}}}{4 \sqrt{2} \pi} c_5 \frac{C_{A^bA^c}^r}{f_r} c^{abc}\ \pi_r^a\ \varepsilon^{\mu \nu \alpha \beta} A^b_{\mu \nu} {A}^c_{\alpha \beta}\,,
\eeq
where
\beq
C_{A^bA^c}^r c^{abc} = d_r \mbox{Tr} [T^a_\pi \{ S^b, S^c \}]\,
\eeq
for complex $r$, and there is an additional factor of $1/2$ for real/pseudo-real representations.

\subsubsection{Singlets}

\begin{table}[tb]
\begin{center}
\begin{tabular}{|l|ccc|ccc|}
\hline
$r$ & coset $\psi$& $C_W^\psi$ & $C_B^\psi$ & coset $\chi$ & $C_G^\chi$ & $C_B^\chi$ \\
\hline
complex & SU(4)$\times$SU(4)/SU(4) & $d_\psi$ & $d_\psi$ & SU(3)$\times$SU(3)/SU(3) & $d_\chi$ & $6 Y_\chi^2 d_\chi$ \\
real & SU(5)/SO(5) & $d_\psi$ & $d_\psi$ & SU(6)/SO(6) & $d_\chi$ & $6 Y_\chi^2 d_\chi$ \\
pseudo-real & SU(4)/Sp(4) & $d_\psi/2$ & $d_\psi/2$ &  SU(6)/Sp(6) & $d_\chi$ & $6 Y_\chi^2 d_\chi$ \\
\hline
\end{tabular}
\caption{Coefficients of the anomalous couplings of the singlets. $d_\psi$ and $d_\chi$ are the dimensions of the representation of the fermions under HC and $Y_\chi$ the hypercharge carried by $\chi$.}
\label{tab:reps}
\end{center}
\end{table}

The coefficients for the anomalous couplings of the two singlets are summarized in Table~\ref{tab:reps}, where we recall that $d_\psi$ and $d_\chi$ are the dimensions of the representation of the fermions under HC.
These numbers, calculated directly from the WZW term, have a simple physical interpretation.
In the EW sector described by $\psi$, up to a factor of $1/2$, the $C_W$ ($C_B$) coefficients count the number of Weyl spinors transforming as SU(2)$_L$ (SU(2)$_R$) doublets:  $d_\psi$ in the SU(4)/Sp(4) coset and $2 d_\psi$ in the other two cases. Furthermore, as the theory is symmetric under the custodial symmetry, the number of doublets is equal, leading to
\beq
C_B^\psi = C_W^\psi\,.
\eeq
Similarly, in the $\chi$ sector, the anomaly of QCD color is equal to half the number of SU(3)$_c$ triplets, which is $2 d_\chi$ in all cases. Furthermore,
\beq
C_B^\chi = 6 Y_\chi^2 C_G^\chi\,.
\eeq
Combining the two relations above, we can see that for both $a_\psi$ and $a_\chi$, the values of the anomalous couplings always satisfy the relation:
\beq
C_W = C_B - 6 Y_\chi^2 C_G\,,
\eeq
which only depends on the model-specific value of the hypercharge $Y_\chi$. This relation will also be respected by the coupling of any linear combination of the two singlets, thus also by the mass eigenstates. As $Y_\chi = 2/3$ or $1/3$, all the models under consideration have anomalous couplings lying on 2 universal lines
\beq
C_W = C_B - \frac{8}{3} C_G\,, \quad C_W = C_B - \frac{2}{3} C_G\,.
\eeq

\subsubsection{Color octet}

The anomalous coupling of the octet with the gluon field strength $G_{\mu\nu}^a$ and the hypercharge field strength $B_{\mu\nu}$ can be computed from Eq.(\ref{eq:WZWna}), and are 
\be \label{eq:WZWoctet}
     {\mathcal{L}} \supset \frac{\alpha_s c_5}{4 \sqrt{2} \pi f_\chi} \left( \frac{d_\chi}{2} d^{abc} \right)\  \pi_8^a \epsilon^{\mu\nu\rho\lambda} G_{\mu\nu}^b G_{\rho\lambda}^c +
      \frac{\sqrt{\alpha_s \alpha_Y} c_5}{4 \sqrt{2} \pi f_\chi} \left(2 d_\chi Y_\chi \delta^{ab} \right)\  \pi_8^a \epsilon^{\mu\nu\rho\lambda} G_{\mu\nu}^b B_{\rho\lambda}\,,
\ee
where $d^{abc} = \frac{1}{4} \mbox{Tr} [\lambda^a \{ \lambda^b, \lambda^c \}]$ and  $Y_\chi$ is the hypercharge assigned to the $\chi$ fermions, in agreement with \cite{Bai:2016czm}. The second term, coupling the color octet to a gluon and hypercharge gauge boson, will thus induce an effective coupling with a photon and one with a $Z$ boson.

Neglecting the mass of the $Z$ boson and using the color factors $(1/8)\sum_{abc} (d^{abc})^2 = 5/3$ and  $(1/8)\sum_{ab} (\delta^{ab})^2 = 1$, we find the following relations between partial widths in the 3 channels
\begin{equation*}
\Gamma_{gg}:\Gamma_{g\gamma}:\Gamma_{gZ}= \frac{1}{2} \frac{5}{3} \alpha_s^2:
4 Y_\chi^2 \alpha_s \alpha:4 Y_\chi^2 \alpha_s \alpha \tan^2\theta_W,
\end{equation*}
with the additional factor of $1/2$ in $\Gamma_{gg}$ being due to the indistinguishability of the gluons. This means that the ratios of branching ratios in di-boson final states only depend on the hypercharge assigned to the $\chi$'s, which has two possible assignments (see Table~\ref{allmodels}). The numerical values are thus reported in Table~\ref{octetBR}, where the coupling constants are evaluated at a mass scale of $1$~TeV. We see that while the decay to a $Z$ boson is always suppressed by a $\tan^2 \theta_W$ factor, the decay into a photon can be sizeable, especially for $Y_\chi = 2/3$, and will lead to interesting phenomenology \cite{Belyaev:1999xe}.

\begin{table}
{\footnotesize
  \begin{tabular}{|c|c|c|}
    \hline
       & $ \frac{ {\mathrm{BR}}(\pi_8\to g\gamma)}{{\mathrm{BR}}(\pi_8\to gg)}  $ & $ \frac{ {\mathrm{BR}}(\pi_8\to g Z)}{{\mathrm{BR}}(\pi_8\to gg)}$  \\
    \hline\hline
    $Y_\chi=  1/3$    &  $0.048$   & $0.014$  \\
    $Y_\chi= 2/3 $    &   $0.19$  &   $0.058$  \\
    \hline
  \end{tabular}}
  \caption{Values of ratios of BRs in di-bosons for the pseudo-scalar octet for a mass of $1$~TeV. The mass
  fixes the dependence due to the running of the strong gauge coupling,
  $\alpha_s(1{\rm TeV}) = 0.0881$ used for this evaluation.}
  \label{octetBR}
  \end{table}

\subsubsection{Top loop effects}\label{sec: top loops effects}

Due to the presence of couplings to fermions, loops of tops contribute to the decays of both the singlets and octet to gauge bosons via triangle loops.
The numerical impact of top loops compared to the WZW interactions crucially depends on the ratio of the couplings, but also on the mass of the pseudo-scalar. In fact, in the limit of large mass, the top loop amplitudes are suppressed by two powers of the top mass over the pseudo-scalar mass: one coming from the coupling itself and the other from a chirality flip of the fermionic line in the loop. Thus, we can expect the loop to become subleading for large masses. The complete results for the top loop amplitudes are reported in Appendix~\ref{app:toploops}.

Another important observation is that top loops are phenomenologically relevant only for large couplings to the top, in which case one would also expect that the decay rate is dominantly into tops. In such a case, the WZW couplings, with top loop corrections, are only important for the production cross section via gluon fusion. To illustrate this fact, we focus on the octet. The correction to the amplitude for gluon fusion production from the top loops from Eq.(\ref{eq:Mtoploop}) gives:
\beq
\mathcal{A} (gg \to \pi_8) = \mathcal{A}_{WZW}\  \left( 1 + \frac{2}{d_\chi} \frac{m_t^2}{m_{\pi_8}^2} f \left( \frac{m_{\pi_8}^2}{m_t^2}\right) \right)\,,
\eeq
where the function $f (x)$ is defined in Eq.(\ref{eq:fx}).
This correction can be compared to the ratio of partial width in tops and gluons (not including top loops):
\beq
\frac{\Gamma (\pi_8 \to t \bar{t})}{\Gamma_{WZW} (\pi_8 \to gg)} = \frac{192 \pi^2}{5 \alpha_s^2 d_\chi^2} \frac{m_t^2}{m_{\pi_8}^2} \sqrt{1-4 \frac{m_t^2}{m_{\pi_8}^2}}\,.
\eeq
Already from the numerical factors involved one can see that the partial width into tops dominates over the one into gluons well before the top loop corrections become relevant.
The same conclusion can be obtained for the singlet, unless the WZW amplitude is small: in this case, however, that particular channel is not relevant for the phenomenology.

\section{Phenomenology}\label{sec:pheno}

We now turn our attention to characterizing the LHC phenomenology of the singlets $a$ and $\eta'$ (that we collectively denote as $\pi_0$ in this section) and of the color octet $\pi_8$.
The experimental results coming from post ICHEP2016 data will be used to derive general constraints on the production cross sections that can be later applied to any of the specific models.

Our goal in this section is to be as model independent as possible. We thus introduce a common notation for the couplings of the various pseudo-scalars to vector bosons, with coefficients denoted by $\kappa_g$, $\kappa_W$, $\kappa_B$, and to tops, with coefficient $C_t$ and perform the analysis with this notation. In Section~\ref{sec:specific pheno} we show how to relate these coefficients with the model-specific ones computed in Section~\ref{sec:PNGBtheory} and obtain model-specific bounds.

\subsection{Phenomenology of the singlet pseudo-scalars} \label{sec:SingPheno}

As we discussed in the previous section, the singlet pseudo-scalars couple to a pair of SM gauge bosons via the WZW anomaly terms, and to a pair of top quarks.
The generic effective Lagrangian for a SM neutral pseudo-scalar $\pi_0$ can be written as 
\begin{align}\label{eq:Lsigma}
\begin{split}
\mathcal{L}_{\pi_0} =&\frac{1}{2}\left(\partial_\mu\pi_0\partial^\mu\pi_0 - M^2_{\pi_0} \pi_0^2\right)+i\ C_t\frac{m_t}{f_\pi}\ \pi_0\ \overline{t}\gamma_5 t\\
&+ \frac{\alpha_s}{8 \pi} \frac{\kappa_g}{f_\pi}\ \pi_0 \left(\epsilon^{\mu\nu\rho\sigma} G^a_{\mu\nu}G^a_{\rho\sigma} + \frac{g^2}{g^2_s}\frac{\kappa_W}{\kappa_g}\epsilon^{\mu\nu\rho\sigma} W^i_{\mu\nu}W^i_{\rho\sigma}  + \frac{{g'}^2}{g^2_s}\frac{\kappa_B}{\kappa_g}\epsilon^{\mu\nu\rho\sigma} B_{\mu\nu}B_{\rho\sigma}  \right)\,,
\end{split}
\end{align}
which is characterized by five parameters: the mass $M_{\pi_0}$, the dimension-full coupling $\kappa_g/f_\pi$ (coefficient of the anomalous coupling to gluons) that controls the production cross section, and the three ratios $C_t/\kappa_g$, $\kappa_B/\kappa_g$ and $\kappa_W/\kappa_g$ which dictate the branching ratios. 

\begin{figure}[t]
\includegraphics[width=0.5\textwidth]{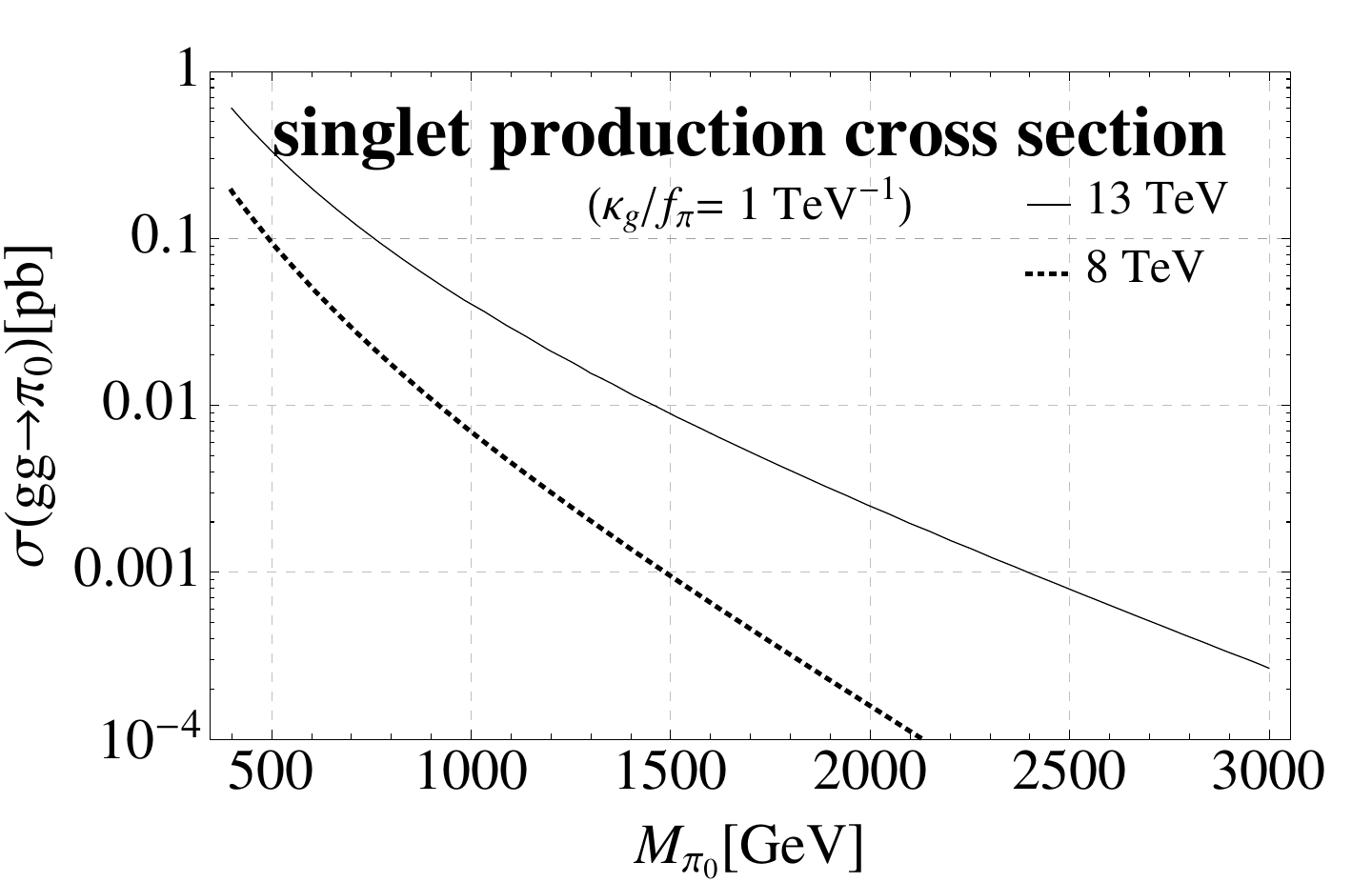} 
\label{fig:cs_sigma}
\caption{Production cross section of a pseudo-scalar $\pi_0$ with coupling $\kappa_g / f_\pi = 1\mbox{ TeV}^{-1}$ from gluon fusion as a function of its mass $M_{\pi_0}$ at the LHC.}
\label{fig:prodX}
\end{figure}

In the following, we will neglect the effect of top loops to the branching ratios into a pair of gauge bosons: the main rationale behind this is that, once such effects are sizeable, the decay is dominated by the $t\bar{t}$ final state, so that searches in di-boson final states become irrelevant. Thus, in this large top coupling limit, the only phenomenologically relevant effect will be on the gluon fusion production. As shown in Appendix~\ref{app:toploops},
 the effect on gluon fusion can be encoded in a $M_{\pi_0}$--dependent shift of the $\kappa_g$ coupling.
  Thus, our analysis can be extended in a straightforward way.

The dominant production channel for $\pi_0$ is gluon fusion\footnote{The only other production channels are vector boson fusion and associated production with gauge bosons or tops. However they are always subdominant.}  .
 In Fig.~\ref{fig:prodX} we show the production cross sections from gluon fusion as a function of $M_{\pi_0}$ at the LHC with 8 and 13 TeV, calculated at leading order (without K-factor) using  {\sc MadGraph 5}~\cite{Alwall:2011uj} and cross-checked against 
{\sc CalcHEP}~\cite{Belyaev:2012qa}. In our analysis we have used the NNPDF23LO (\verb|as_0130_qed|) PDF set~\cite{Ball:2012cx}
and the QCD scale naturally chosen to be the mass of the resonance.
We would like to note that although we evaluate cross sections at LO, one can re-scale them 
to known higher order corrections, which, for example for CP-Even Higgs boson production, are determined up to N$^3\mbox{LO}$ in QCD (see e.g.\cite{Anastasiou:2016hlm}
for review and references there in). Since in our signal simulation we do not include correction factors for higher order QCD corrections, the estimate of the LHC potential to probe the theories under study is conservative.

In Fig.~\ref{fig:prodX}, the coupling to gluons is fixed to $\kappa_g / f_\pi = 1$~TeV$^{-1}$, and the production cross section scales like $(\kappa_g / f_\pi)^2$.

The singlet pseudo-scalars decay to either di-boson via the WZW interactions or to $t\bar{t}$. The partial widths are related to the parameters in the Lagrangian in Eq.(\ref{eq:Lsigma})  as \cite{Cacciapaglia:2015nga} 
\beq
\Gamma(\pi_0\rightarrow gg) &=&\frac{\alpha_s^2 \kappa_g^2 M_{\pi_0}^3}{8 \pi^3 f_\pi^2}  \, , \label{eq:Gammagg}\\
\Gamma(\pi_0 \rightarrow WW) &=&\frac{\alpha^2 \kappa_W^2 M_{\pi_0}^3}{32\pi^3 f_\pi^2 \sin^4 \theta_W}\ \left(1- 4 \frac{m_W^2}{M_{\pi_0}^2}\right)^{\frac{3}{2}} \, ,\\
\Gamma(\pi_0 \rightarrow ZZ) &=&\frac{\alpha^2 \cos^4\theta_W \left(\kappa_W+\kappa_B\tan^4\theta_W\right)^2 M_{\pi_0}^3}{64\pi^3 f_\pi^2\sin^4 \theta_W}\  \left(1- 4 \frac{m_Z^2}{M_{\pi_0}^2}\right)^{\frac{3}{2}} \, ,\\
\Gamma(\pi_0\rightarrow Z\gamma) &=&\frac{\alpha \alpha\cos^2\theta_W\left(\kappa_W-\kappa_B\tan^2\theta_W\right)^2 M_{\pi_0}^3}{32 \pi^3 f_\pi^2\sin^2 \theta_W}\ \left(1- \frac{m_Z^2}{M_{\pi_0}^2}\right)^3 \, ,\\
\Gamma(\pi_0 \rightarrow \gamma\gamma) &=&\frac{\alpha^2 \left(\kappa_W+\kappa_B\right)^2 M_{\pi_0}^3}{64 \pi^3 f_\pi^2} \, ,\label{eq:Gammagamgam}\\
\Gamma(\pi_0 \rightarrow t\bar{t}) &=&\frac{3C_t^2 M_{\pi_0}}{8\pi}\frac{m_t^2}{f_\pi^2}\ \left(1-4\frac{m_t^2}{M_{\pi_0}^2}\right)^{1/2} \, ,\label{eq:Gammatt}
\eeq
where $\theta_W$ is the Weinberg angle.  Decays into other SM fermions are negligible, since they are suppressed by the fermion masses.
As the couplings are typically small, we expect the total width to be always small. To give a numerical estimate, the partial widths in gluons and tops (that are typically dominant) evaluate to:
\beq \label{eq:totwidthsinglet}
\Gamma (gg) \sim  0.04~\mbox{GeV}~\left( \frac{1~\mbox{TeV}}{f_\pi/\kappa_g}\right)^2 \left(\frac{M_{\pi_0}}{1~\mbox{TeV}} \right)^3\,, \quad \Gamma (t\bar{t}) \sim  0.4~\mbox{GeV}~\left( \frac{1~\mbox{TeV}}{f_\pi/C_t}\right)^2 \left(\frac{M_{\pi_0}}{1~\mbox{TeV}} \right)\,.
\eeq

It is instructive to split the decay modes into the final state $t\bar{t}$ and into di-boson final states. Furthermore, we will use ratios of branching ratios, which depend only on few of the couplings, to characterize the decay pattern of the singlets.   As a starter, the ratio
\beq
BF_{tt/gg} \equiv  \frac{\Gamma(\pi_0\rightarrow t\bar{t})}{\Gamma(\pi_0\rightarrow gg)} = \left(\frac{\alpha_s^2 }{3 \pi^2 }\right)^{-1}\frac{C^2_t}{\kappa_g^2}\frac{m^2_t}{M_{\pi_0}^2}\left(1-4 \frac{m_t^2}{M_{\pi_0}^2}\right)^{1/2}
\label{eq:sigttogg}
\eeq
only depends on the ratio $C_t/\kappa_g$, and on the mass $M_{\pi_0}$: this is mainly due to the fact that the partial width in $t\bar{t}$ scales linearly with the scalar mass versus the cubic power in di-boson partial widths. Therefore, the relevance of the top final states decreases for increasing $\pi_0$ mass. 
We also define di-boson ratios 
\beq
BF_{XY/\rm{bosons}}\equiv \frac{\Gamma (\pi_0\rightarrow XY)}{\Gamma (\pi_0 \rightarrow {\rm bosons})}\,, \qquad  \mbox{with} \quad XY= gg, WW, ZZ, Z\gamma, \gamma\gamma\,.
\eeq
These ratios depend on the coupling ratios $\kappa_{B}/\kappa_g$ and $\kappa_{W}/\kappa_g$, while the dependence on the mass is weak and only entering through kinematic phase space due to the non-zero masses of the $W$ and $Z$ bosons and the logarithmic running of the couplings (in particular, the QCD one).
We will thus use the ratios defined above to characterize the decay rates in a model-independent way.

\begin{figure}[t]
\begin{center}
\begin{tabular}{cc}
\includegraphics[width=.45\textwidth]{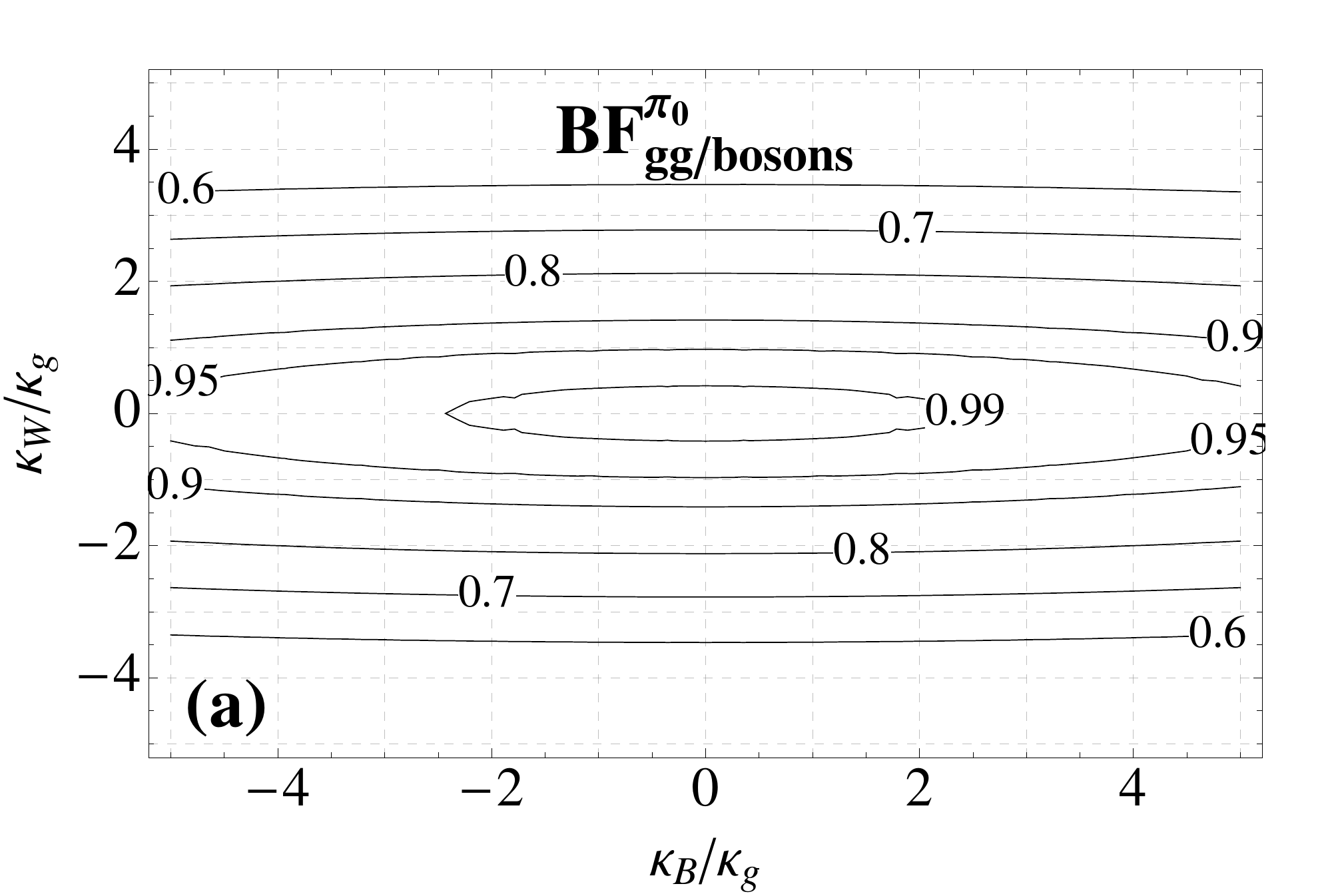} &
\includegraphics[width=.45\textwidth]{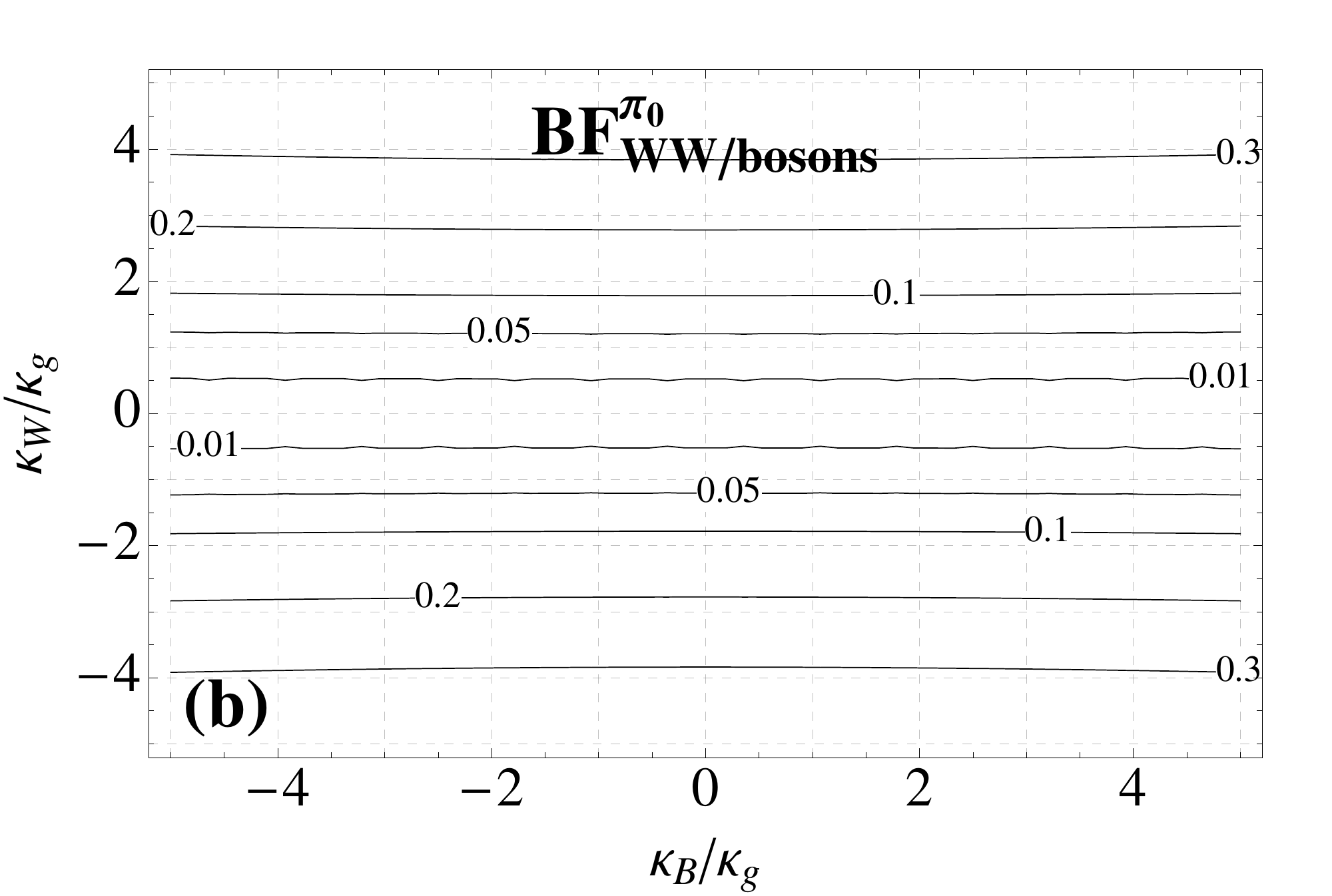} \\
\includegraphics[width=.45\textwidth]{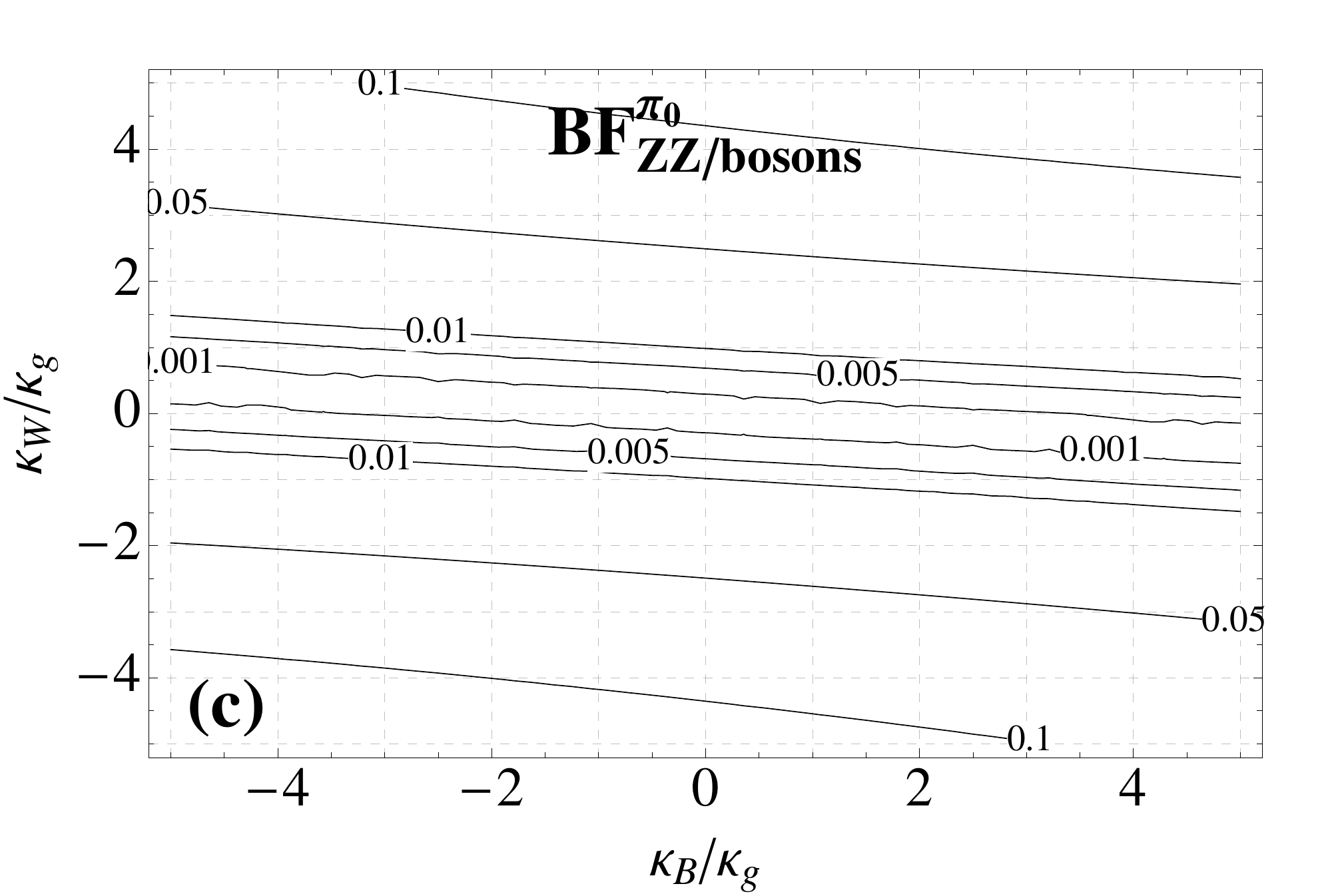} &
\includegraphics[width=.45\textwidth]{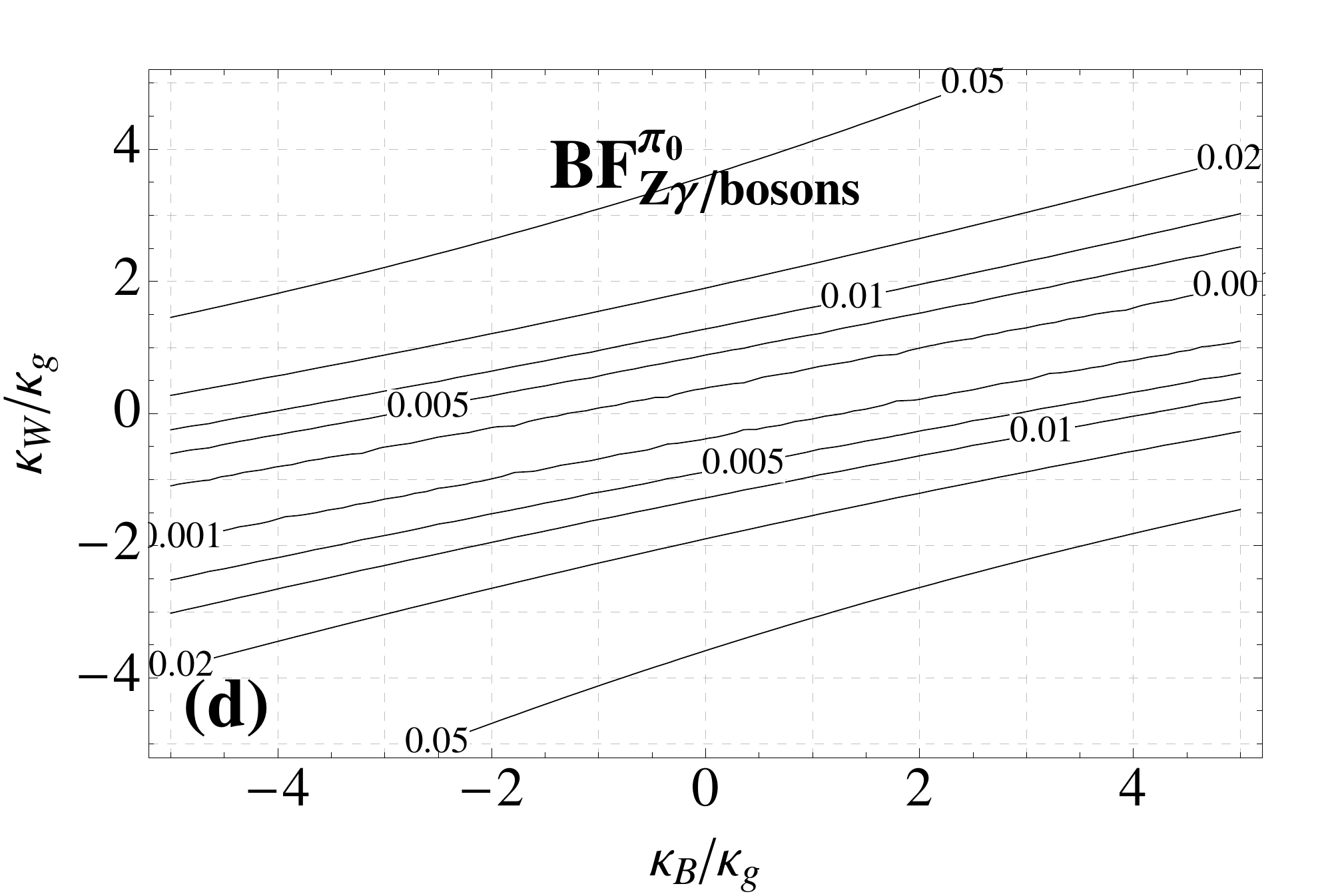} \\
\includegraphics[width=.45\textwidth]{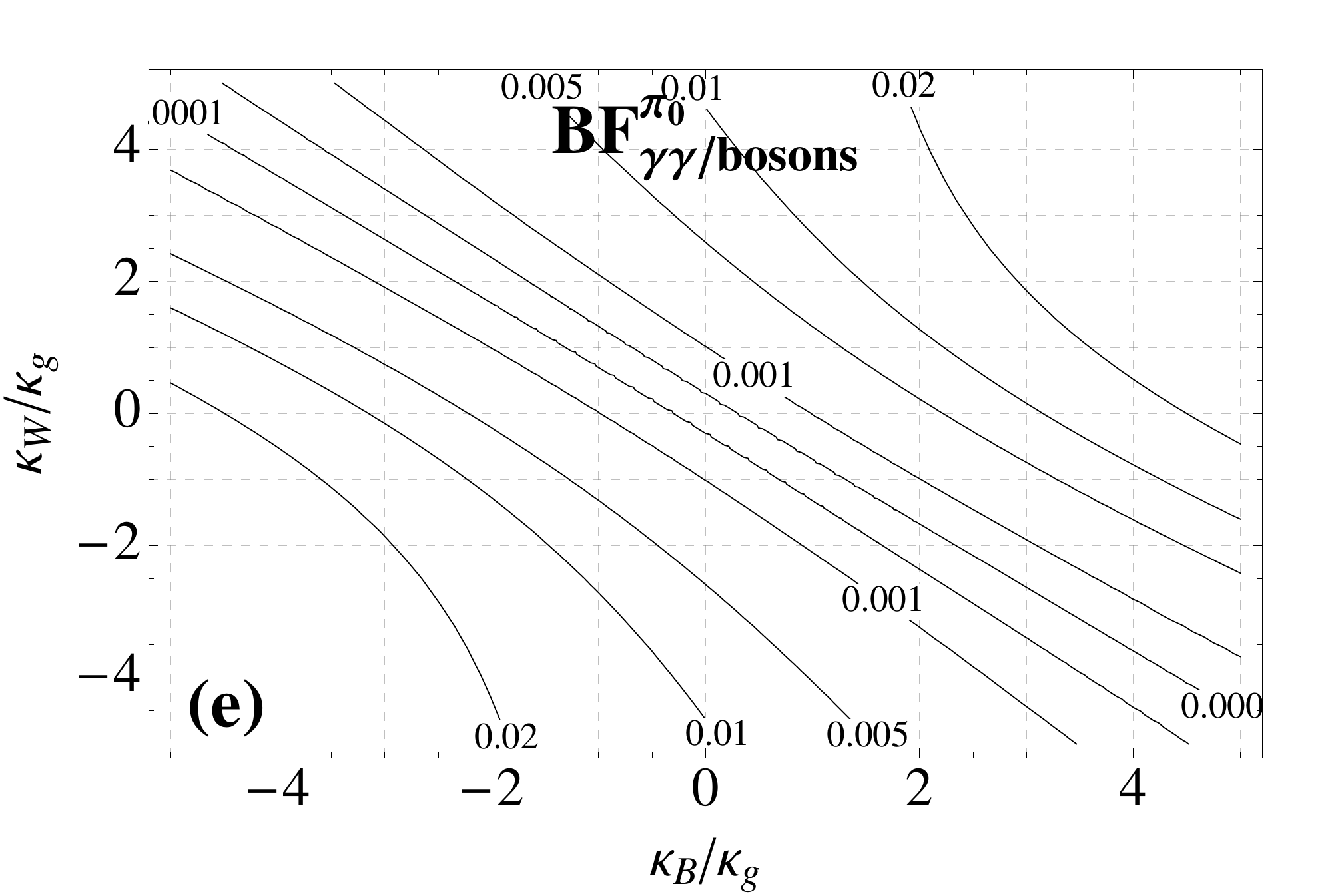} &
\includegraphics[width=.45\textwidth]{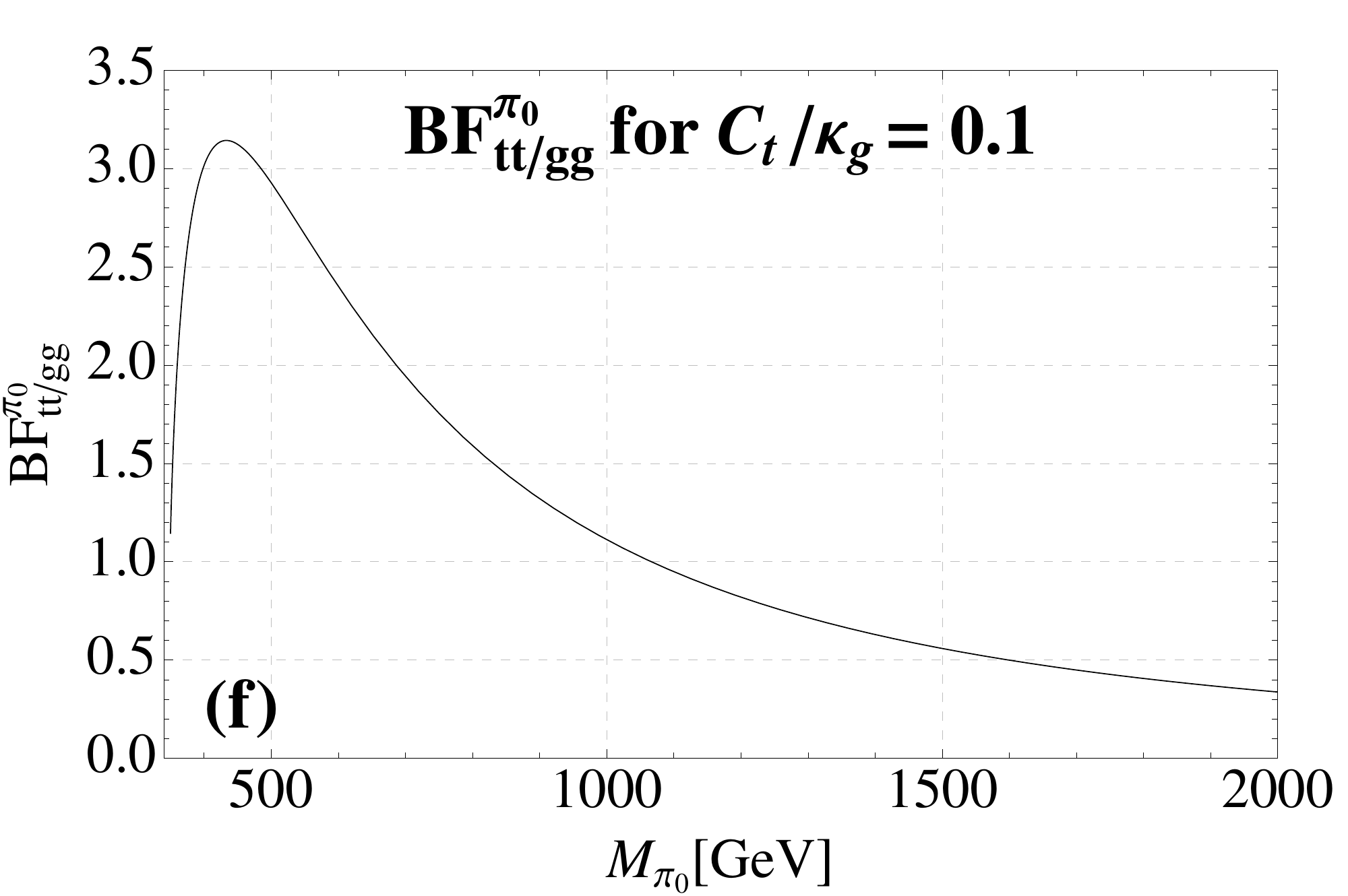} 
\end{tabular}
\end{center}
\caption{Ratios of the $\pi_0$ decay channels. In (a) - (e) we show $BF_{XY/\rm{bosons}}$ in the $\kappa_W/\kappa_g$ vs. $\kappa_B/\kappa_g$ plane, evaluated at  $M_{\pi_0} = 1$~TeV (the mass-dependence mainly enters through the running of $\alpha_s$).
In (f),  $BF_{tt/gg}$ is plotted as a function of $M_{\pi_0}$ for  $C_t / \kappa_g = 0.1$ ($BF_{tt/gg}$ scales like $\left(C_t / \kappa_g\right)^2$).}
\label{fig:BR}
\end{figure}
For illustration and later use, in Fig.~\ref{fig:BR} (a) - (e) we show the ratios $BF_{XY/\rm{bosons}}$ in the $\kappa_W/\kappa_g$ vs. $\kappa_B/\kappa_g$ plane, while Fig.~\ref{fig:BR} (f) shows the branching fraction $BF_{tt/gg}$ as a function of $M_{\pi_0}$. 
The first plots, (a) - (e), depend on $M_{\pi_0}$ via the running of $\alpha_s$ (the weak couplings are fixed to their values at the $Z$ pole for simplicity, as their running up to the TeV scale is mild), thus the plots refer to a mass $M_{\pi_0} = 1$~TeV. The mass dependence can be disentangled by absorbing the running coupling in the definition of $\kappa_g$, so that the ratios at a different mass can be obtained by rescaling
\beq
\frac{\kappa_{W/B}}{\kappa_g} \to \frac{\kappa_{W/B}}{\kappa_g} \left(\frac{\alpha_s (1~\mbox{TeV})}{\alpha_s (M_{\pi_0})} \right)\,.
\eeq
Figs.~\ref{fig:BR} show that the $gg$ channel dominates the di-boson branching fractions, followed by $WW$ which becomes increasingly important for increasing $\kappa_W/\kappa_g$. The (smaller) branching fractions of $ZZ$, $Z\gamma$, and $\gamma\gamma$ increase along the directions $|\kappa_W+\kappa_B\tan^4\theta_W|$, $|\kappa_W-\kappa_B\tan^2\theta_W|$, and $|\kappa_B+\kappa_W|$.  The magnitude of the branching fraction into tops is mainly controlled by $C_t / \kappa_g$. As the $t\bar{t}$ partial width scales with $M_{\pi_0}$ while all di-boson partial widths scale with $M^3_{\pi_0}$, the branching fraction into $t\bar{t}$ is reduced at high masses $M_{\pi_0} \gg 2 m_t$ and kinematically suppressed near the threshold $M_{\pi_0} = 2 m_t$.

\subsubsection{Experimental bounds from di-boson and $t\bar{t}$ resonance searches}

Both ATLAS and CMS presented numerous searches for di-jet, $WW$, $ZZ$, $Z\gamma$, and di-photon resonances in the high mass region. The list of searches we include into our study is summarized in Table~\ref{tab:searches}. 
Where possible, we directly use the bounds on the production cross section times branching ratio ($\sigma\times BR$) into the respective channel given by ATLAS and CMS. In several studies (in particular for di-jet searches and partially for $Z\gamma$ and $\gamma\gamma$ searches), some results were  presented  in terms of fiducial cross sections or in terms of cross section times acceptance. In Appendix \ref{app:Exp}, we  summarize the assumptions made in order to extract the bounds from ATLAS and CMS studies for the model discussed in this article.

\begin{table}[h]
\begin{center}
\begin{tabular}{| c |c|c|c|c|}
\hline
&\multicolumn{2}{c|}{8 TeV}& \multicolumn{2}{c|}{13 TeV}\\
\hline
channel & ATLAS & CMS  &  ATLAS& CMS \\  \hline
$gg$                       &\cite{Aad:2014aqa}
                               &\cite{CMS:2015neg,Khachatryan:2015sja}
                               &\cite{ATLAS:2016xiv,ATLAS:2016lvi}
                               &\cite{CMS:2016wpz}\\
 $\gamma\gamma$&\cite{Aad:2015mna}
                               &\cite{CMS:2016owr}
                               &\cite{ATLAS:2016eeo}
                               &\cite{CMS:2016owr}\\
$WW$                    &\cite{Aad:2015agg,Aad:2015ufa,Aad:2015owa}
                               &\cite{Khachatryan:2014gha,Khachatryan:2014hpa}
                               &\cite{ATLAS:2016kjy,ATLAS:2016cwq,ATLAS:2016yqq}
                               &\cite{CMS:2015nmz,CMS:2016jpd,CMS:2016pfl}\\
$ZZ$                      &\cite{Aad:2015kna,Aad:2014xka,Aad:2015owa}
                               &\cite{Khachatryan:2014gha,Khachatryan:2014hpa}
                               &\cite{ATLAS:2016oum,ATLAS:2016npe,ATLAS:2016bza,ATLAS:2016yqq}
                               &\cite{CMS:2015nmz,CMS:2016tio,CMS:2016ilx}\\
$Z\gamma$           & 
                               &\cite{CMS:2015lza,CMS:2016mvc}
                               &\cite{ATLAS:2016lri}
                               &\cite{CMS:2016pax,CMS:2016cbb}\\
$t\bar{t}$                &\cite{Aad:2015fna}
                               &\cite{Khachatryan:2015sma}
                               &\cite{ATLAStt13}
                               &\cite{CMS:2016zte}\\                               
\hline
\end{tabular}
\end{center}
\caption{List of di-boson and $t\bar{t}$ searches included in our analysis. For a more detailed discussion see Appendix \ref{app:Exp}.} 
\label{tab:searches} 
\end{table}

\begin{figure}[t]
\begin{tabular}{cc}
\includegraphics[width=.45\textwidth]{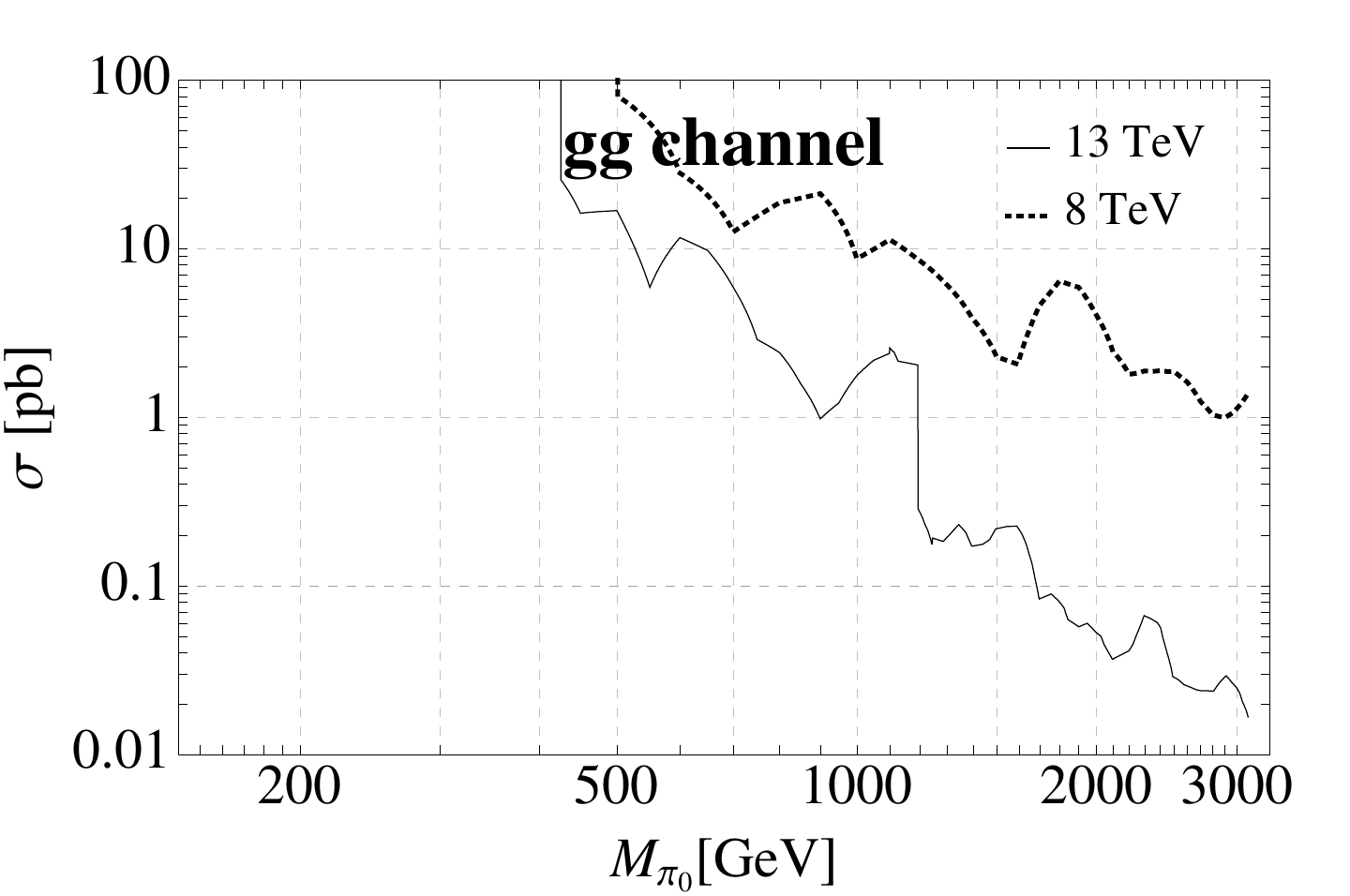} & 
\includegraphics[width=.45\textwidth]{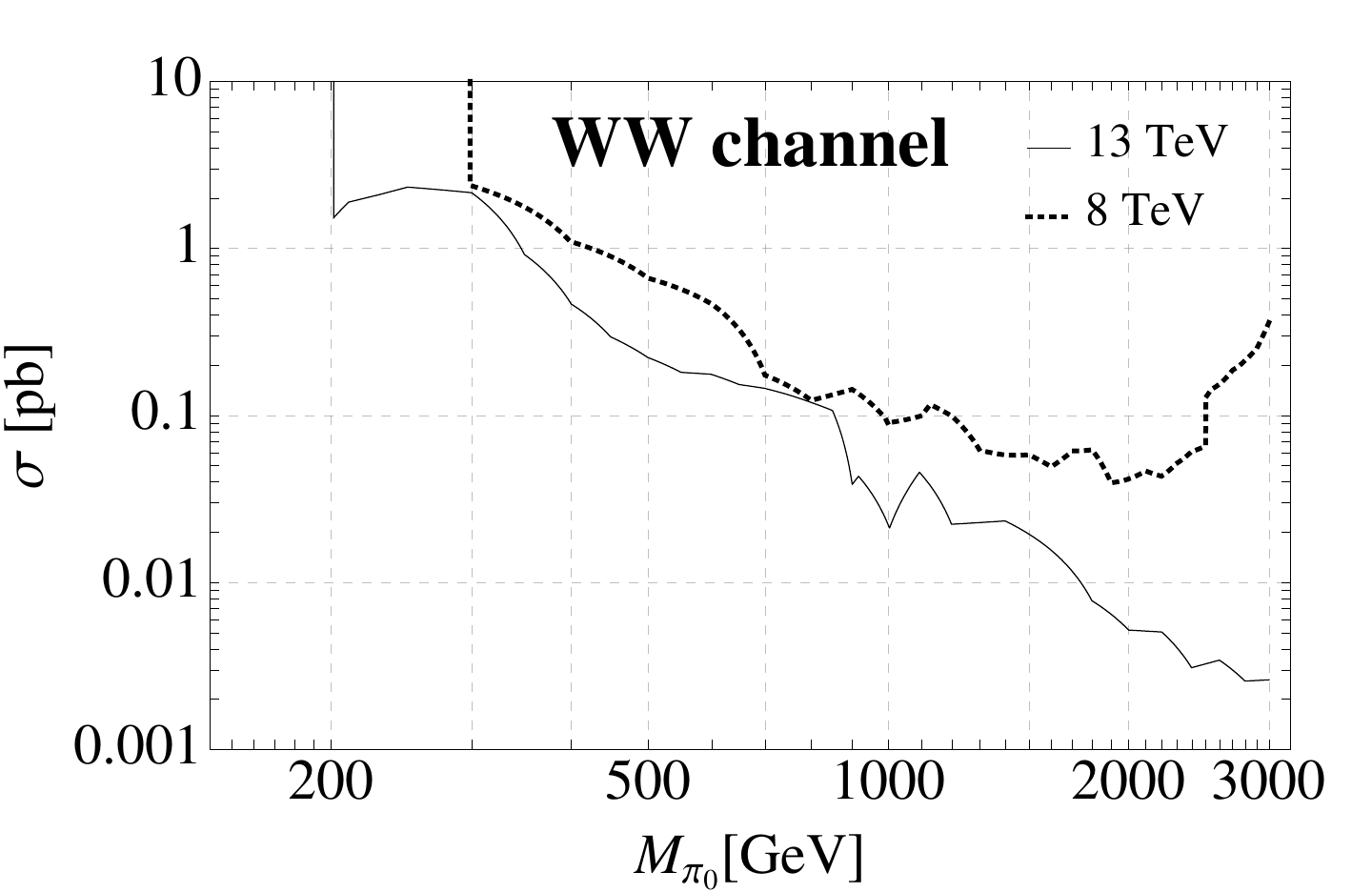} \\
\includegraphics[width=.45\textwidth]{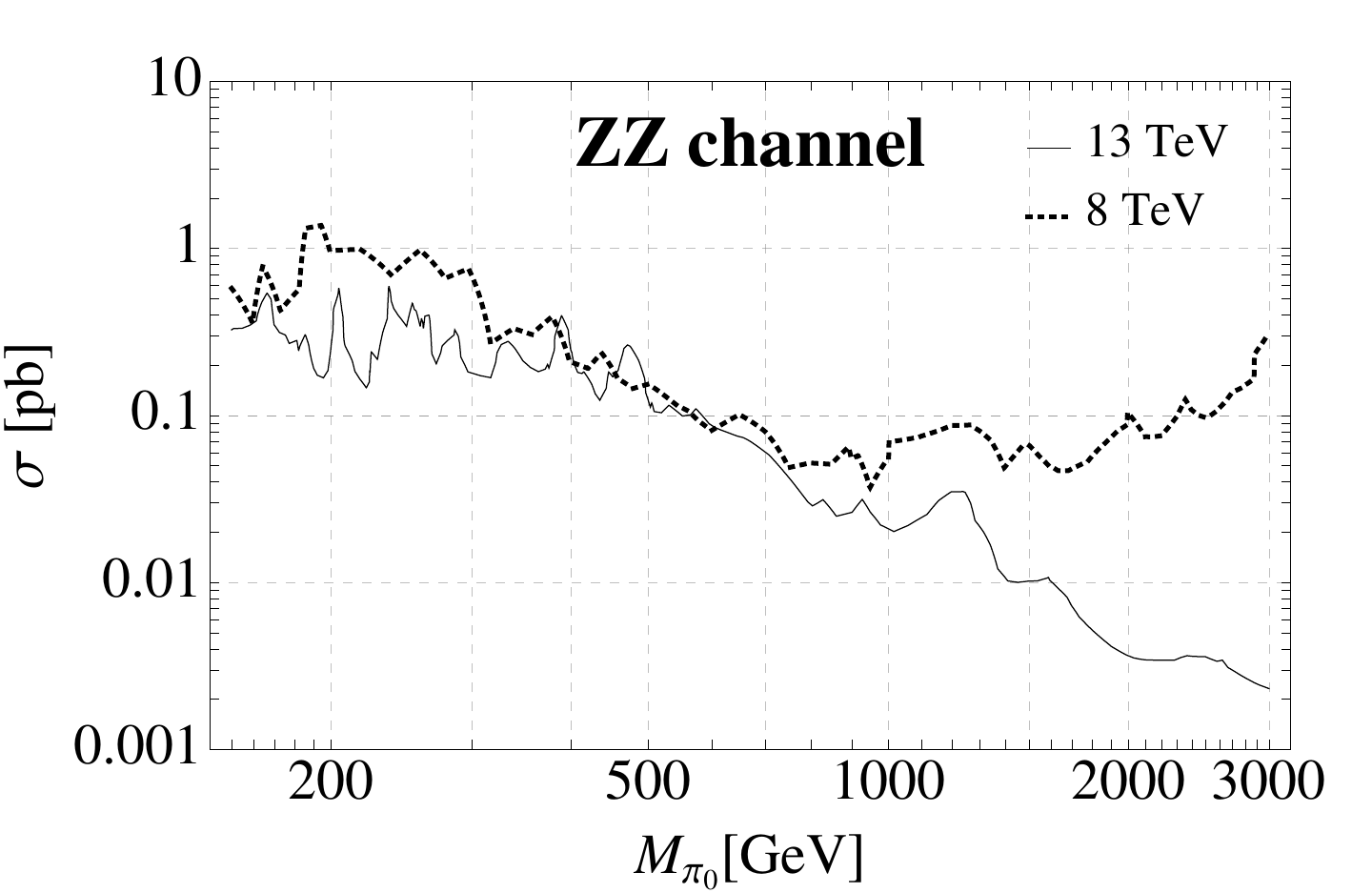} &
\includegraphics[width=.45\textwidth]{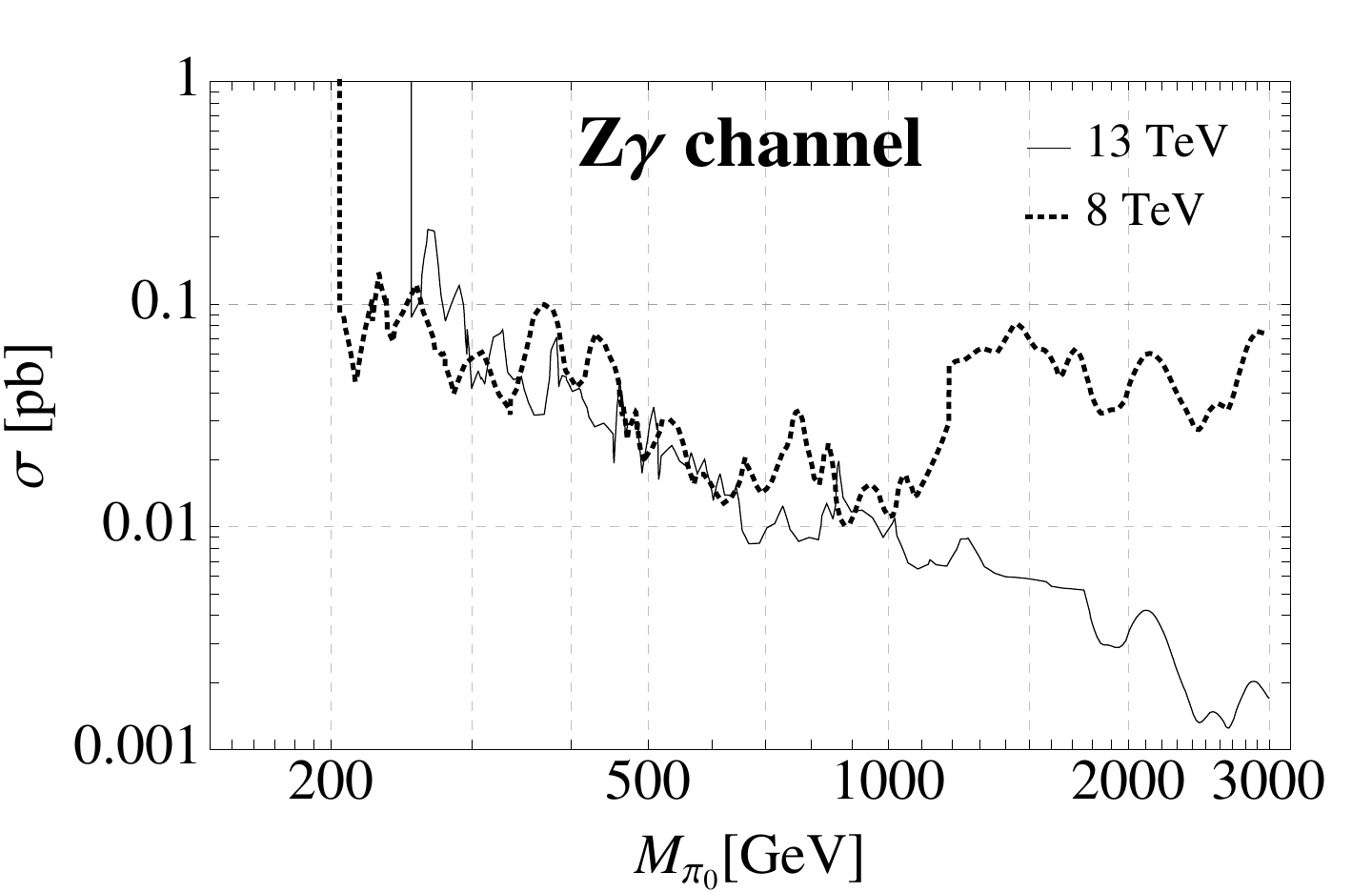} \\ 
\includegraphics[width=.45\textwidth]{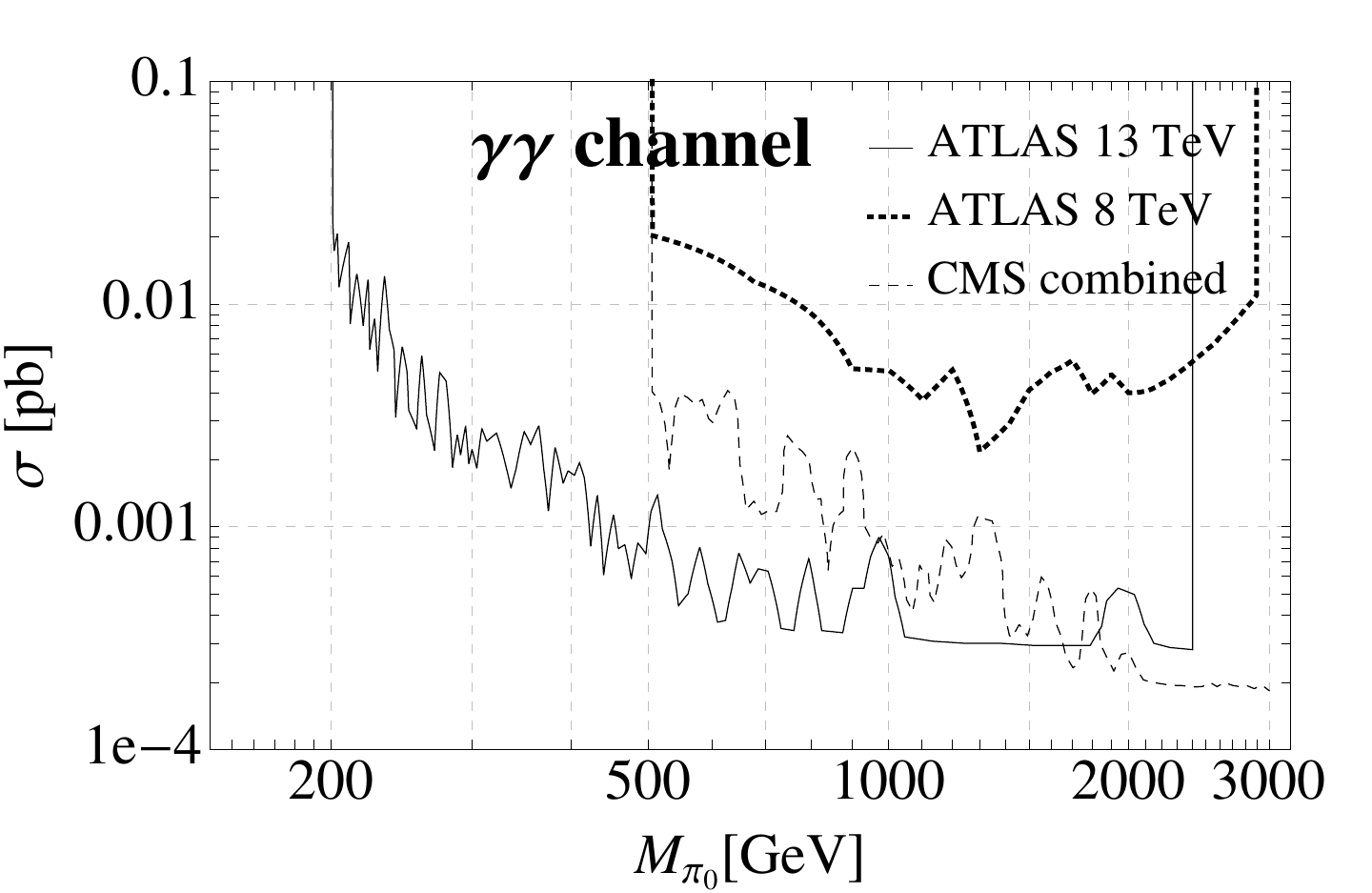} & 
\includegraphics[width=.45\textwidth]{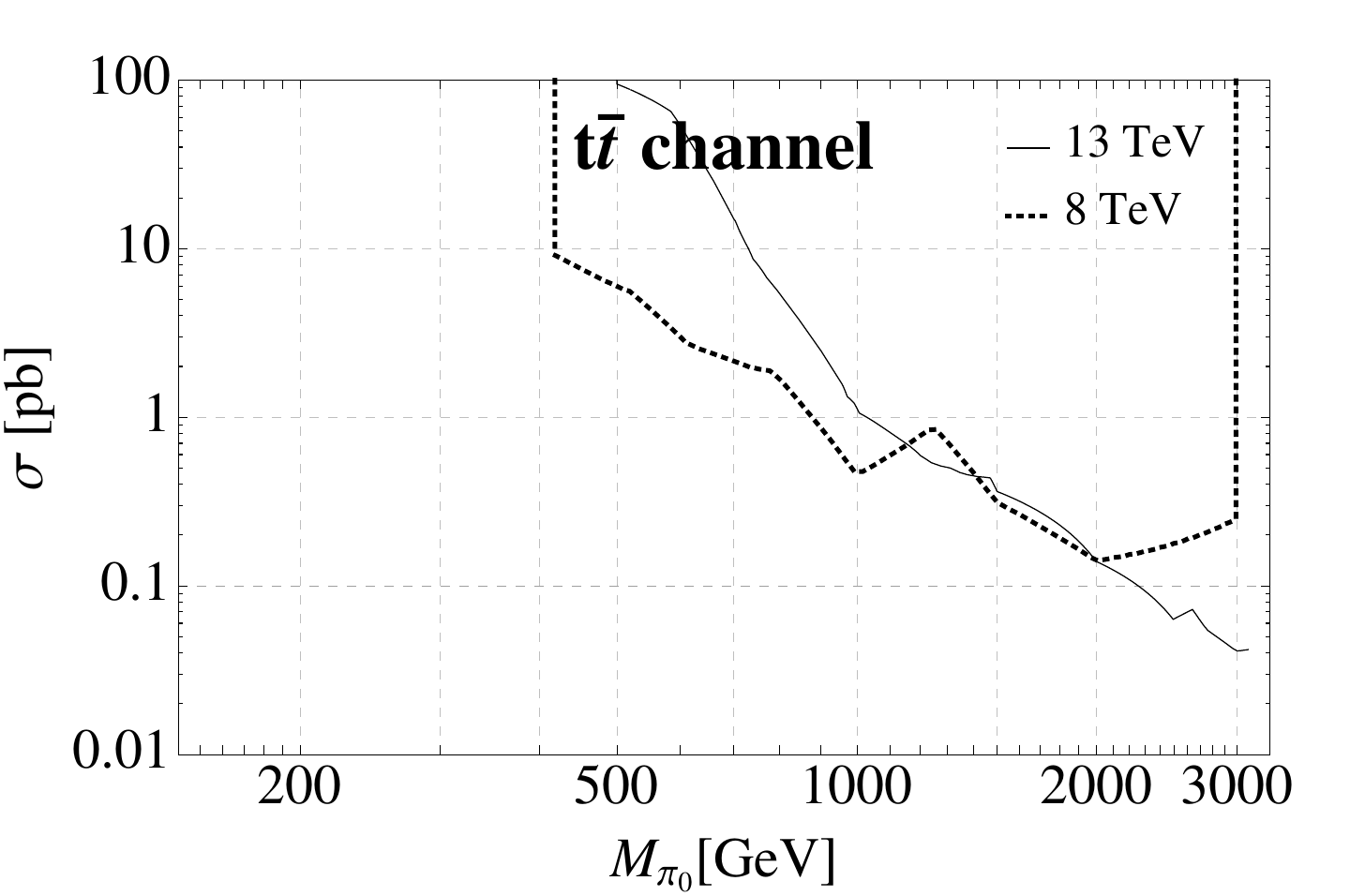} 

\end{tabular}
\caption{Bounds on the di-boson and $t\bar{t}$ channels from 13 TeV searches and 8 TeV searches on the 13 TeV production cross section times branching ratio. 8 TeV bounds have been rescaled by the ratio of 13 TeV / 8 TeV production cross section for gluon fusion in order to allow direct comparison. For the di-photon channel, we show the combined results from 13 and 8 TeV determined by CMS as well as the (still separate) 8 and 13 TeV search results by ATLAS.}
\label{fig:bds}
\end{figure}

We aim at presenting collective bounds for the different di-boson and $t\bar{t}$ final states from pNGB decays. For searches in a given channel at $\sqrt{s}=13$~TeV, we do not perform a combination of the ATLAS  and CMS searches but simply use the strongest bound obtained for a given $M_{\pi_0}$. To include Run I bounds, we analogously take the strongest bound at $M_{\pi_0}$ in each channel, and rescale the cross section by a factor $\sigma(gg\rightarrow\pi_0)_{13} / \sigma(gg\rightarrow\pi_0)_8$. The resulting constraints on the $\sigma\times BR$ at 13 TeV for the $gg$, $WW$, $ZZ$, $Z\gamma$, $\gamma\gamma$, and $t \bar{t}$ channels are shown in Fig.~\ref{fig:bds} \footnote{For the di-photon channel, CMS performed a combination of the 8 TeV and 13 TeV bounds, so for this channel we give the ATLAS bounds from 13 and 8 TeV data and the combined bound from CMS.}.

\subsubsection{Model-independent bounds on the singlet pseudo-scalar parameter space}
 
The experimental constraints shown in Fig.~\ref{fig:bds} translate into bounds for the still allowed production cross section as a function of ($M_{\pi_0}$, $\kappa_W/\kappa_g$, $\kappa_B/\kappa_g$, $C_t/\kappa_g$) via the branching fractions following from Eqs.~(\ref{eq:Gammagg}-\ref{eq:Gammatt}), as exemplified in Fig.~\ref{fig:BR}. Using Fig.~\ref{fig:prodX}, the bound on the production cross section translates into a bound on the coupling to gluons $\kappa_g/f_\pi$.

To simplify the impact of the multi-dimensional parameter space, it is useful to split the final states into two categories: di-boson and tops ($t\bar{t}$). The advantage is that the ratios between di-boson modes only depend on two ratios of couplings (and very mildly on the mass), while the rate of $t\bar{t}$ final states can be expressed in terms of $C_t/\kappa_g$. We thus define the following strategy apt to explore, in a way which is as model independent as possible, the parameter space of this class of models:
\begin{itemize}
\item[-] define the cross section in a specific di-boson final state as:
\beq
\sigma \times BR (\pi_0 \to XY) = (\sigma \times BR_{\rm bosons}) \times BF_{XY/\rm{bosons}}\,;
\eeq

\item[-]  from the above, one can extract a bound on $\sigma \times BR_{\rm bosons}$ as a function of the mass and the two ratios of couplings $\kappa_W/\kappa_g$ and $\kappa_B/\kappa_g$;

\item[-] for each value of $C_t/\kappa_g$, the function  $BF_{tt/gg}$ can be used to calculate the cross section in $t\bar{t}$ final state, as
\beq
\sigma \times BR (\pi_0 \to t\bar{t}) = (\sigma \times BR_{\rm bosons}) \times BF_{gg/\rm{bosons}}\times BF_{tt/gg}\,,
\eeq
matching the di-boson bound, which can be directly compared to the bound from $t\bar{t}$ searches as shown in Fig.~\ref{fig:bds}.

\end{itemize}
The latter step allows to determine whether the strongest bound comes from di-boson searches, or from $t\bar{t}$. Note, however, that this approach is only valid in the narrow width approximation, which is always true in this class of models where the couplings are small, as suppressed by a loop factor in the case of WZW interactions, or a ration $m_t/f_\pi$ for top couplings, as shown in Eq.(\ref{eq:totwidthsinglet}).

\begin{figure}[h]
\begin{tabular}{cc}
\includegraphics[width=0.5\textwidth]{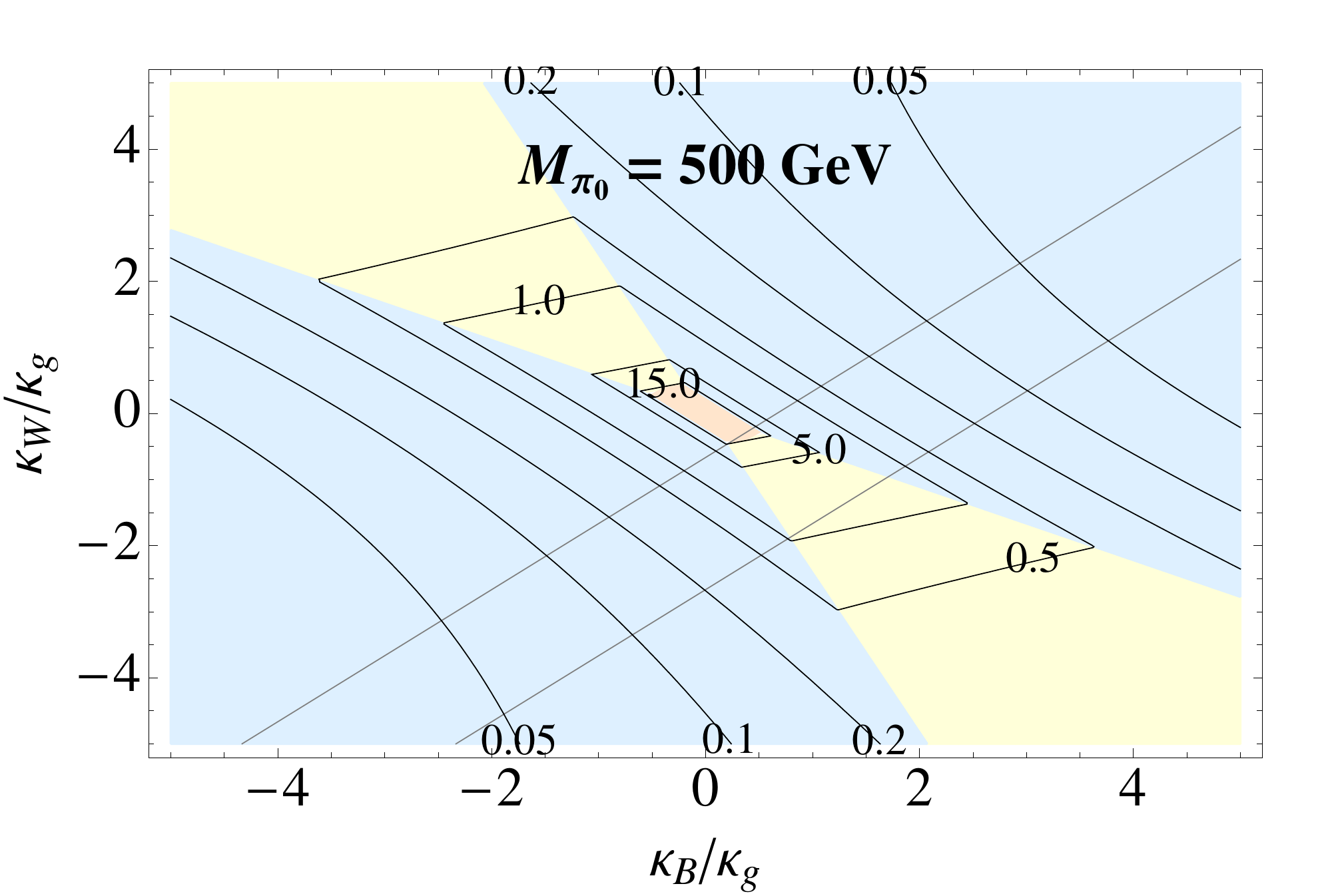} & 
\includegraphics[width=0.5\textwidth]{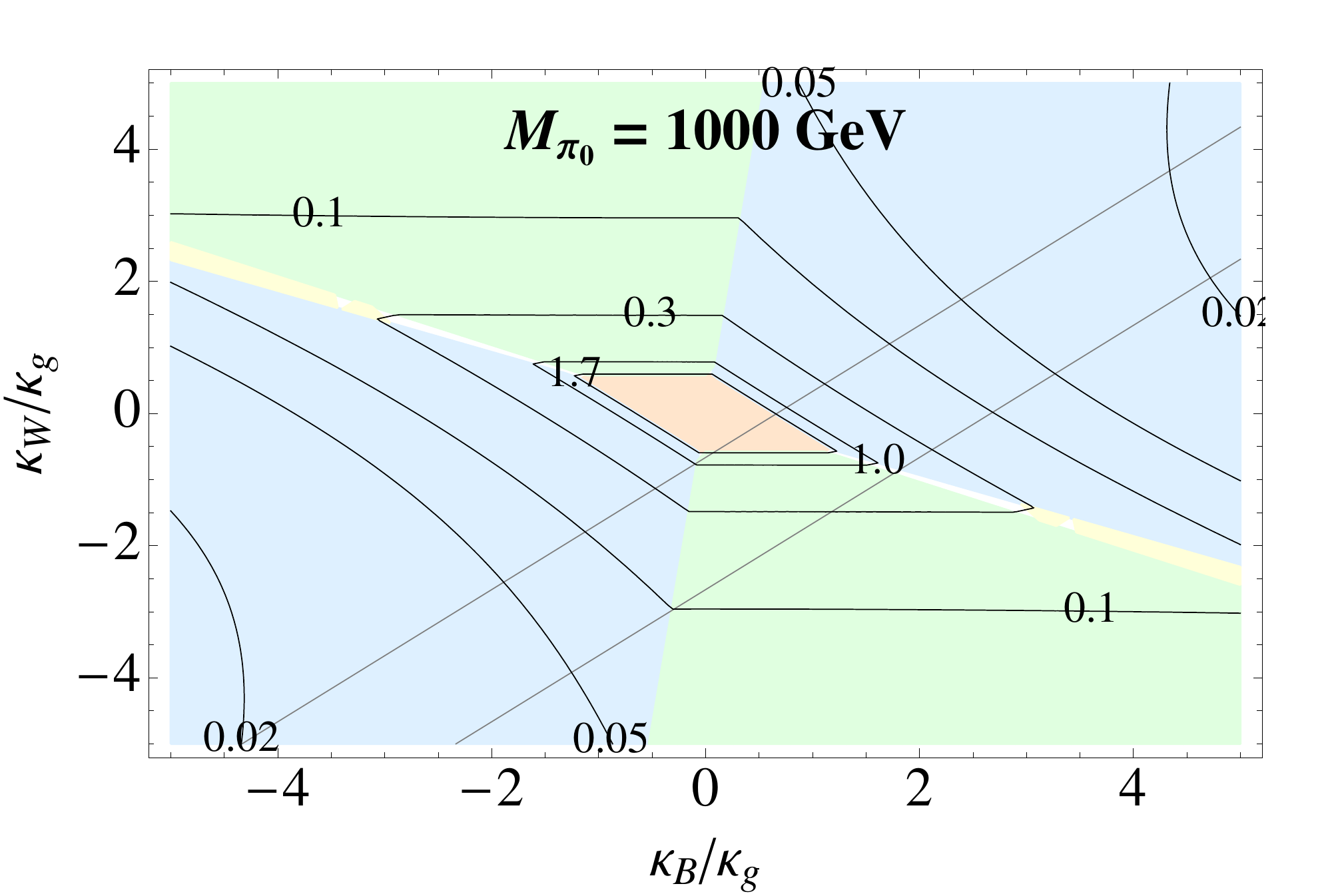} \\
\includegraphics[width=0.5\textwidth]{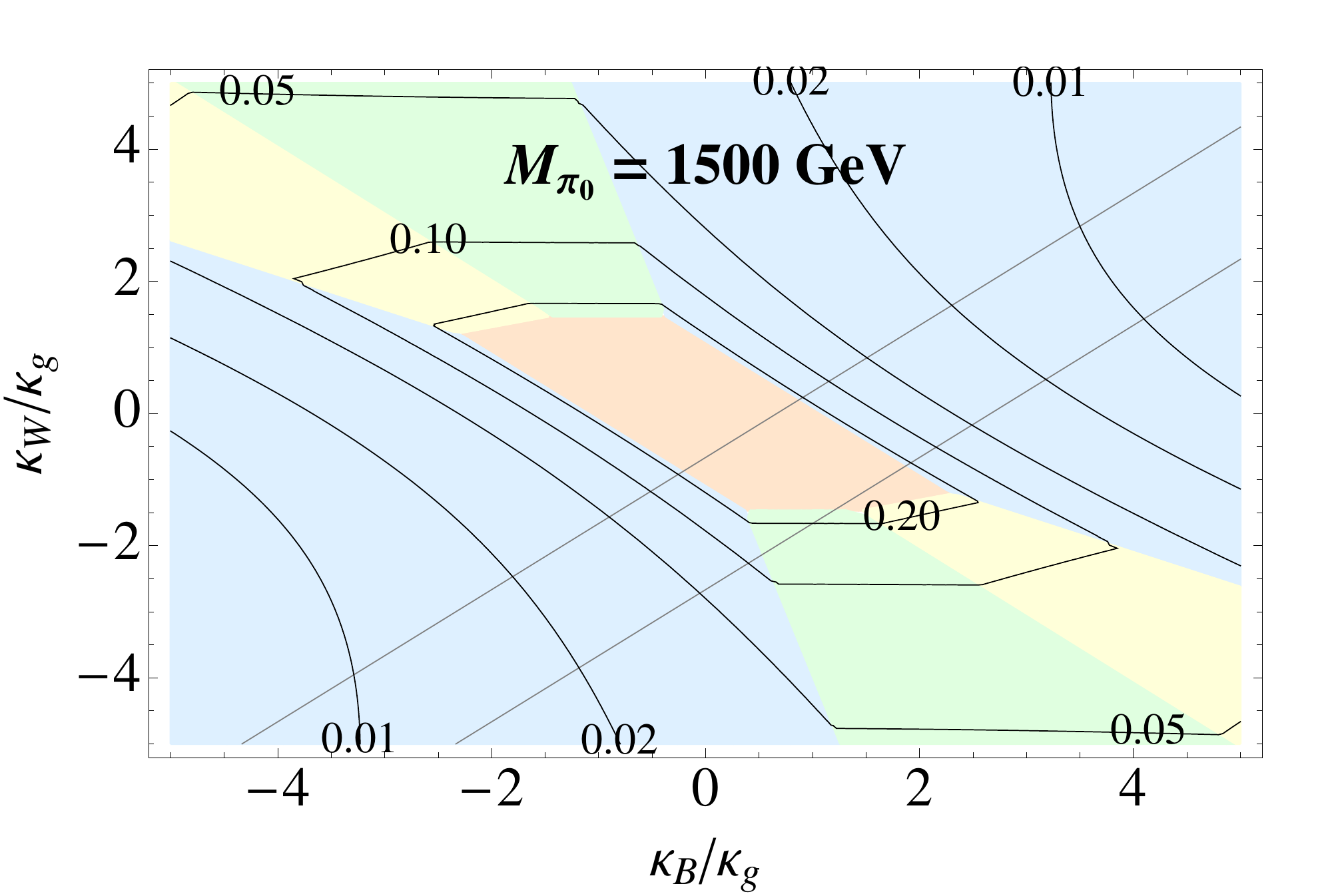} &
\includegraphics[width=0.5\textwidth]{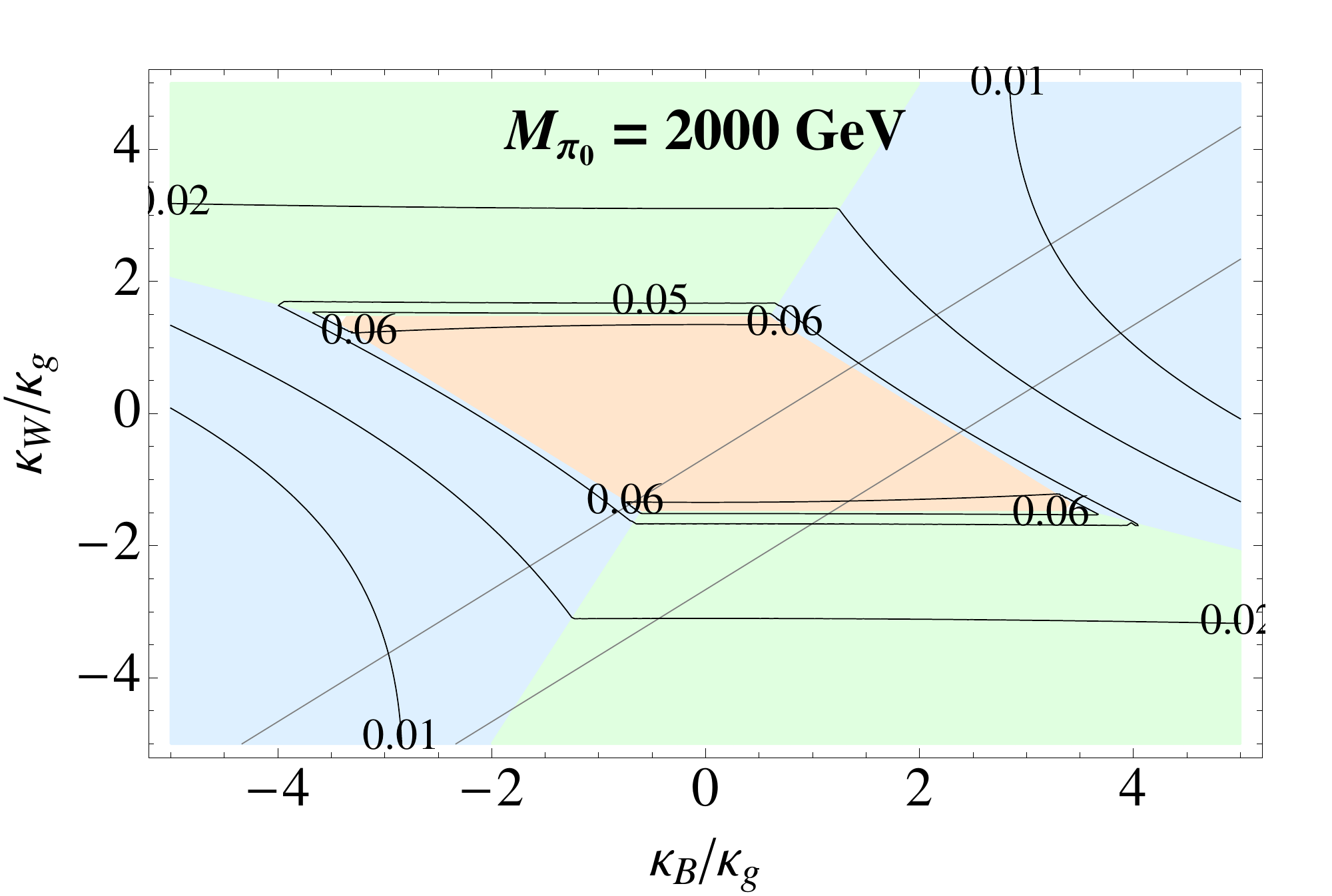} 
\end{tabular}
\caption{Combined bounds on the production cross section at 13 TeV times branching ratio into di-bosons ($\sigma_{13}\times BR_{\rm{bosons}}$ [pb]) in the $\kappa_W/\kappa_g$ vs. $\kappa_B/\kappa_g$ plane for $M_{\pi_0} = 500$, $1000$, $1500$, $2000$~GeV. The contours give the bounds in pb. The colored areas indicate the decay channel that, with current data, yields the strongest constraint: $gg$ (orange), $WW$ (green), $Z\gamma$ (yellow), or $\gamma\gamma$ (blue). The two grey diagonals indicate the lines on which the SM singlets $a$ and $\eta'$ of the models discussed in Sec. \ref{sec:models} lie.}
\label{fig:kbkWMI}
\end{figure}

To map out the model parameter space, let us first consider bounds for fixed mass $M_{\pi_0}$. In Fig.~\ref{fig:kbkWMI} we show the bounds on $\sigma_{13}\times BR_{\rm{bosons}}$ in the $\kappa_W/\kappa_g$ vs. $\kappa_B/\kappa_g$ plane for various resonance masses. The colored regions tag the decay channel that, with current data, yields the strongest bound at a given parameter point. At  $\kappa_W/\kappa_g=\kappa_B/\kappa_g=0$, the branching ratios in all di-boson channels, apart from $gg$, are zero, thus strongest bound around the origin arises from the $gg$ channel (in orange). For increasing $|\kappa_{B,W}/\kappa_g|$, the bound on $\sigma_{13}\times BR_{ \rm{bosons}}$ initially becomes marginally weaker because of a depletion in the leading $gg$ channel. For further increased $|\kappa_{B,W}/\kappa_g|$, channels other than $gg$ become the most constraining ones, at which point the bound becomes stronger again, being dominated by EW boson final states. We see that along the direction $\kappa_B \sim \kappa_W$, it is $\gamma \gamma$ that dominates the constraints (in blue), while along the orthogonal direction, where the coupling to photons partially cancels, the $WW$ (green) and/or $Z\gamma$ (yellow) channels take over the lead.

Fig.~\ref{fig:kbkWMI} quantifies the bounds for any model described by the effective Langrangian in Eq.\eqref{eq:Lsigma}. 
As outlined in Sec.\ref{sec:models}, the models considered in this article predict SM singlets $a$ and $\eta'$ for which 
\begin{equation}
\kappa_W = \kappa_B -  6 Y_\chi^2 \kappa_g \ {\rm with \ } Y_\chi = 1/3  \ {\rm or \ } 2/3
\label{eq:beta}
\end{equation}
depending on the hypercharge of $\chi$: the two grey diagonal lines in Fig.~\ref{fig:kbkWMI} mark these model lines for reference.

As becomes clear from Fig.~\ref{fig:kbkWMI}, all di-boson channels (apart from $ZZ$) yield the dominant constraint in some portion of the parameter space. Fig.~\ref{fig:kbkWMI} only indicates the channel setting the bound, but through Fig.~\ref{fig:BR}, or equivalently Eqs.~(\ref{eq:Gammagg}-\ref{eq:Gammagamgam}), and the experimental bounds shown in Fig.~\ref{fig:bds}, the relevance of each decay channel at a given parameter point $M_{\pi_0}, \kappa_B/\kappa_g,\kappa_W/\kappa_g$ can easily be obtained.

As an example of how to use these results
 in application to a specific model, let us consider the point $(\frac{\kappa_B}{\kappa_g },\frac{\kappa_W}{\kappa_g }) =(2.1,-0.55)$ at $M_{\pi_0} = 1$~TeV~\footnote{This sample point corresponds to the pseudo-scalar $a$ in model M9 from Table~\ref{tab:reps}, in the decoupling limit of the $\eta'$ mass.}. From Fig.~\ref{fig:kbkWMI} (top right), the constraint on $\sigma_{13}\times BR_{\rm{bosons}}$ reads  as 0.3~pb. Multiplying 0.3~pb by $BF^\sigma_{XY/\rm{bosons}}= (98\%,\, 1\%,\, 0.14\%,\, 0.9\% )$ for $XY=(gg,WW,ZZ,Z\gamma)$  (extracted from Fig.~\ref{fig:BR}) one obtains a signal cross sections of $(290\,\rm{ fb}, 3\,\rm{ fb}, 0.4\,\rm{ fb}, 2.7\,\rm{ fb})$ for the respective final states. These values are a factor of $(6,\, 6.5,\, 50,\, 3.5)$ respectively lower than the cross section bound for $M_{\pi_0} = 1$~TeV in Fig.~\ref{fig:bds}, showing how close each bound is to the limit. Following this universal  recipe different  models from Table~\ref{tab:reps} with different   mass values can be easily tested using the information from Figs.~\ref{fig:BR}-\ref{fig:kbkWMI}.

\begin{figure}[t]
\begin{tabular}{cc}
\includegraphics[width=0.5\textwidth]{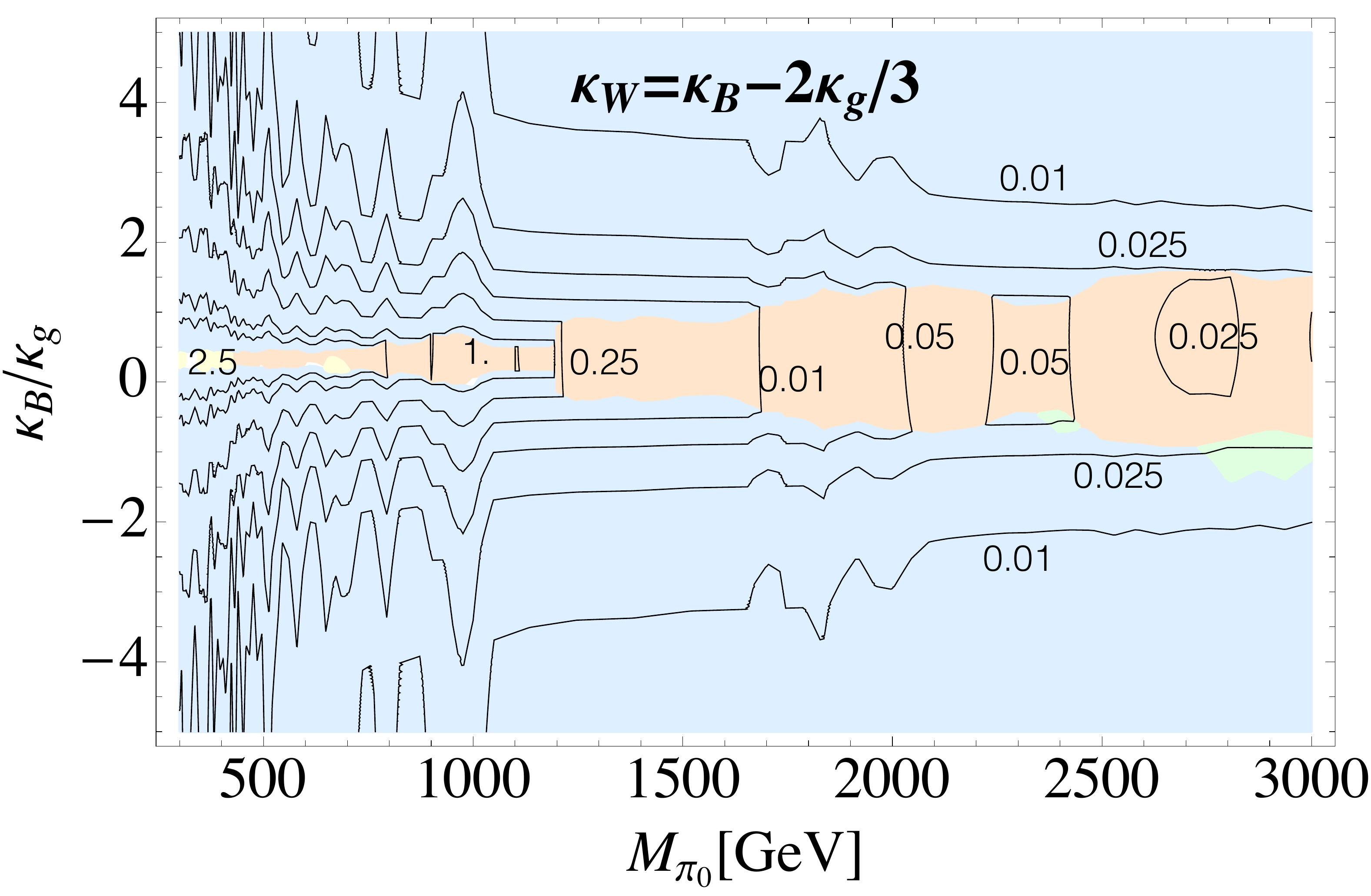} & 
\includegraphics[width=0.5\textwidth]{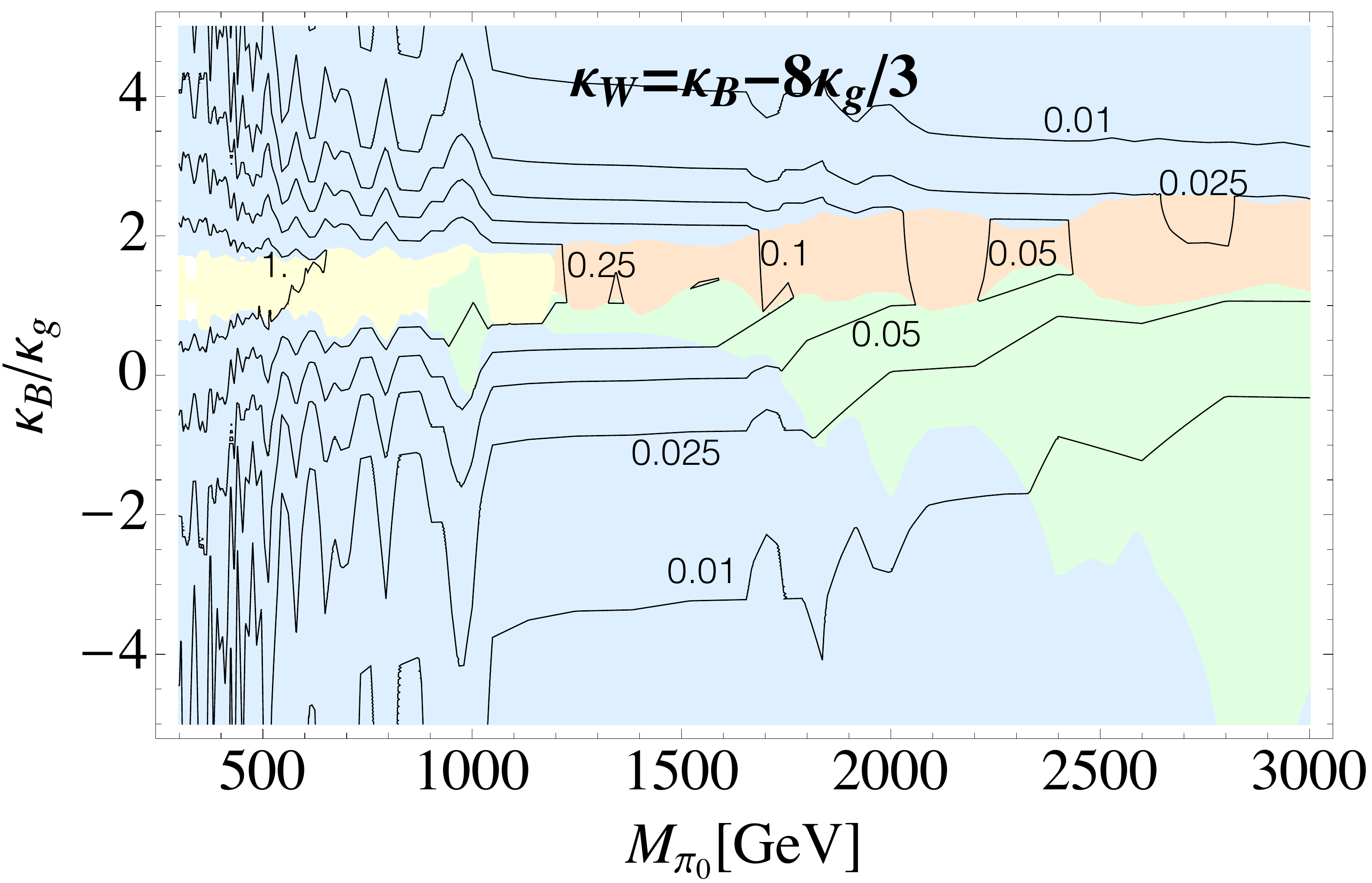} \\
\end{tabular}
\caption{Bounds on $\sigma_{13}\times BR_{\rm{bosons}}$ (in pb) in the mass vs. $\kappa_B/\kappa_g$ plane. We present the results  for two particular lines in the $\kappa_B/\kappa_g$ vs. $\kappa_B/\kappa_g$ parameter plane. The relation  $\kappa_W = \kappa_B - \frac{2}{3}\kappa_g$ (left plot) is realized for $a$ and $\eta'$ of all models with $Y_\chi = 1/3$ discussed in Sec.~\ref{sec:models}, while the relation $\kappa_W = \kappa_B - \frac{8}{3}\kappa_g$ (right plot) is realized for $a$ and $\eta'$ of all models with $Y_\chi = 2/3$.
The colored areas indicate the decay channel which with current data yields the strongest constraint: $gg$ (orange), $WW$ (green),   $Z\gamma$ (yellow), or $\gamma\gamma$ (blue).
The strongly varying bounds in the area below $\sim 1$~TeV are a direct consequence of the strong variation with mass of the experimental constraints (mainly the di-photon) as evident in Fig.~\ref{fig:bds}. }
\label{fig:kapvsM}
\end{figure}

As outlined in Sec.~\ref{sec:models}, the models considered in this article predict SM singlets $a$ and $\eta'$ whose couplings lie on two lines depending on the hypercharge of the $\chi$'s (shown by the grey diagonals in Fig.~\ref{fig:kbkWMI}).  In order to present bounds for different resonance masses than the ones given in Fig.~\ref{fig:kbkWMI}, we give results along the above lines. The bound on $\sigma_{13}\times BR_{\rm{bosons}}$  as a function of $M_{\pi_0}$ and $\kappa_B/\kappa_g$ is shown in Fig.~\ref{fig:kapvsM}, with the same color code as above. 

\bigskip

\begin{figure}[t]
\begin{tabular}{cc}
\includegraphics[width=0.45\textwidth]{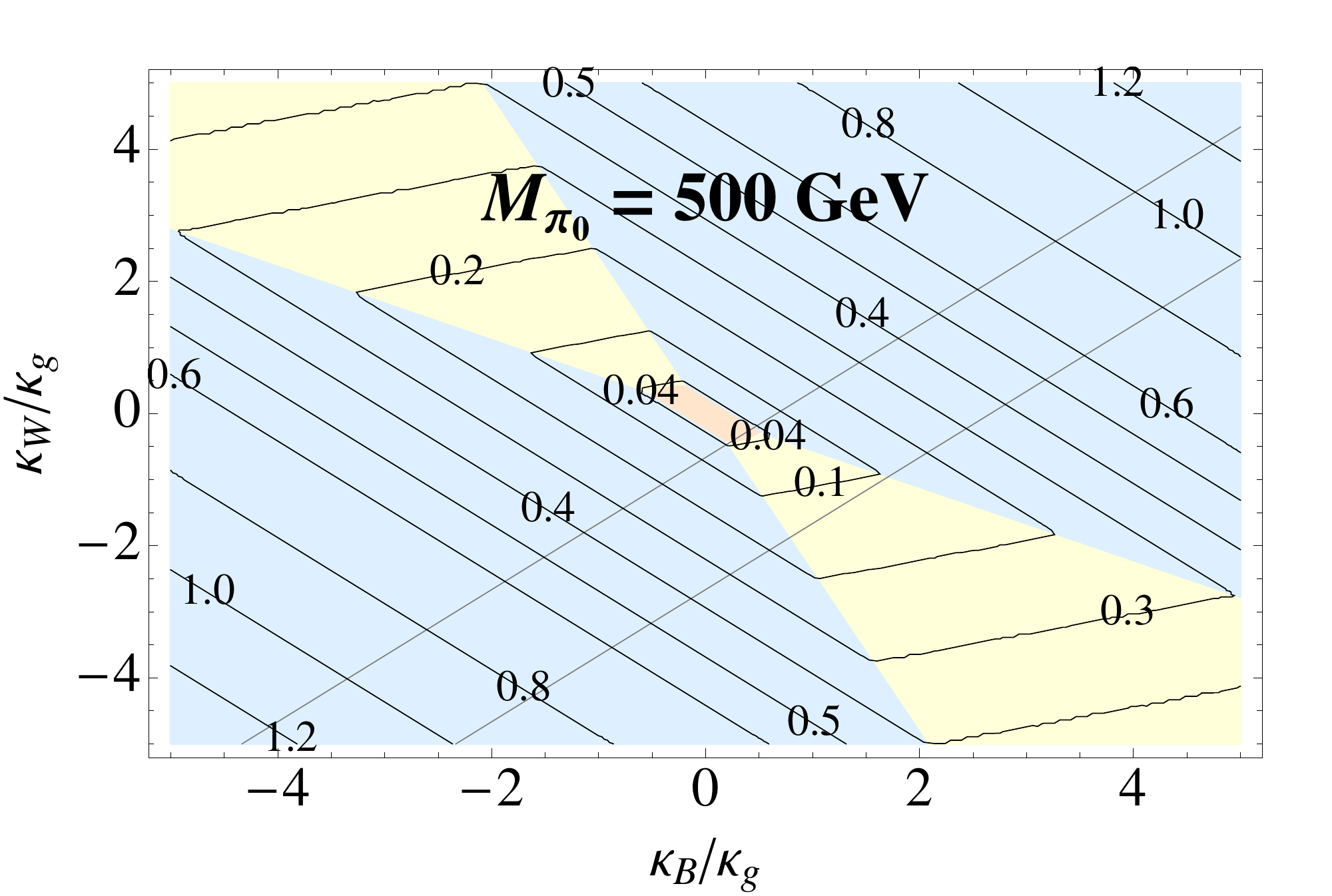} & 
\includegraphics[width=0.45\textwidth]{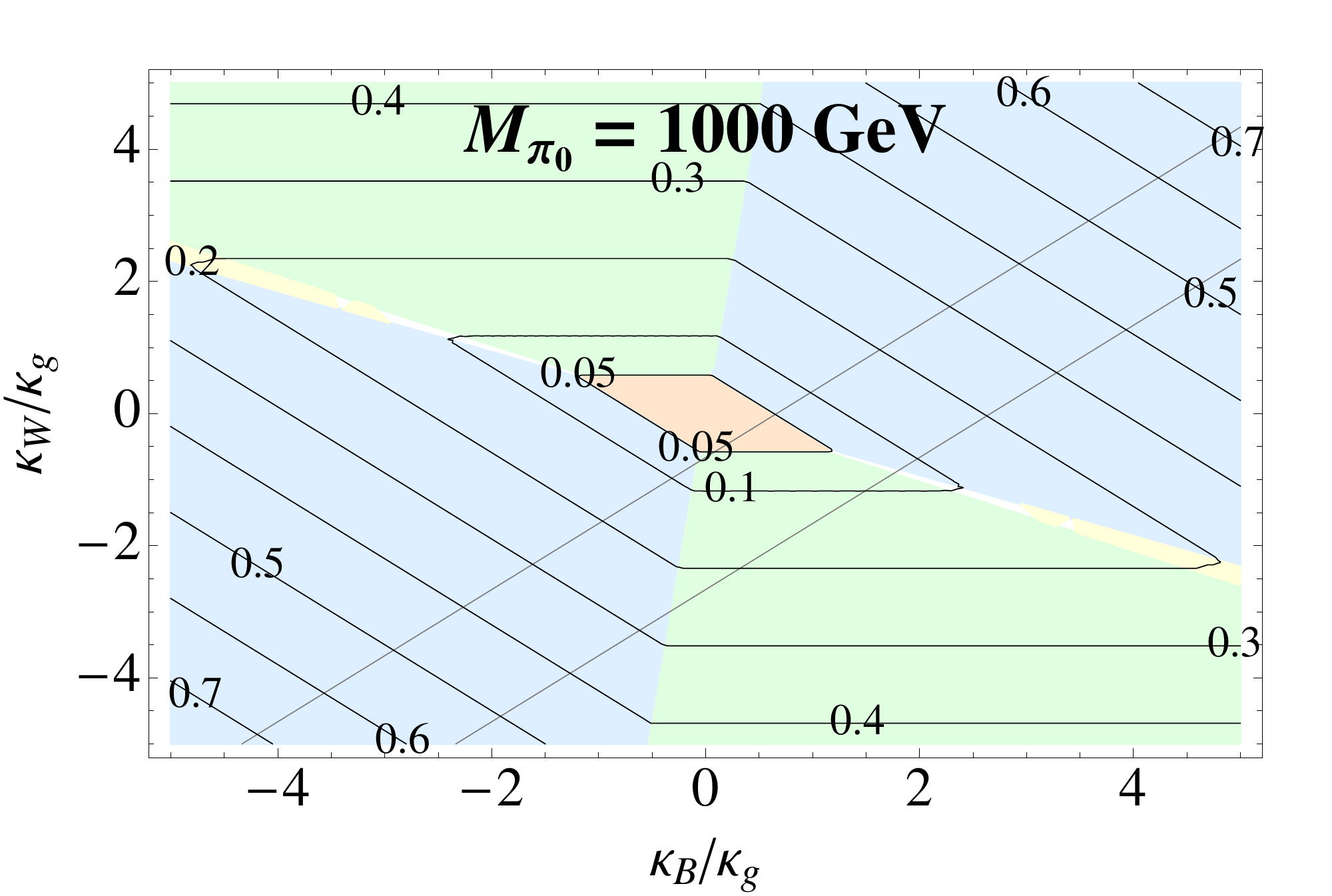} \\ 
\includegraphics[width=0.45\textwidth]{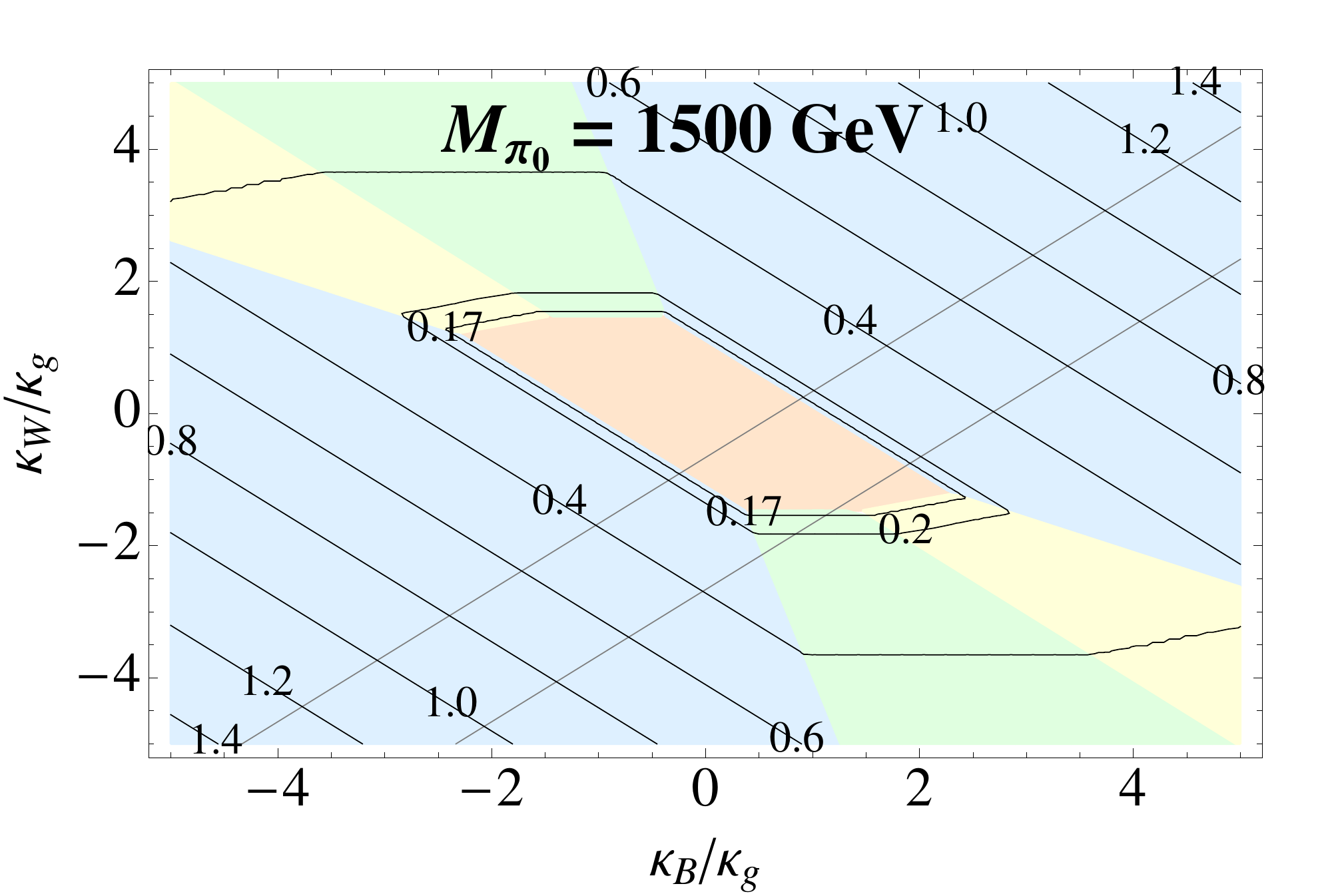} &
\includegraphics[width=0.45\textwidth]{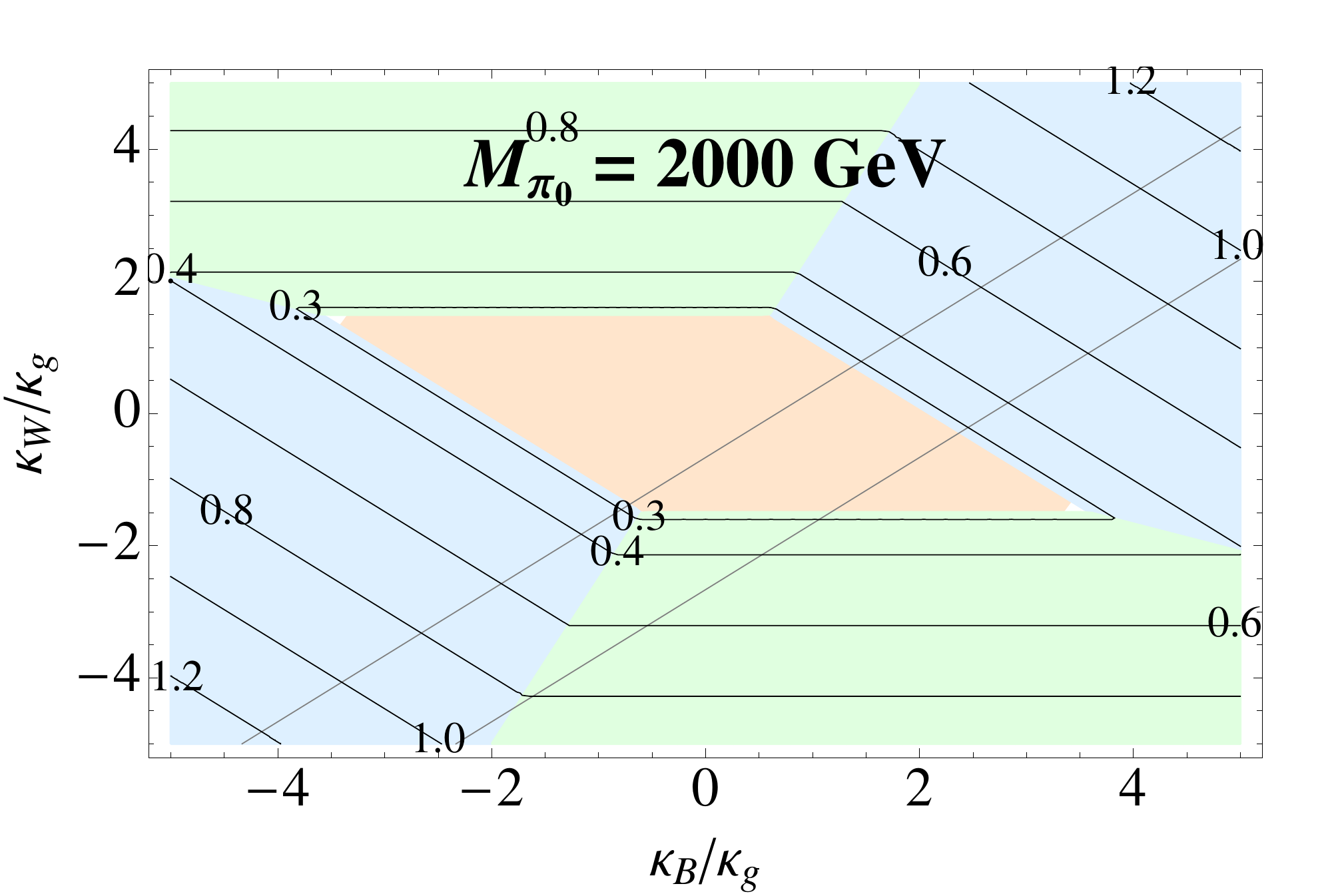} \\ 
\includegraphics[width=0.45\textwidth]{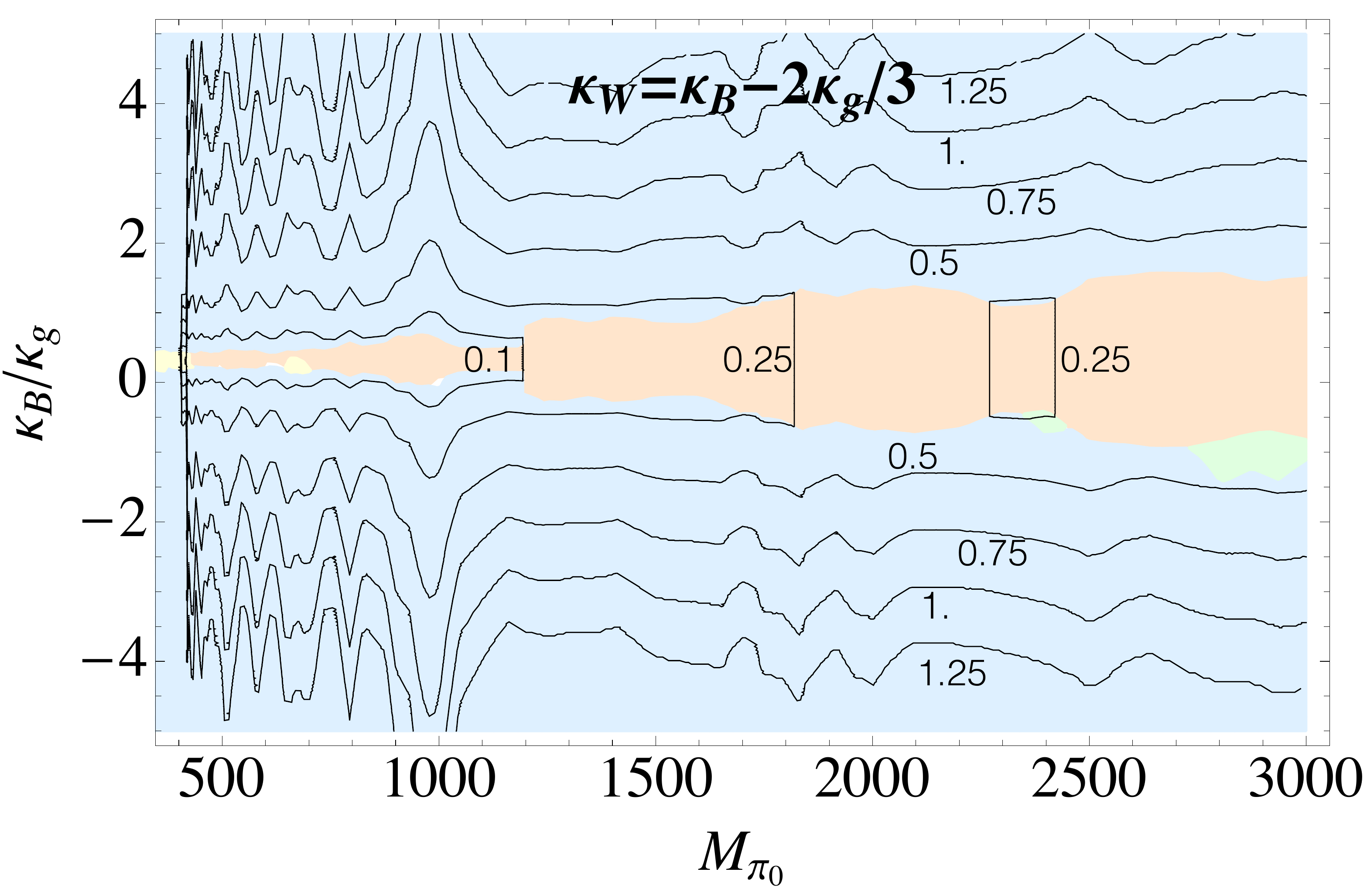} & 
\includegraphics[width=0.45\textwidth]{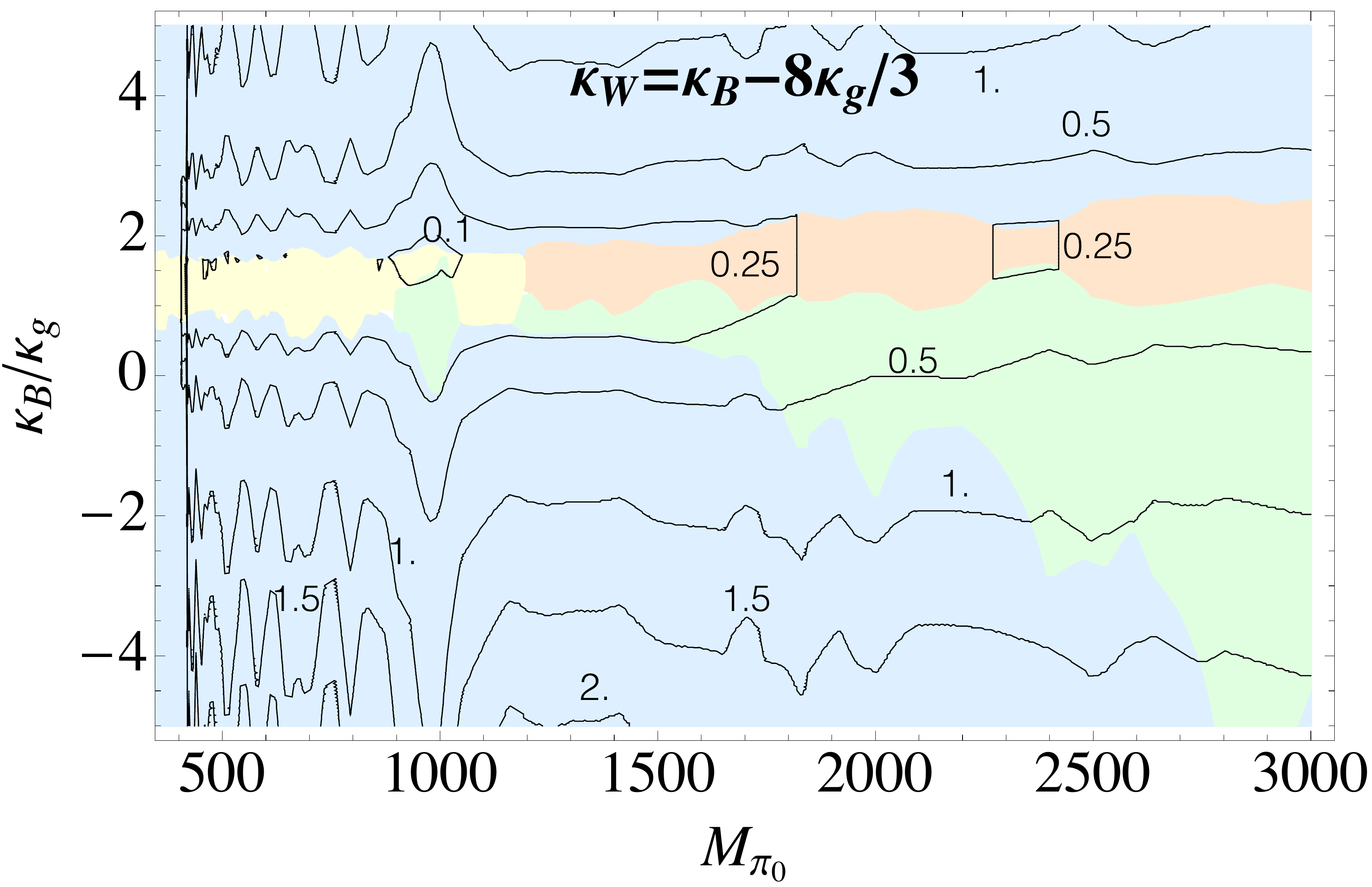} 
\end{tabular}
\caption{Values of $C_t / \kappa_g$ at which $t\bar{t}$ searches start to yield strongest constraint on the production cross section in pb. The top four figures show the results in the $\kappa_W/\kappa_g$ vs. $\kappa_B/\kappa_g$ plane for the same masses $M_\sigma$ as in Fig.~\ref{fig:kbkWMI}. The two figures on the bottom show the results in the $\kappa_B/\kappa_g$ vs. $M_\sigma$ plane, along the same two lines in the  $\kappa_W/\kappa_g$ vs. $\kappa_B/\kappa_g$ parameter plane as in Fig.~\ref{fig:kapvsM}.}
\label{fig:cok}
\end{figure}

In addition to the di-boson bounds presented in Figs.~\ref{fig:kbkWMI} and \ref{fig:kapvsM}, $t\bar{t}$ resonant searches provide a further constraint, that depends on the precise value of the ratio of couplings $C_t/\kappa_g$: we present here a simple way to extract the bound on the cross section. The $t\bar{t}$ constraint dominates over the di-boson bounds if
\beq
\left(\sigma_{13}\times BR_{t\bar{t}}\right)_{\rm{exp}} <
\left(\sigma_{13}\times BR_{\rm{bosons}}\right)_{\rm max} \times BF_{gg/\rm{bosons}} \times BF_{tt/gg},
\label{eq:cok}
\eeq
where $\left(\sigma_{13}\times BR_{t\bar{t}}\right)_{\rm{exp}}$ is shown in Fig.~\ref{fig:bds} (f) and the value of $\left(\sigma_{13}\times BR_{\rm{bosons}}\right)_{\rm max} $ can be extracted from Figs.~\ref{fig:kbkWMI} and \ref{fig:kapvsM}. The values of $BF_{gg/\rm{bosons}}$ and $BF_{tt/gg}$ are shown in Fig.~\ref{fig:BR}, and  $BF_{tt/gg}$ is the only quantity that depends on $C_t/\kappa_g$ (scaling quadratically with it).
Thus, given a set of values of the couplings, one can easily extract the dominant bound.
To quantify the relevance of top decays, following Eq.~\eqref{eq:cok} we determined the minimum value of $C_t/\kappa_g$ as a function of $\kappa_B/\kappa_g$ and $\kappa_W/\kappa_g$, $M_{\pi_0}$ above which the decay into tops yields the strongest constraint. The results are shown in Fig.~\ref{fig:cok} for a set of sample masses (4 plots on the top), and projected along the two model lines (two plots on the bottom).
 The plot shows that in the regions where final states with EW bosons dominate the bound, the $t\bar{t}$ final state overcomes the constraint for large values of the top couplings with $C_t \sim \kappa_g$. On the other hand, in the central region where $gg$ drives the bound, much smaller values of the top coupling are enough to drive the constraint.

\subsection{Phenomenology of the color octet}

\begin{figure}[t]
\begin{tabular}{cc}
\includegraphics[width=0.5\textwidth]{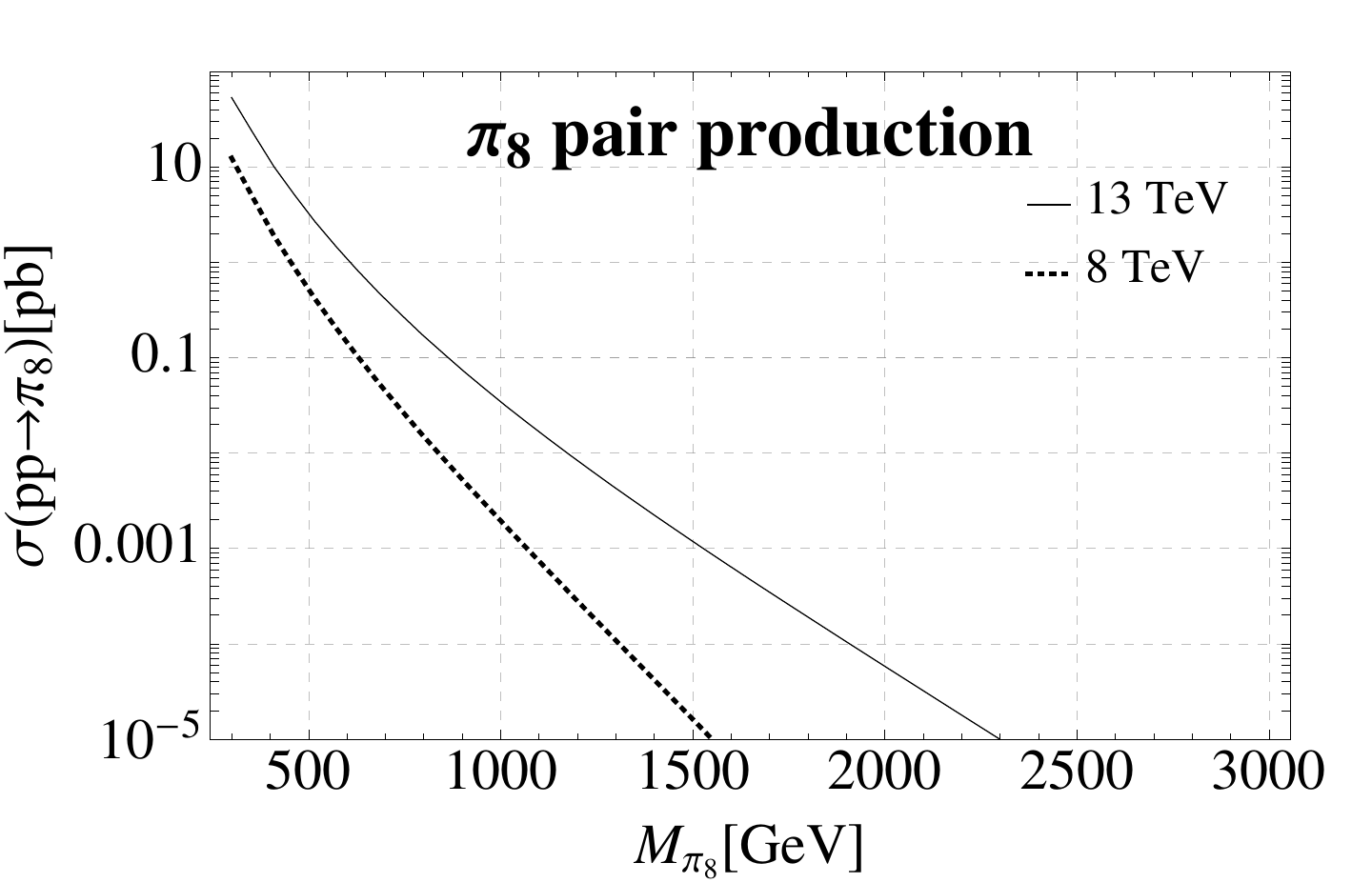} 
\includegraphics[width=0.5\textwidth]{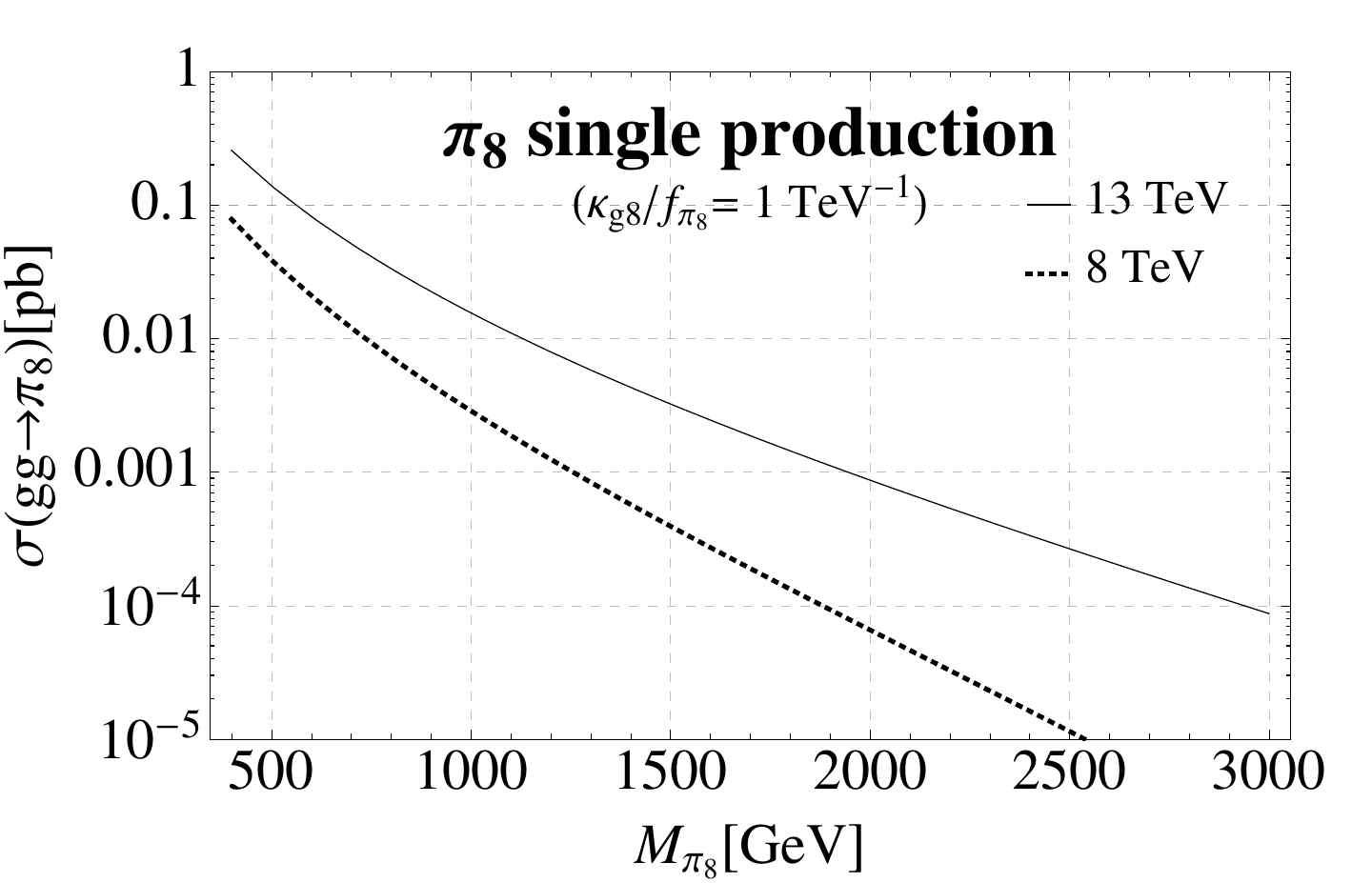} 
\label{fig:cs_octet}
\end{tabular}
\caption{Production cross sections of the color octet pNGB as a function of $M_{\pi_8}$ at 8 and 13 TeV LHC. Left: Pair-production via QCD interactions. Right: single-production (through gluon fusion) for $\kappa_{g8}/f_{\pi_8} = 1$~TeV$^{-1}$. Single production scales with $(\kappa_{g8}/f_{\pi_8})^2$.}  
\label{fig:octProd}
\end{figure}

The color octet $\pi_8$, which is present in all models discussed in Sec. \ref{sec:models}, can be described by the effective Lagrangian
\begin{multline}
\mathcal{L}_{\pi_8} = \frac{1}{2} (D_\mu \pi_8^a)^2- \frac{1}{2} m_{\pi_8}^2 (\pi_8^a)^2 + i\ C_{t8} \frac{m_t}{f_{\pi_8}} \pi_8^a\ \bar{t}\gamma_5 \frac{\lambda^a}{2} t\\
 + \frac{\alpha_s \kappa_{g8}}{8 \pi f_{\pi_8}}\pi_8^a\ \epsilon^{\mu\nu\rho\sigma}\left[ \frac{1}{2}d^{abc}\ G^b_{\mu\nu}G^c_{\rho\sigma}+\frac{{g'} \kappa_{B8}}{{g_s}  \kappa_{g8}}\ G^a_{\mu\nu}B_{\rho\sigma}  \right],
\label{eq:PhiLag}
\end{multline}
where the covariant derivative contains QCD interactions with gluons. In the models discussed in this article ($f_{\pi_8} = f_\chi$), matching with Eq.s (\ref{eq:WZWoctet}) and (\ref{eq:8top}), the coefficients are equal to
\beq\label{eq:kappas8}
\kappa_{g8} = c_5 \sqrt{2} d_\chi\,, \quad \kappa_{B8} = c_5 2\sqrt{2} d_\chi Y_\chi \,, \quad C_{t8} = c_5 n_\chi \sqrt{2}\,.
\eeq
The octet $\pi_8$ is produced at the LHC in pairs via QCD interactions or singly via gluon fusion\footnote{Single production through gluon-photon fusion would require very large hypercharge of the constituent fermions, so we neglect it, here. Single production from $t\bar{t}$ fusion is also suppressed by the need of creating top pairs from gluon splittings as well as by the additional $(m_t/f_{\pi_8})^2$ suppression from \eqref{eq:PhiLag}.}. The production cross section at the LHC for 8 and 13 TeV are shown in Fig.~\ref{fig:octProd}. Like for the singlet, we calculated the cross section at leading order  (without K-factor)  using  {\sc MadGraph 5} and cross-checked against 
{\sc CalcHEP} both with the NNPDF23LO (\verb|as_0130_qed|) PDF set,
and the QCD scale set to  the mass of the resonance.

The partial widths of $\pi_8$ from the Lagrangian (\ref{eq:PhiLag}) are given by:
\beq
\Gamma_{gg}  &= & \frac{5 \alpha_s^2 \kappa_{g8}^2 M_{\pi_8}^3}{768 \pi^3 f_{\pi_8}^2},\\
\Gamma_{g \gamma} &=& \frac{\alpha \alpha_s \kappa_{B8}^2 M_{\pi_8}^3}{128 \pi^3 f_{\pi_8}^2}, \\
\Gamma_{g Z} &=& \frac{\alpha \alpha_s \tan^2 \theta_W\ \kappa_{B8}^2 M_{\pi_8}^3}{128 \pi^3 f_{\pi_8}^2} \left(1- \frac{m_Z^2}{M_{\pi_8}^2}\right)^3,\\
 \Gamma_{tt} &=& \frac{C_{t8}^2 M_{\pi_8}}{16\pi} \frac{m_t^2}{f_{\pi_8}^2} \left(1-4\frac{m_t^2}{M_{\pi_8}^2}\right)^{1/2}\,. 
 \label{eq:octPW}
\eeq
Like for the singlets, the total width is always small, as numerically shown for the di-gluon partial width below:
\beq
\Gamma^{\pi_8} (gg) \sim 2~\mbox{MeV}~ \left( \frac{1~\mbox{TeV}}{f_{\pi_8}/\kappa_{g8}} \right)^2 \left( \frac{M_{\pi_8}}{1~\mbox{TeV}} \right)^3\,.
\eeq
In all the models under study in this paper, the ratio of the two WZW couplings only depends on the hypercharge $Y_\chi$, that can take two values $Y_\chi= 1/3$ or $2/3$, depending on the model.
Instead of a complete model independent analysis, we will impose this constraint that fixes the ratios of decay rate in the bosons. In analogy with the singlets, we can thus define
\beq
BF^{\pi_8}_{g\gamma/gg}\equiv\frac{Br(\pi_8\rightarrow \gamma g)}{Br(\pi_8\rightarrow g g)} = \frac{24\alpha}{5\alpha_s} Y_\chi^2\,,\quad BF^{\pi_8}_{gZ/gg} = \tan^2 \theta_W\ BF^{\pi_8}_{g\gamma/gg} \left(1- \frac{m_Z^2}{M_{\pi_8}^2}\right)^3\,.
\eeq
Besides a mild mass dependence, the rate into a $Z$ is suppressed by the Weinberg angle: numerical values of these ratios are reported in Table~\ref{octetBR}.
The decay in tops, however, strongly depends on the ratio $C_{t8}/\kappa_{g8}$ which is very dependent on the details of the model, and on the mass of the octet:
\beq
BF^{\pi_8}_{tt/gg}\equiv\frac{Br(\pi_8 \rightarrow t\bar{t})}{Br(\pi_8 \rightarrow gg)} = \frac{48\pi^2}{5 \alpha_s^2}\frac{C_{t8}^2}{\kappa_{g8}^2}\frac{m_t^2}{M_{\pi_8}^2}\left(1-4 \frac{m_t^2}{M_{\pi_8}^2}\right)^{1/2}\, .  \label{eq:ggott}
\eeq
In the following we will treat the ratio
${C_{t8}}/{\kappa_{g8}}$
as a free parameter. For the models in Section~\ref{sec:PNGBtheory}, this ratio is always smaller than 1 and it vanishes when the couplings to tops is absent.

\subsubsection{Searches and bounds for pair-produced color octets}

Spin zero color octets have already received some attention in the literature \cite{Krasnikov:1995kn,Manohar:2006ga,Gresham:2007ri}, as they arise in other models like sgluons in extended supersymmetry \cite{Choi:2008ub,Plehn:2008ae}, and they are copiously produced at hadronic colliders. Their decays lead, in general, to several final states due to the four allowed decay modes:  $t\bar{t}$, $gg$, $g\gamma$ and $gZ$.
However, most of these final states are not explicitly searched for in the ATLAS and CMS exotics searches, with two exceptions. 
The search for pair produced resonances with each one decaying into two jets done by CMS with 8 TeV data \cite{Khachatryan:2014lpa} and ATLAS with 13 TeV data \cite{ATLAS:2016sfd} can be straightforwardly reinterpreted to cover the $(gg)(gg)$ final state \footnote{For the 8 TeV search, CMS presents bounds for an inclusive search for R-parity violating decays of pair produced squarks as well as for a coloron (a scalar octet) search. We use the latter, which yields weaker bounds, such that the exclusions quoted here are conservative. We remark that the 13 TeV search \cite{ATLAS:2016sfd} yields stronger bounds than both searches at 8 TeV.}. Analogously, ATLAS has searches for scalar color octets producing a 4-top final state in the 8 TeV data, both in the same-sign di-lepton channel \cite{Aad:2015gdg} and in the  lepton-plus-jets final state \cite{Aad:2015kqa} \footnote{A first search with the 13 TeV data has been published \cite{ATLAS4t1l13}, however not presenting the case of the color octet. Thus, we cannot use directly these results.}.  

\begin{figure}[t]
\begin{tabular}{cc}
\includegraphics[width=0.5\textwidth]{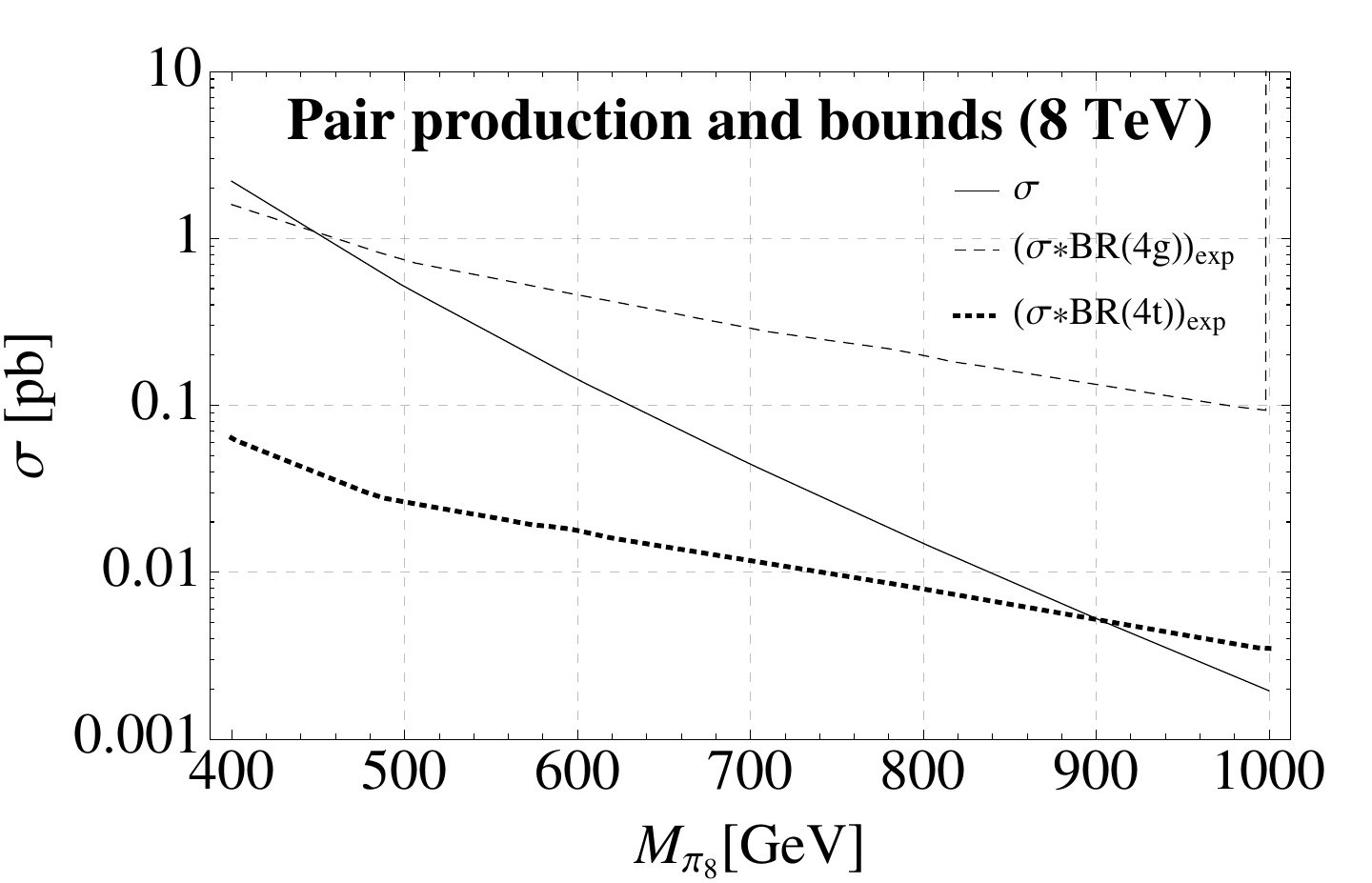} &
\includegraphics[width=0.5\textwidth]{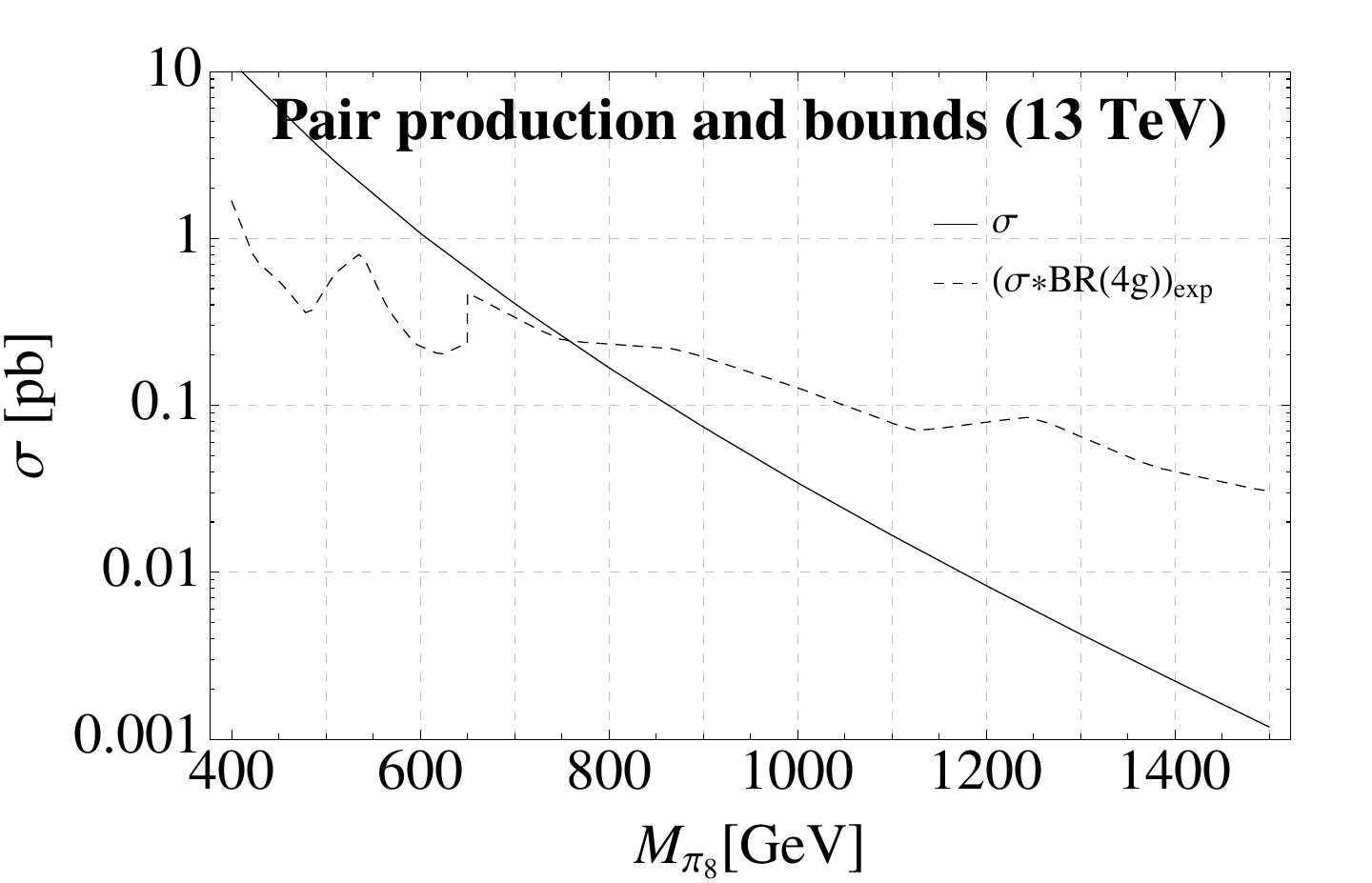} 
\end{tabular}
\caption{Color octet pair production cross section (from Ref.~\cite{Cacciapaglia:2015eqa}) and the current bounds on $\sigma\times BR(pp\rightarrow \bar{t}t\bar{t}t)$ \cite{Aad:2015kqa,Aad:2015gdg}  and $\sigma\times BR(pp\rightarrow 4 j)$ \cite{Khachatryan:2014lpa} for $\sqrt{s}= 8$~TeV (left) and $\sigma\times BR(pp\rightarrow 4 j)$ \cite{ATLAS:2016sfd} for $\sqrt{s}= 13$~TeV (right).}
\label{fig:octetp1}
\end{figure}

A direct comparison of cross sections can be seen in Fig.~\ref{fig:octetp1} where we show the pair production cross sections at 8 and 13 TeV, together with the bounds on the cross section times branching ratios in the two covered final states, $(\sigma \times BR(4g))_{\rm exp}$ and $(\sigma \times BR(4t))_{\rm exp}$ respectively.
The bound on the 4-top final state can be directly compared to the production cross section for large $C_{t8}/\kappa_{g8}$, for which the $BR$ in $t\bar{t}$ is nearly 100\%, excluding masses below $880$~GeV. On the other hand, in the absence of top couplings, the bound on 4-jet depends on the BR which depends on $Y_\chi$ \footnote{We give branching ratios for a reference mass $M_{\pi_8} = 1$~TeV, here.}:
\beq
BR_{gg} (Y_\chi = 2/3) \approx 75\%\,, \qquad BR_{gg} (Y_\chi=1/3) \approx 94\%\,.
\eeq 

\begin{figure}[t]
\begin{tabular}{cc}
\includegraphics[width=0.5\textwidth]{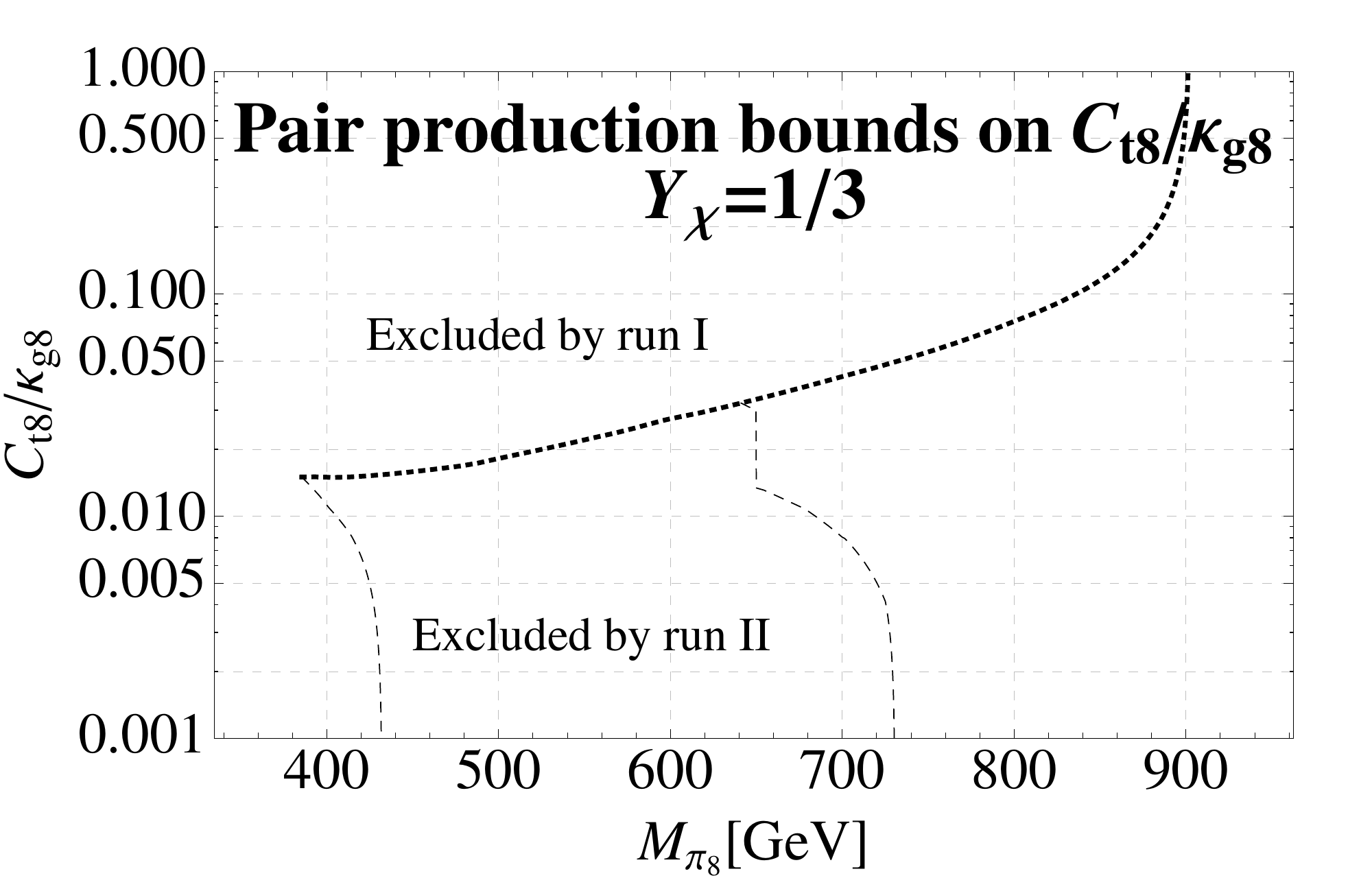} &
\includegraphics[width=0.5\textwidth]{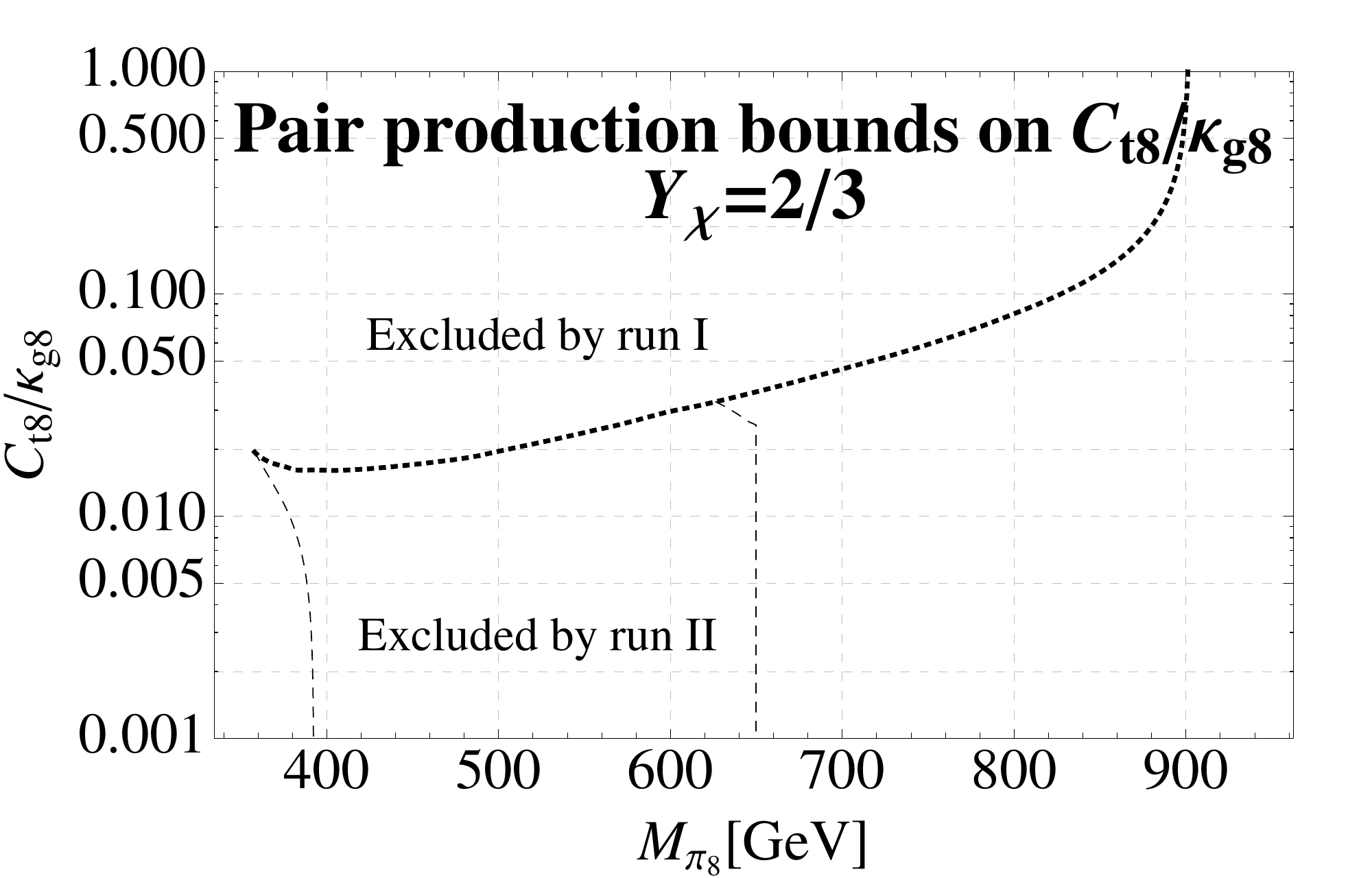} 
\end{tabular}
\caption{Bounds on $C_{t8}/\kappa_{g8}$ as a function of $M_{\pi_8}$ from $4t$ and $4j$ searches. The parameter regions excluded by 8 and 13 TeV data are also shown, as they do not depend on the overall value of the couplings.}
\label{fig:octetp2}
\end{figure}

A recast of the pair production bounds as a function of $C_{t8}/\kappa_{g8}$ is shown in Fig.~\ref{fig:octetp2}. For vanishing  top coupling, the lower bound on the mass from the 13 TeV search gives $650$ ($700$) GeV for $Y_\chi = 2/3$ ($1/3$).

\bigskip

We wish to point out that pair produced color octets have a large number of additional final states after decay, some of which promise a competitive sensitivity. Possible final states are all combinations of $t\bar{t}$, $gg$, $g\gamma$ and $gZ$. The $4g$ and $4t$ channels are covered by current ATLAS and CMS searches which have been used above in order to obtain constraints on the parameter space. The $(gg)(t\bar{t})$ channel can be searched for in a single-lepton search similar to the $4t$ search \cite{Aad:2015kqa}, or in a search for two leptonically decaying tops and two jets (which however suffers from a lower branching ratio of the tops into leptons). Such a search would yield additional bounds which can be relevant if the octet decay into $t\bar{t}$ and $gg$ are comparable.

A very interesting option is to search for $(g\gamma)(gg)$, i.e. a di-jet resonance with the same invariant mass as a photon-jet resonance. As compared to the two di-jet channel $(gg)(gg)$, the cross section of the $(g\gamma)(gg)$ channel is only reduced by $\sigma(pp\rightarrow \pi_8\pi_8\rightarrow \gamma ggg) /   \sigma(pp\rightarrow \pi_8\pi_8\rightarrow 4g) \approx 0.4$ (0.1) for  $Y_\chi = 2/3$ (1/3) \footnote{We use the branching ratios for an octet mass of 1 TeV given in Table \ref{octetBR}.}. However, the background of the process can be vastly reduced due to the photon in the final state.

Finally, the decay $\pi_8\rightarrow gZ$ would also allow to search for, e.g., $(gZ_{ll})(gg)$ or $(gZ_{ll})(t\bar{t})$ with leptonic $Z$'s. These channels promise very low background, but the signal cross section is reduced as compared to the $(g\gamma)(gg)$ and $(g\gamma)(t\bar{t})$ states, by a factor of $BR(\pi_8\rightarrow g Z) \times BR(Z\rightarrow ll)/ BR(\pi_8\rightarrow g \gamma) \approx 0.02$.

\subsubsection{Searches and bounds for a singly produced color octet}

\begin{figure}[t]
\begin{tabular}{cc} 
\includegraphics[height=0.25\textheight]{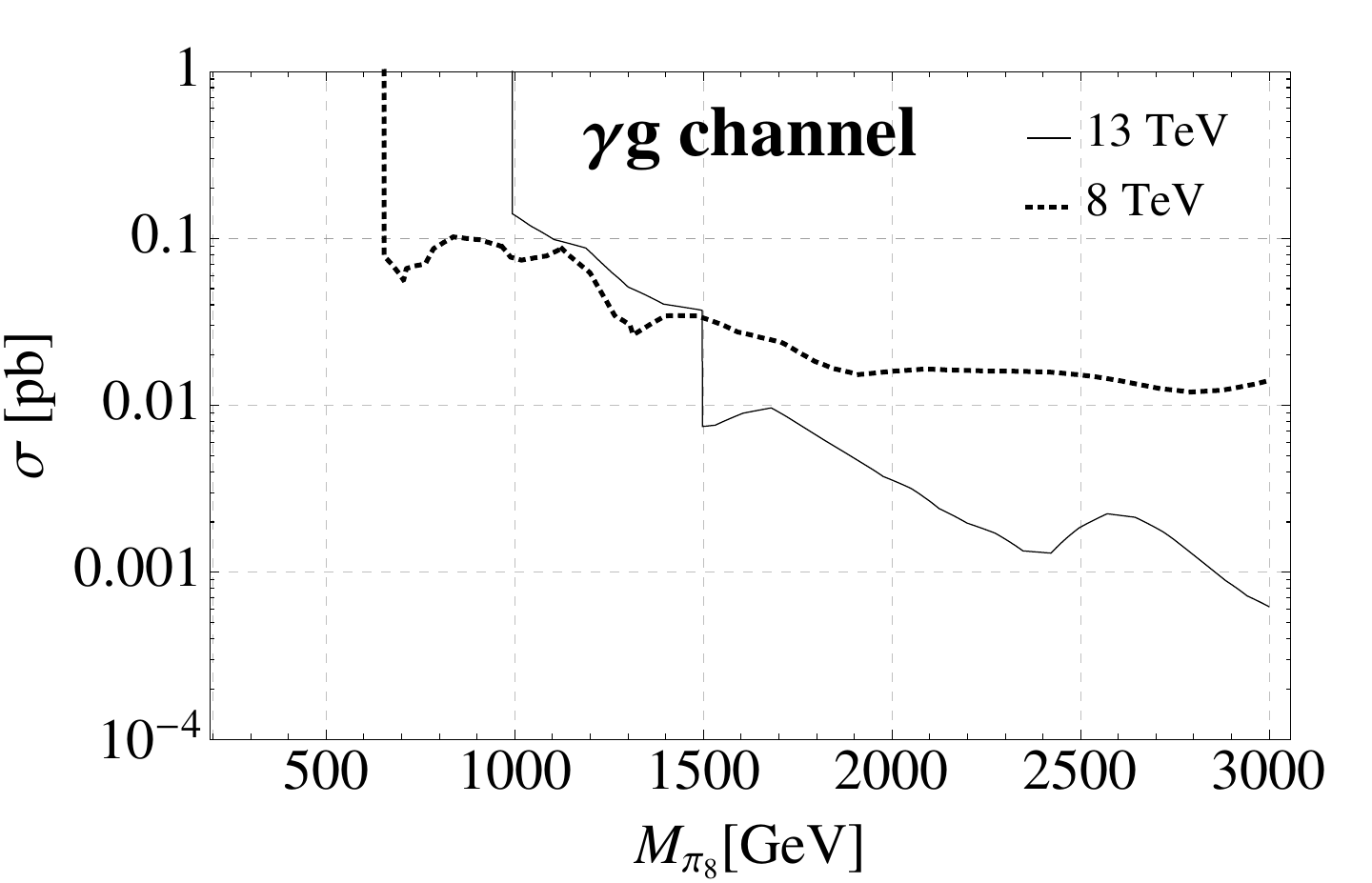}
\end{tabular}
\caption{
Bounds from excited quark searches (final state:  $\gamma \,j$) at 13 TeV \cite{Aad:2015ywd,CMS:2016qtb} and 8 TeV \cite{Aad:2013cva,Khachatryan:2014aka} which we use to constrain the $\gamma g$ channel of color-octet decay.
The 8 TeV limits are
rescaled to the value of the 13 TeV cross section times branching ratio (8 TeV bounds have been appropriately rescaled by the ratio of 13 TeV / 8 TeV production cross section for gluon fusion).}
\label{fig:bdsoct}
\end{figure}

The color octet $\pi_8$ can be singly produced via gluon fusion, with cross section at $\sqrt{s}=8$~TeV and $13$~TeV shown in Fig.~\ref{fig:octProd}. The possible final states for single-production are $t\bar{t}$, $gg$, and $g\gamma$, and $gZ$.  The bounds on the di-jet, and $t\bar{t}$ resonances are shown in Fig.~\ref{fig:bds} (e) and (f), and can be reused here as the kinematical differences due to the color of the resonance are subleading. 
The contribution of a color octet to $t\bar t$ production was also analyzed in~\cite{Eichten:1994nc} in the similar context of multiscale technicolor.
The $g\gamma$ final state offers a clean channel due to the presence of an energetic photon~\cite{Belyaev:1999xe}.
While dedicated searches are not available, one can easily adapt searches for excited quarks by both ATLAS \cite{Aad:2013cva,Aad:2015ywd} and CMS \cite{Khachatryan:2014aka,CMS:2016qtb} where the gluon is replaced by a light quark jet. The bounds from these searches are summarized in Fig.~\ref{fig:bdsoct}, where we take the strongest bound from all searches for a given mass $M_{\pi_8}$ expressed in terms of the cross section at $\sqrt{s}=13$~TeV.
The $gZ$ final state can be constrained by two published searches: the $Z_{\rm had} j$ search by CMS \cite{Khachatryan:2014hpa} at 8~TeV, and by mono-jet searches (sensitive to invisible decays of the $Z$, $Z_{\rm inv} j$) by ATLAS \cite{Aad:2015zva,Aaboud:2016tnv} and CMS \cite{Khachatryan:2014rra,Aaboud:2016tnv} at both 8 and 13~TeV. In Ref.~\cite{Carpenter:2015gua}, a recast of the 8 TeV searches in di-boson channels for a color octet scalar are presented, showing that, for the $Zg / gg$ branching ratios in the models considered in this article, the $Z g$ channel yields subleading bounds as compared to the $gg$ and $g\gamma$ channels. We therefore do not include the $Zg$ channel in our analysis.

\begin{figure}[t]
\begin{tabular}{cc}
\includegraphics[width=0.5\textwidth]{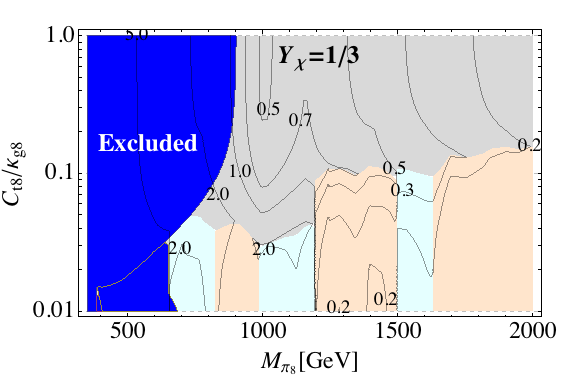} & 
\includegraphics[width=0.5\textwidth]{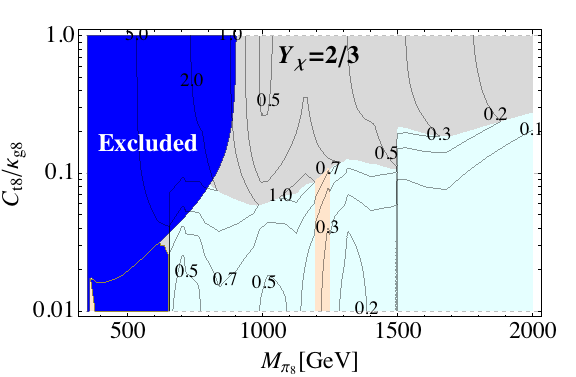} \\
\end{tabular}
\caption{Bounds on the production cross section (in pb) in the  $C_{t8}/\kappa_{g8}$ vs. $M_{\pi_8}$ plane for $Y_\chi=1/3$ (left) and $Y_\chi=2/3$ (right). The blue region is excluded by octet pair production searches. The currently strongest bounds arise from the $t\bar{t}$ channel (gray regions), $gg$ channel (orange regions) and the $g\gamma$ channel (light-cyan regions).}
\label{fig:kapoctvsM}
\end{figure}

To combine the constraints on singly produced color octets, we follow a strategy similar to the one designed for the singlet pseudo-scalars in Sec.~\ref{sec:SingPheno}: the analysis is simpler because the color octet bosonic ratios  $BF^{\pi_8}_{g\gamma/gg}$ and $BF^{\pi_8}_{gZ/gg}$ in Eq.~\eqref{eq:ggott} are fixed  up to a discrete choice $Y_\chi=1/3$ or $2/3$. For these two choices, we can directly translate the bounds on the $g\gamma$ channel (Fig.~\ref{fig:bdsoct}) and the $gg$ and $t\bar{t}$ channels (Fig.~\ref{fig:bds}) into bounds on the $\pi_8$ production cross section as a function of the mass and the model parameters $C_{t8}/\kappa_{g8}$, as shown in Fig.~\ref{fig:kapoctvsM}. 

\section{Implications for composite models}\label{sec:specific pheno}

After presenting general results in Section \ref{sec:pheno}, we want to look back at the models introduced in Section \ref{sec:models} and study how the present searches can constrain the presence of the singlet and octet pseudo-scalars, and what are the prospects at the LHC Run 2. Instead of looking at all models, we will derive general properties before focusing on a few interesting cases.
The first step is to connect the general Lagrangians used in the previous section with the couplings derived in Section \ref{sec:PNGBtheory}: the two scales $f_\pi$ and $f_{\pi_8}$ that we used to normalize the couplings are arbitrary, thus allowing us to chose the most convenient normalization.

For the singlets, the most natural choice is to normalize $f_\pi = f_\psi$ for both mass eigenstates, as this is the scale most directly connected to the EW symmetry breaking and thus to the fine tuning in the Higgs sector. Furthermore, the couplings will depend on the mixing angle $\alpha$ that defines the mass eigenbasis.
For the lightest singlet $a$, the couplings to field strengths in Eq.~(\ref{eq:Lanom_singlet}) are mapped to the model independent parametrization Eq.~(\ref{eq:Lsigma}) by
\beq
\kappa_A =  c_5 \frac{f_\psi}{f_{a_\psi}} \left( C_A^{\psi}\ \cos\alpha + \frac{f_{a_\psi}}{f_{a_\chi}} C_A^{\chi}\ \sin\alpha \right) \qquad \mbox{with} \quad A = g,~W,~B\,.
\eeq
Similarly, the coupling to tops is matched by
\beq
C_t = c_5  \frac{f_\psi}{f_{a_\psi}} \left(n_\psi \cos\alpha + n_\chi \frac{f_{a_\psi}}{f_{a_\chi}}\sin\alpha\right)\,,
\eeq
where we recall that $n_{\psi/\chi}$ are the U(1)$_{\psi/\chi}$ charges associated to the top mass operator.
The couplings of $\eta'$ are obtained from the same formulas with the replacement $\alpha\rightarrow\alpha+\pi/2$.
Concerning the octet pseudo-scalar, it is convenient to normalize $f_{\pi_8} = f_\chi$, as this is the only scale directly connected to it. The couplings to the field strengths in  Eq.~(\ref{eq:PhiLag}) have already been identified in Eq.~(\ref{eq:kappas8}).

Once a specific model is chosen, the group theory factors are fixed, however the Chiral Lagrangian contains other unknown parameters: 4 decay constants $f_\psi$, $f_{a_\psi}$, $f_\chi$ and $f_{a_\chi}$, the singlet mass induced by the anomaly $M_A$, 2 explicit fermion masses $m_\psi$ and $m_\chi$ (in the following, as explained in Section \ref{sec:PNGBtheory}, we will work at $m_\psi = 0$), and the loop corrections to the octet mass (from QCD and tops).
In addition, we have a discrete choice of $n_\psi$ and $n_\chi$ associated to the operator that generates the top partners.
However, not all these parameters are on the same footing:
\begin{itemize}

\item[-] the decay constants, the anomaly mass $M_A$ and the loop corrections to the pNGB masses are dynamical quantities, in the sense that they can be calculated if the underlying dynamics is solved (on the Lattice, one may potentially compute all the ratios of these dimensional quantities, so that only a single tuneable scale remains);

\item[-] the underlying fermion masses are free parameters, external to the dynamics, and can assume any value as long as the Chiral expansion does not break down;

\item[-] the charges $n_\psi$ and $n_\chi$ are determined by the UV physics generating the partial compositeness couplings.

\end{itemize}
While no Lattice data is available, we will reduce the number of unknown parameters by imposing some reasonable relations between the decay constants:
\begin{enumerate}

\item we impose the ``large-$N_c$'' relation between the decay constants of the singlets and non-abelian pNGBs: $f_{a_\psi} = \sqrt{N_\psi} f_\psi$ and $f_{a_\chi} = \sqrt{N_\chi} f_\chi$;

\item we fix the ratio of the two remaining decay constants to be equal: $f_\chi = f_\psi$.
\end{enumerate}

Regarding the second relation, this simple choice is not entirely justified on dynamical grounds\footnote{We thank Michele Frigerio for pointing out this argument to us.}, and it is only chosen for its simplicity. The effect of making any other choice for this ration can be easily inferred by rescaling the couplings of the effective theory.

One could use an argument based on the MAC hypothesis~\cite{Raby:1979my} to estimate the ratio of the scales where the two condensates occur~\cite{Eichten:1981mu}. This argument has been used in~\cite{Ferretti:2014qta} to estimate the ratio
in the case of the $SU(4)$ hypercolor theory. A similar estimate for all models M1 to M12 yields the following ratios
\begin{equation}
\begin{tabular}{|c|c|c|c|c|c|c|c|c|c|c|c|c|}
\hline
&M1&M2&M3&M4&M5&M6&M7&M8&M9&M10&M11&M12\\
\cline{2-13}
$f_\psi/f_\chi$& 1.4 & 0.75 & 0.73 & 1.3 & 2.8 & 1.9 & 0.58 & 0.38 & 2.3 & 1.7 & 0.52 & 0.38  \\
\hline
\end{tabular}\,.
\end{equation}
As this arguments are semi-quantitative at best, we do not use these numbers in the paper, but only present them to show that it reasonable to expect the ratios to be of order one. 
Similar estimates have been performed earlier in the context of multi-scale walking technicolor~\cite{Lane:1989ej, Eichten:1994nc}.

Besides one decay constant $f_\psi$, setting the scale of condensation, the other 3 mass parameters can be traded for the 3 mass eigenvalues $m_a$, $m_{\eta'}$ and $m_{\pi_8}$. The mixing angle $\alpha$ between the two singlets is then related to the mass eigenvalues (and the value of $\zeta$) by Eq.~(\ref{eq:alphazeta}).

\bigskip
A first phenomenological observation is that the mass splitting $m_{\eta'}^2-m_a^2$ is constrained by Eq.~(\ref{eq: mass splitting}) and models with small $|\tan\zeta|$ predict a large mass splitting and vice versa. To better quantify this effect, Eq.~(\ref{eq: mass splitting}) that contains the minimum mass splitting, can be used to extract the maximal ratio of the two masses (achieved at minimal splitting with $\alpha = \zeta/2$):
\beq
\left. \frac{m_a}{m_{\eta'}} \right|_{\rm max} = \sqrt{\frac{1-\cos \zeta}{1+\cos \zeta}} = \left| \tan \frac{\zeta}{2} \right|\,.
\eeq
Numerical values of $\tan \zeta$ and the quantity in the above equation for the 12 models under consideration, and under our ansatz on the decay constants, are reported in Table \ref{tab:models1}.
In models with small $\tan \zeta$, like for instance M2, M7, M8 and M12, therefore, the light singlet tends to be substantially lighter than the second one and the octet. Another consideration is that the largest $m_a$ mass is correlated to the $\chi$-mass by $m_{a} \leq m_{a_\chi} \sin \zeta$: for the lighter singlet to be in the TeV range, one would thus need the mass generated by the $\chi$ to be in the multi TeV scale, implying that $\chi $ tends to behave like a heavy flavor - a fundamental fermion with sizeable mass compared to the condensation scale - and the chiral Lagrangian description needs to be modified. Those are qualitative arguments, but they tend to point towards a situation where only one of the two singlets ($\eta'$) may be relevant at the LHC.
\begin{table}[tb]
\begin{tabular}{|c||c|c|c|c|c|c|c|c|c|c|c|c|}
\hline
 & M1 & M2 & M3 & M4 & M5 & M6 & M7 & M8 & M9 & M10 & M11 & M12 \\
\hline
$- \tan \zeta$ & 0.91 & 0.45 & 0.91 & 1.82 & 1.82 & 1.29 & 0.32 & 0.41 & 3.26 & 3.26 & 0.82 & 0.38 \\
$\left. \frac{m_a}{m_{\eta'}} \right|_{\rm max}$ & 0.39 & 0.22 & 0.39 & 0.59 & 0.59 & 0.49 & 0.16 & 0.20 & 0.74 & 0.74 & 0.36 & 0.18 \\
\hline
\end{tabular}
\caption{Values of $\tan \zeta$ and of the maximum ration of the light/heavy singlet masses for the 12 models under consideration, assuming $f_\chi = f_\psi$. Note that $\tan \zeta$ is proportional to $f_\chi/f_\psi$.} \label{tab:models1}
\end{table}
In models with large $\tan \zeta$, like M4, M5, M6, M9 and M10, on the other hand, the two mass eigenvalues can be close to each other, and one can easily have a situation where both lie in the mass range where the LHC is sensitive.

Another more general consideration involves the value of the coupling to tops, for both singlets and octet. If such couplings are large, then the most sensitive final state for their detection at the LHC is in di-top (or 4-tops for pair produced octets), and the LHC cannot be very sensitive to the di-boson final states. On the other hand, models with small coupling to tops have a better chance to be detected in the di-boson final states, from which more information can be extracted.

In the following, instead of studying all the models, we will focus on two sample cases: they are chosen in such a way that the symmetries at low energy are the same, so that they can be described by the same low energy effective theory, and they have small couplings to tops.
Nevertheless, they differ in the value of $\tan \zeta$, which substantially affects the spectra and couplings of the singlets.
Complete tables reporting the numerical values of the couplings for all models can be found in Appendix~\ref{sec:modeltables}.

\subsection{Two explicit examples}

The two models we focus on are M8 and M9 (see Table~\ref{allmodels}), where M8 was first introduced in \cite{Barnard:2013zea} and its phenomenology partially studied in \cite{Cacciapaglia:2015eqa}. The two models are based on very different underlying gauge theories, however the global symmetry breaking pattern is the same,
\beq
\mathcal{G}/\mathcal{H} = \frac{SU(4)\times SU(6)\times U(1)}{Sp(4)\times SO(6)}\,,
\eeq
so that they can be described, at low energy, by the same chiral effective Lagrangian. To completely specify the phenomenology, the operators that couple to the top must also be chosen, in order to fix the charges of the top mass operators under the two global U(1)'s: we then focus on the 2 cases $(n_\psi, n_\chi) = (\pm 2, 0)$ and $(0, \pm 2)$. This choice allows to compare the case where the octet couples to tops (the latter) versus a situation where such coupling is absent (the former). Furthermore, we checked that top loops are always small corrections for masses above $500$~GeV in those two cases.
For completeness, we would like to specify the representation under the global symmetries of the chosen top partners in the various cases: note that the sign of the charge, now, matters while it is irrelevant for the phenomenology. The transformation properties of the top partner operators, $\mathcal{O}_1$ and $\mathcal{O}_2$, associated to the 4 charge choices are summarized in Table \ref{tab:toppartners}. Note that in both cases, the bound states contain 2 $\psi$'s and one $\chi$, and that either operator can be associated with the left-handed or right-handed tops, according to the transformation properties under the EW gauge group. Interestingly, in all cases the right-handed top can couple to a singlet of Sp(4) and the left-handed one to a 5-plet.

\begin{table}[tb]
\begin{tabular}{|l|c|c|c|}
\hline
$(n_\psi, n_{\chi})$ & $\mathcal{O}_{1/2}$ & SU(4)$_{U(1)_\psi, U(1)_{\chi}}$ & Sp(4) \\
\hline
$(2,0)$ & $\psi \psi \chi$  &  $(6 \oplus 10)_{2,1}$ &  $1 \oplus 5 \oplus 10$  \\
            & $\psi \bar{\psi} \bar{\chi}$ & $(1 \oplus 15)_{0,-1}$ & $1 \oplus 5 \oplus 10$ \\
\hline
$(-2,0)$ & $\bar{\psi} \bar{\psi} \chi$ & $(6)_{-2,1}$ & $1 \oplus 5$  \\
            & $\psi \bar{\psi} \bar{\chi}$ & $(1 \oplus 15)_{0,-1}$ & $1 \oplus 5 \oplus 10$ \\
\hline
$(0,2)$ & $\psi \psi \chi$  &  $(6 \oplus 10)_{2,1}$ &  $1 \oplus 5 \oplus 10$  \\
            & $\bar{\psi} \bar{\psi} \chi$ & $(6)_{-2,1}$ & $1 \oplus 5$  \\
\hline
$(0,-2)$ & $\psi \bar{\psi} \bar{\chi}$ & $(1 \oplus 15)_{0,-1}$ & $1 \oplus 5 \oplus 10$ \\
\hline
\end{tabular}
\caption{Representations of the top partners corresponding to the four choices of charges studied in this section. Either operator can be associated to $t_L$ or $t_R$. When only one operator is shown, both top chiralities are associated to it.} \label{tab:toppartners}
\end{table}

Once the gauge theory and the top partner operators are specified, the couplings of the singlets and octet can be calculated: as an example, we provide the numerical values in Table \ref{tab:numvalues}, where the couplings of the singlets are provided for a mixing angle $\alpha = \zeta$ (which corresponds to the decoupling of $\eta'$) and $\alpha = \zeta/2$ (corresponding to the minimal splitting). We recall that the couplings are normalised to $f_\psi$ for both singlets, and to $f_\chi$ (here set equal to $f_\psi$) for the octet. The table clearly shows that the two models give rise to very different values of the couplings, thus providing an handle apt to distinguish the two if a signal is detected. Armed with the values of the couplings in the Table, one can go back to the plots of the previous section and reconstruct the best constraint for each mass point.
In the following, we will put together all the constraints, and extract a lower bound on the decay constant $f_\psi$.

\begin{table}[tbh]
\begin{tabular}{|c|c|ccc|cc|}
\hline
Model &&$\kappa_g$&$\frac{\kappa_W}{\kappa_g}$&$\frac{\kappa_B}{\kappa_g}$&$\frac{C_t}{\kappa_g}\ (2,0)$&$ \frac{C_t}{\kappa_g}\ (0,2)$\\
\hline
M8&$a$&$-0.77 (-0.39)$&$-1.2 (-2.5)$&$1.5 (0.17)$& $-1.2 (-2.5)$&$0.40 (0.40)$\\
 &$\eta'$&$1.9 (2.0)$&$0.20 (0.096)$&$2.9 (2.8)$&$0.20 (0.0.96)$&$0.40 (0.40)$\\
 & $\pi_8$ & $7.1$ & $0$ & $1.3$ & $0$ & $0.40$ \\
\hline
M9&$a$&$-4.3 (-2.7)$&$-0.55 (-2.4)$&$2.1 (0.26)$&$-0.068 (-0.30)$& $0.18 (0.18)$\\
 &$\eta'$&$1.3 (3.6)$&$5.8 (1.3)$&$8.5 (4.0)$& $0.73 (0.16)$& $0.18 (0.18)$\\
 & $\pi_8$ & $16.$ & $0$ & $1.3$ & $0$ & $0.18$ \\
\hline
\end{tabular}
\caption{Couplings for the two models discussed in the text in the limit $\alpha = \zeta$ (and in parenthesis the values for $\alpha =
\zeta/2$).} \label{tab:numvalues}
\end{table}

\begin{figure}[t]
\begin{tabular}{cc}
\includegraphics[width=0.5\textwidth]{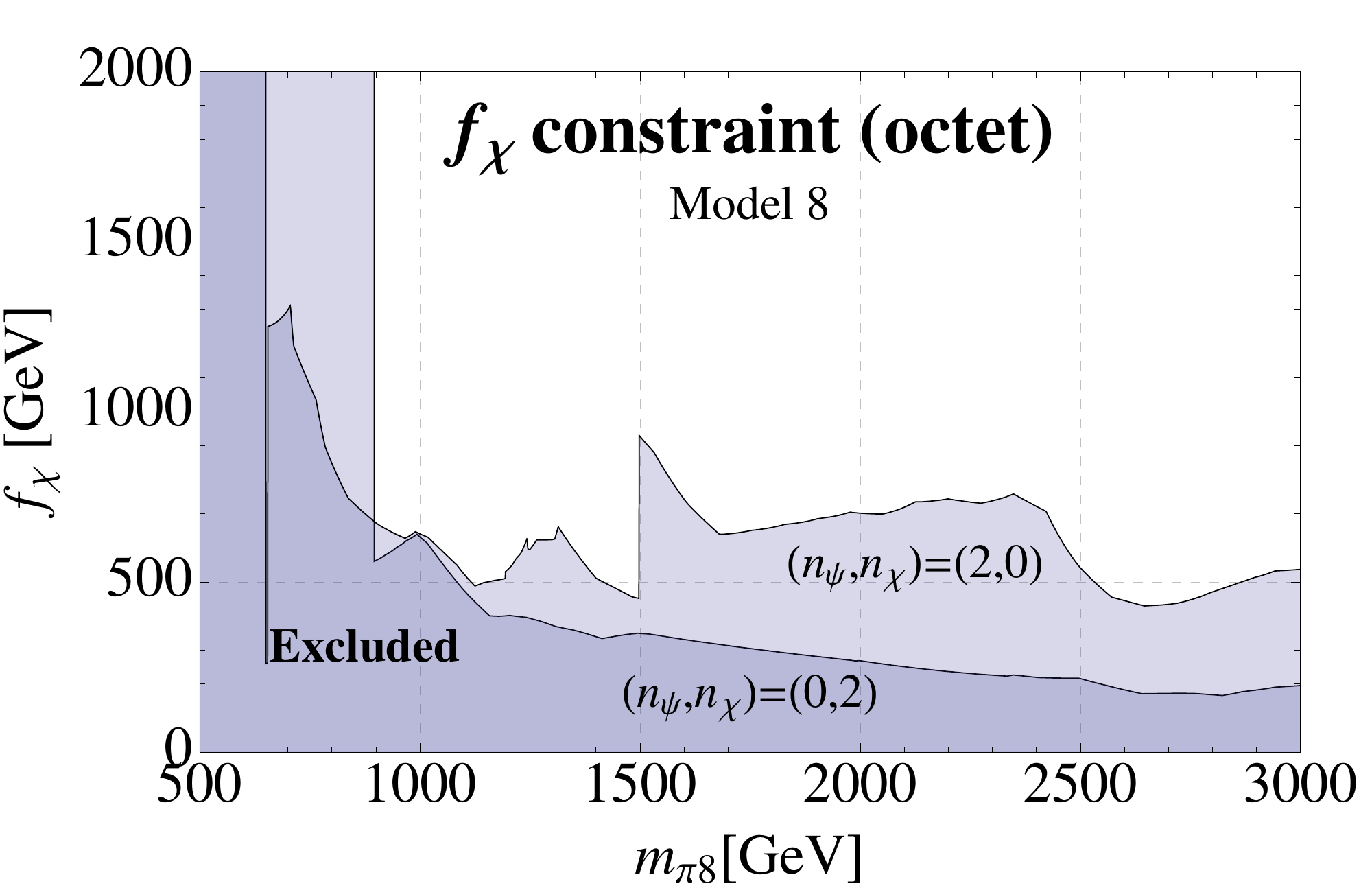} &
\includegraphics[width=0.5\textwidth]{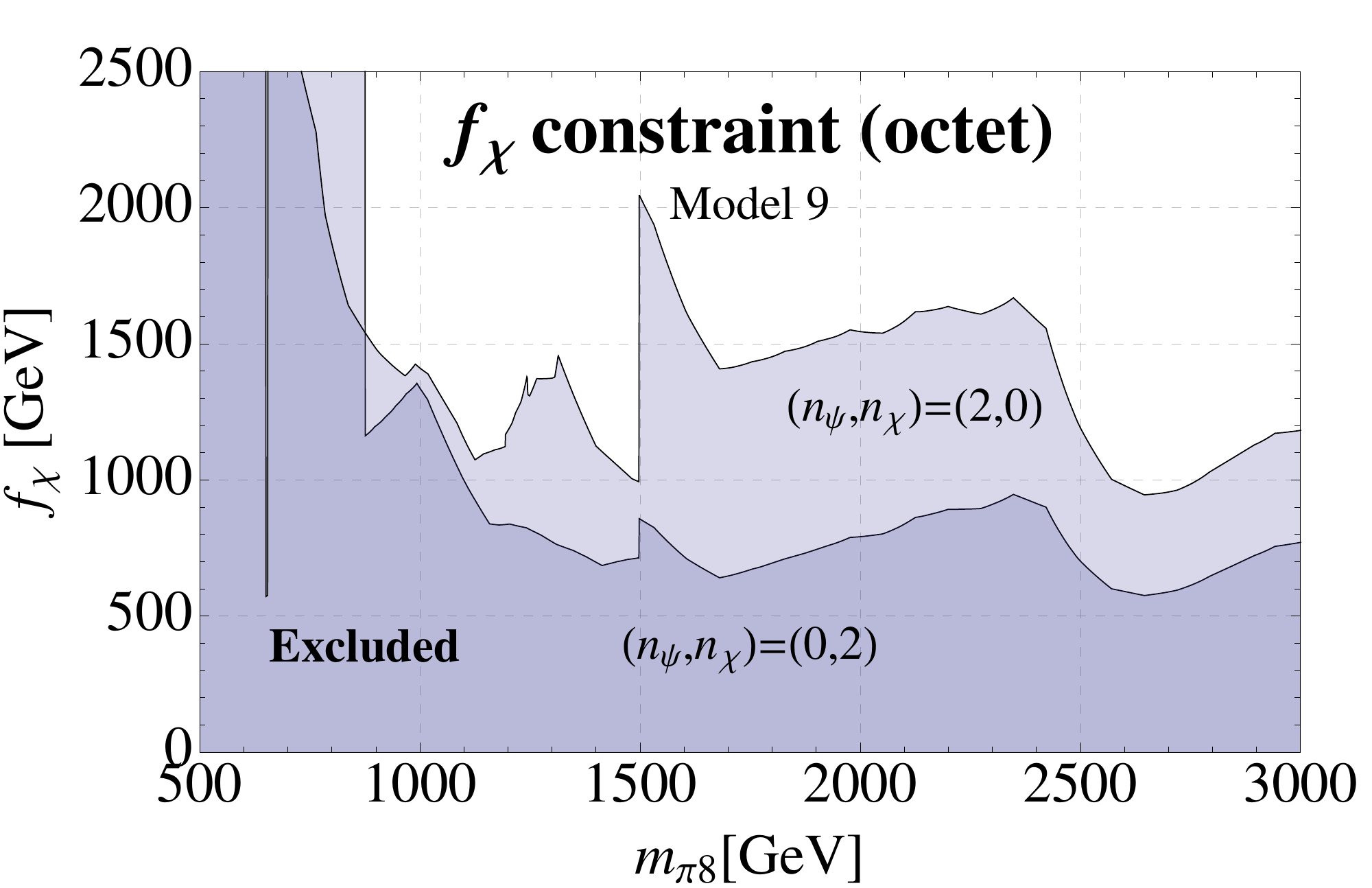} \\
\end{tabular}
\caption{Lower bounds on $f_\chi$ from LHC searches sensitive to the octet, in the case of model M8 (Left) and M9 (Right). The two lines correspond to the two choices of charges $(2,0)$ and $(0,2)$.}
\label{fig:octet}
\end{figure}

We start by discussing the octet in the two models: combining all the searches described in Section \ref{sec:pheno}, one can extract a lower bound on $f_\chi$ as a function of the octet mass. The final result is shown in Fig.~\ref{fig:octet}, Left plot for model M8, and Right plot for model M9. The first feature we observe is that the constraint is much stronger for M9, due to the larger coupling to gluons of the octet, as shown in Table~\ref{tab:numvalues}. The two charge assignments also bear very different features. For $(2,0)$ the coupling to tops vanishes and the bound is dominated by the di-boson final states: pair production searches (di-jet pairs - shown by the vertical line to the left) exclude masses below $640$~GeV. For masses above $650$~GeV, the constraint is given by $g\gamma$ searches, which start at this threshold as shown in Fig.~\ref{fig:bdsoct}.
In the case $(0,2)$, couplings to tops are present and affect the bounds. First of all, the lower bound from pair production on the mass is stronger, as coming from 4-top searches (we observe a lower bound of $880$~GeV). At higher masses, the bound from single production crucially depends on the model. For M8, which has larger couplings to tops w.r.t. the gluon couplings, the $t\bar{t}$ final state dominates over the whole mass range, providing weak bounds on $f_\chi$. On the contrary, for M9, the $t\bar{t}$ final states only dominated up to $1500$~\GeV, above which the $g\gamma$ final states dominates again: the weaker bound w.r.t. the $(2,0)$ case is due to a depletion of the signal because of a non-zero BR into tops.
On general grounds, we see that the constraint on the decay constant is always comparable if not stronger than the typical lower bound $f \geq 800$~GeV from EW precision tests. This comparison, however, is only valid if $f_\psi = f_\chi$.
While here we consider only the octet, these two models also feature a charged sextet in the spectrum: its phenomenology has been studied in detail in \cite{Cacciapaglia:2015eqa}. The sextet mainly couples to right-handed tops, and it is expected to be slightly lighter that the octet.  It only affects searches for 4-tops, which yield a lower bound on the mass of the order of $1$~TeV, thus stronger that the octet one, while the other octet final states are not affected by the presence of the sextet.

\begin{figure}[t]
\begin{tabular}{cc}
\includegraphics[width=0.5\textwidth]{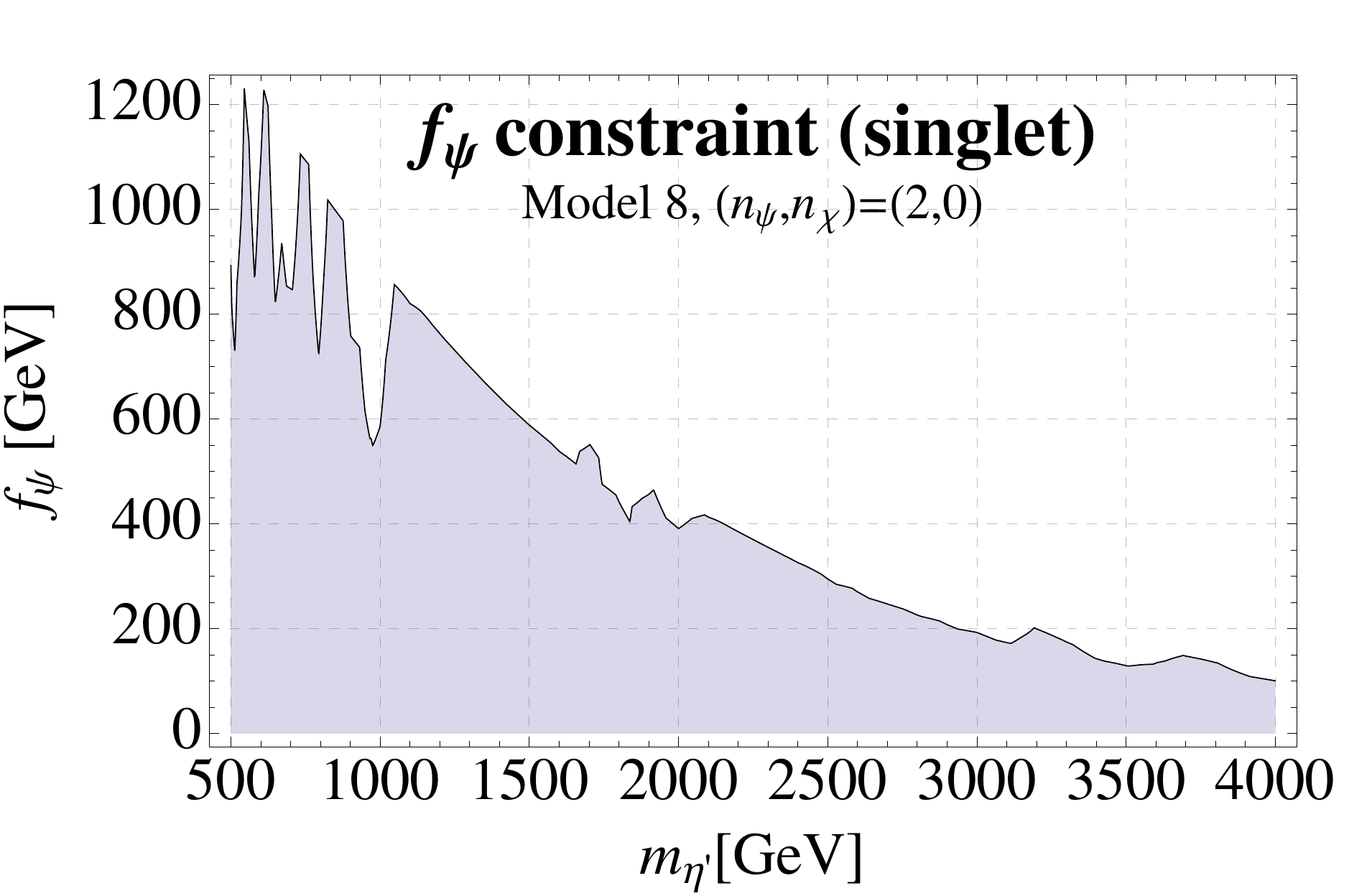} &
\includegraphics[width=0.5\textwidth]{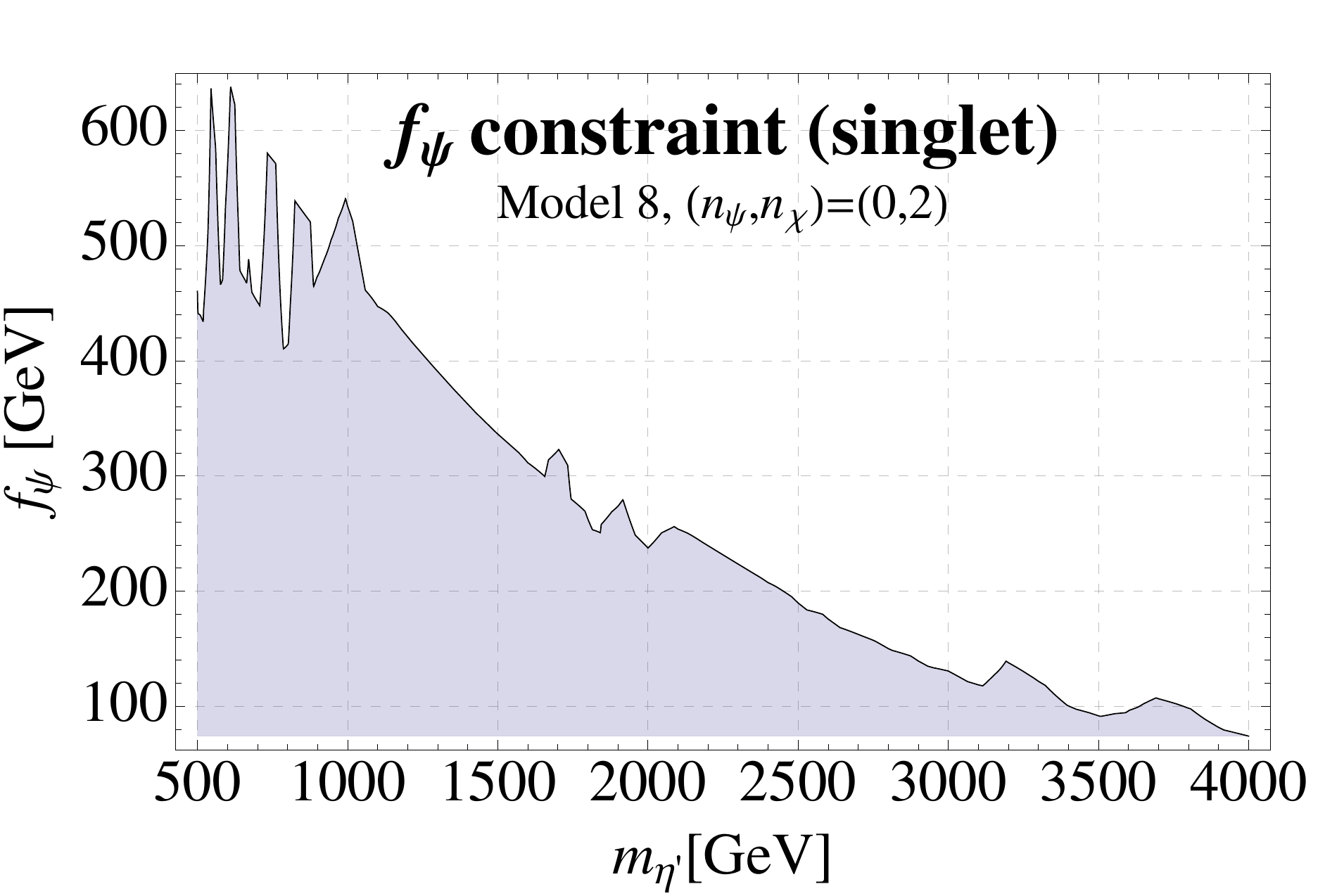} \\
\includegraphics[width=0.5\textwidth]{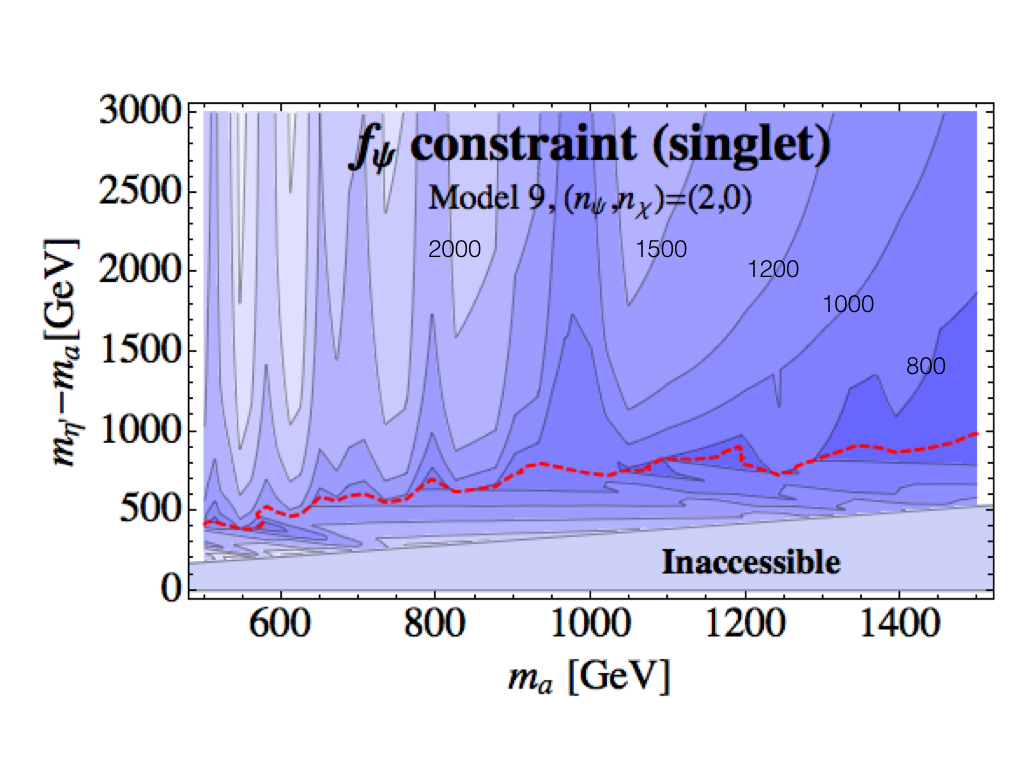} &
\includegraphics[width=0.5\textwidth]{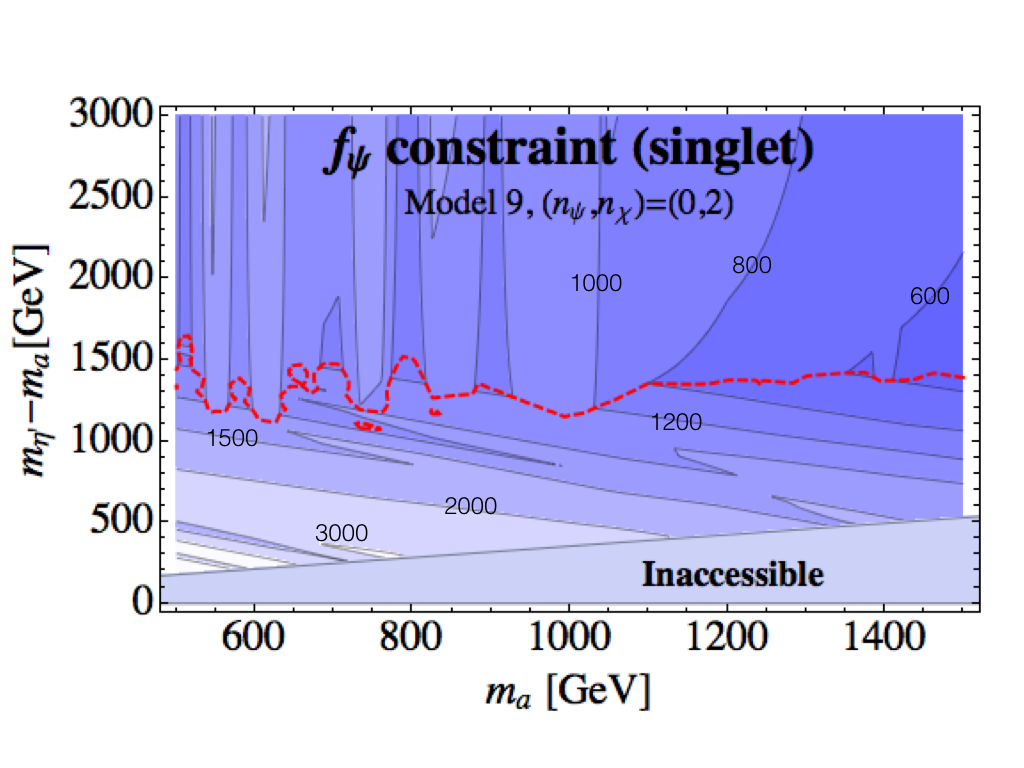} \\
\end{tabular}
\caption{Lower bound on $f_\psi$ (in GeV) from LHC searches sensitive to singlets. Top row shows results for model M8, as a function of $m_{\eta'}$. Bottom row for model M9, as a function of the two masses $m_a$ and $m_{\eta'}$. The Left (Right) plots corresponds to charges $(\pm 2,0)$ ($(0,\pm 2)$). The red line in the M9 plots delimits the region where the bound is driven by the $\eta'$ (below) from the $a$ one (above).}
\label{fig:singlets}
\end{figure}

The channels that can give a direct probe of the fine tuning in the EW scale are the singlets, which can directly feel the value of $f_\psi$. Also, for singlets, the two models appear rather different due to the value of $\tan \zeta$: in particular, for M8, the lighter singlet is always expected to be much lighter that the second one, as $m_a \lesssim m_{\eta'}/5$. Furthermore, due to the large coupling to tops, as shown in Table \ref{tab:numvalues}, it will dominantly lead to $t\bar{t}$ final states with a weak bound on $f_\psi$ of the order of $200$ GeV for a mass below $1$~TeV.
We thus decided to focus on the signatures generated by the heavier singlet $\eta'$: in the top row of Fig.~\ref{fig:singlets} we show the results for the two charge assignments, assuming decoupling limit (i.e. $\alpha = \zeta$ and $m_a \ll m_{\eta'}$).
From the ratio values in Table~\ref{tab:numvalues} and Fig.~\ref{fig:cok} (bottom right), we see that for $(n_\psi, n_\chi)= (\pm 2,0)$ the most constraining channel is $\gamma\gamma$ in the entire mass range $500 \mbox{ GeV} < m_{\eta'} <  4000 \mbox{ GeV}$, while for $(n_\psi,n_\chi)= (0,\pm 2)$, the $\gamma\gamma$ and the $t\bar{t}$ channels yield comparable bounds.
Closer investigation shows that the current constraint from the $t\bar{t}$ channel dominates only in the mass regime $900 \mbox{ GeV} < m_{\eta'} <  1100 \mbox{ GeV}$. Overall, the bound on $f_\psi$ is weaker for the $(n_\psi, n_\chi)= (0,\pm 2)$ charge assignment, which has a larger coupling to tops, because the larger branching ratio into $t\bar{t}$ reduces the dominant $\gamma \gamma$ one.
The plots also show that the bounds tend to be below the EW precision test ones, except for very light masses.

In the case of model M9, we can study the constraint as a function of the two masses, as they are allowed to be close. We recall that the mixing angle $\alpha$ depends on the values of the two masses, thus the couplings are not fixed over the parameter space, making the interpretation of the results more difficult. The result is shown in the bottom row of Fig.~\ref{fig:singlets}, where we present the lower bounds on $f_\psi$ as a function of the two masses. Like for M8, we see that the bound is weakened in the model with larger top couplings, i.e. $(0,\pm 2)$. Nevertheless, in both cases, the bounds are stronger that the ones expected from EW precision tests in a wide portion of the parameter space.  The general behavior of the bound is difficult to read, because it comes from a complicated interplay of many factors. One general remark is that the mixing angle varies from the value $\alpha = \zeta/2$ near the border of the inaccessible region, where the lightest singlet couples dominantly to the SU(2) gauge bosons ($W$'s), while  for heavy $\eta'$ the coupling to the hypercharge becomes dominant. As an example, we would like to discuss what happens for the $(\pm 2,0)$ case at $m_{a} = 1000$~GeV. From the bottom-right plot in Fig.~\ref{fig:cok}, we see that near the inaccessible boundary, where $\kappa_B/\kappa_g \sim 0$, the bound on $a$ is dominated by the $t\bar{t}$ final state due to the fairly large value of the top coupling, as $|C_t/\kappa_g| = 0.3$. This shows that the strong bounds on $f_\psi$ observed in the plot are driven by the $\eta'$, which has large coupling to gluons and large BR in di-photons (due to the large $\kappa_B$).

Moving away at larger $m_{\eta'}$, $a$ takes over with weaker bounds due to the fact that the coupling to $t\bar{t}$ is still large and the region near $1$~TeV shows a very sensitive island to low values of it (with $C_t/\kappa_g$ below $0.1$, see Fig.~\ref{fig:cok}). For increasing values of $m_{\eta'}$, the bound on $f_\psi$ from $a$ gets stronger due to the increase in the value of the coupling to gluons $\kappa_g$, thus explaining the presence of a minimum around $m_{\eta'} - m_a \sim 500$~GeV. The fact that the bound at $m_a = 1$~TeV is weaker that other mass values around it is due to the fact that the top coupling remains close to the critical value shown in Fig.~\ref{fig:cok}. A red line in Fig.~\ref{fig:singlets} shows the watershed dividing the region where the bound is driven by the $\eta'$ (below the red line) and the one driven by $a$ (above).

A final word is necessary on the arbitrary parameters that we fixed in order to obtain simple results. While the relation between the $f_{a_r}$ and $f_r$ ($r=\psi, \chi$) is somewhat justified, there is no underlying reason why $f_\chi = f_\psi$. The decay constants in the two sectors can, in principle, be different. We checked that, varying this ratio, our numerical result do not change qualitatively but there are $\mathcal{O} (1)$ changes in the numerical values of the bound on $f_\psi$, due to the change in the couplings. This ambiguity can, however, be fixed if the models under study is studied on the Lattice: in this case, the ratios between the various decay constants can be calculated, and a more solid prediction can be obtained for each model. The plots we present in this section are, therefore, just a numerical example. New plots following any Lattice input can be easily obtained following the recipe presented in this paper.

\section{Conclusions}\label{sec:conclusions}

We investigated the dynamics of  a specific class of Composite Higgs Models with top partial compositeness, constructed via ordinary four-dimensional gauge theories with fermionic matter belonging to two different irreps of the hypercolor group. These models give rise to EW cosets beyond the ``minimal'' $SO(5)/SO(4)$ type and thus contain additional pNGBs carrying EW charges. Furthermore, additional colored pNGBs arise from the need of introducing hyperquarks carrying ordinary color in order to construct top-partners. Two more pseudo-scalars arise from the breaking of the two chiral global $U(1)$ symmetries associated to the two hyperquarks.

In our choice of models of this type, we were guided by the need to preserve both the custodial symmetry of the Higgs sector as well as the one protecting the $Z\to b_L ~ \bar b_L$ branching ratio. As discussed in Section~\ref{sec:models}, we focused on models that are likely to lie outside of the conformal window and that can be brought into it from strong coupling at energies above the confinement scale $\Lambda$.

We identified a set of three pseudo-scalars,  the two  singlets $a$ and $\eta'$ with respect to the SM groups and a color octet $\pi_8$, that are present in all models in this class.
Their dynamics is controlled by model-specific group-theory data and a few phenomenological parameters  such as the hyperfermion masses and the pNGBs decay constants. In particular, the couplings to gauge bosons are determined by the coefficients of the WZW anomalies, which are sensitive to the microscopic details of the model. 

One of the most striking signals from these preudo-scalars are di-boson signatures
which is one of the main focuses of the paper.
We have performed a complete analysis of the constraints from di-boson and di-top final states (also including pair production for the octet)
using post-ICHEP2016 LHC data and the respective experimental results, and formulated a model independent strategy to 
combine known bounds and establish new limits on models under study. 

Following our recipe, formulated in Section IV (with the concrete example given in Section IV.A.2)
we applied the bounds to the models under consideration, giving numerical results in the case of two of them (M8 and M9 in Table~\ref{allmodels}). 
We found that present LHC data already sets important constraints on the condensation scale which are stronger than the typical bounds from EW precision tests, thus demonstrating that the direct search for additional pNGBs with di-boson and di-top signatures
in models of partial compositeness can be the first probe for such models. The fact that the couplings are predictive and sensitive to the underlying model makes these channels attractive. We should remark that the presence of these signatures is common to all models of partial compositeness based on a gauge-fermionic underlying theory.

The analysis of the post-ICHEP2016 data, and the framework we have developed in Section \ref{sec:pheno} can be 
used in a straightforward way (including an update with the new coming data) to any model containing pseudo-scalar singlets or octets
at the TeV scale.


\section*{Acknowledgements}
The authors would like to thank Y.Bai, V.Barger, J.Berger and M.Frigerio for useful comments and correspondence, and are grateful to the Mainz Institute for Theoretical Physics (MITP) for its hospitality and its partial support during the initial stages of this project.
TFs work was supported by the Basic Science Research Program through the National Research Foundation of Korea (NRF) funded by the ministry of Education, Science and Technology (No. 2013R1A1A1062597)  and by IBS under the project code, IBS-R018-D1.
AB acknowledges partial support from the STFC grant ST/L000296/1, the NExT Institute , Royal Society Leverhulme Trust Senior Research Fellowship LT140094 and
Soton-FAPESP grant.
GC and HC acknowledge partial support from the Labex-LIO (Lyon Institute of Origins) under grant ANR-10-LABX-66 and FRAMA (FR3127, F\'ed\'eration de Recherche ``Andr\'e Marie Amp\`ere''). GF would like to thank SISSA for the kind hospitality and partial support while this work was finalized. HS has received funding from the European Research Council (ERC) under the European Union's Horizon 2020 research and innovation programme (grant agreement No 668679).

\appendix
\section{Interpretation of ATLAS and CMS searches used in this article}\label{app:Exp}

In  this appendix, we  summarize the experimental searches used in this article in order to constrain the diboson channels, the $t\bar{t}$ channel, and the octet pair production channels and detail the assumptions made in order to extract the bounds for the models discussed.

\subsection{Diboson and $t\bar{t}$ searches}
ATLAS and CMS published a large number of searches for $jj$, $WW$, $ZZ$, $Z\gamma$, $\gamma\gamma$, $j\gamma$, and $t\bar{t}$ resonance searches at run I (with a center of mass energy of $\sqrt{s}=8$~TeV) and run II (with a center of mass energy of $\sqrt{s}=13$~TeV).  The constraints  are typically given as bounds on the (folded) production cross section as a function of the resonance mass. In those cases, we directly use the bound on the folded cross section, implicitly assuming that the acceptances and efficiencies of the pseudo-scalar SM singlet resonance $\pi_0$ and the color octet $\pi_8$ are comparable to the ones of the sample model used in the respective experimental study. We do not perform explicit recasts of the various searches. In cases in which bounds are given as constraints on a fiducial cross section or on cross section times acceptance times efficiency, we estimate the acceptance and efficiency following the information provided in the respective articles and list our assumptions below. Finally, in some cases, studies give bounds on the cross section in a particular final state after the decays of the SM gauge bosons. In these cases we rescale results with the appropriate SM gauge boson branching ratios.

\subsubsection*{8 TeV searches}

\begin{figure}[t]
\begin{tabular}{cc}
\includegraphics[width=.45\textwidth]{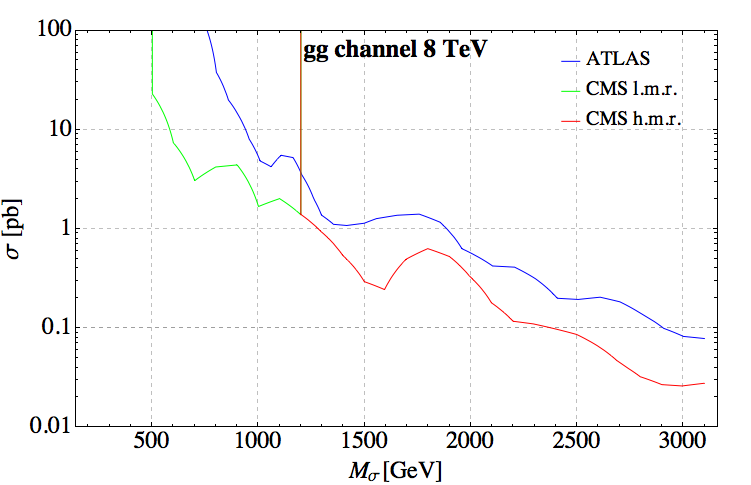} & 
\includegraphics[width=.45\textwidth]{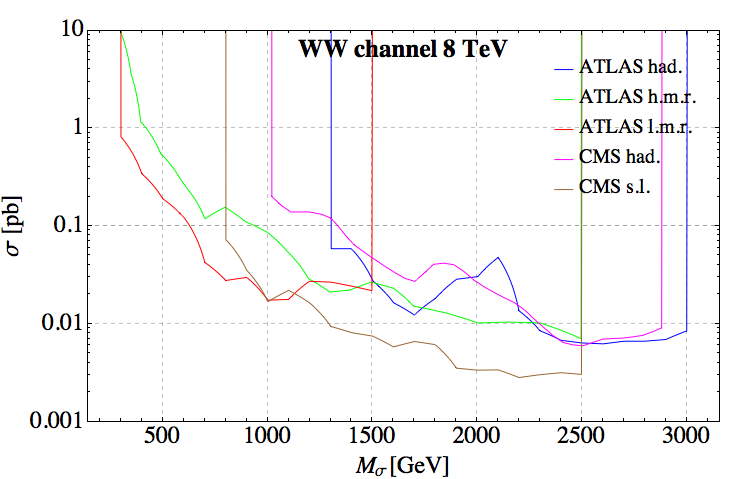} \\
\includegraphics[width=.45\textwidth]{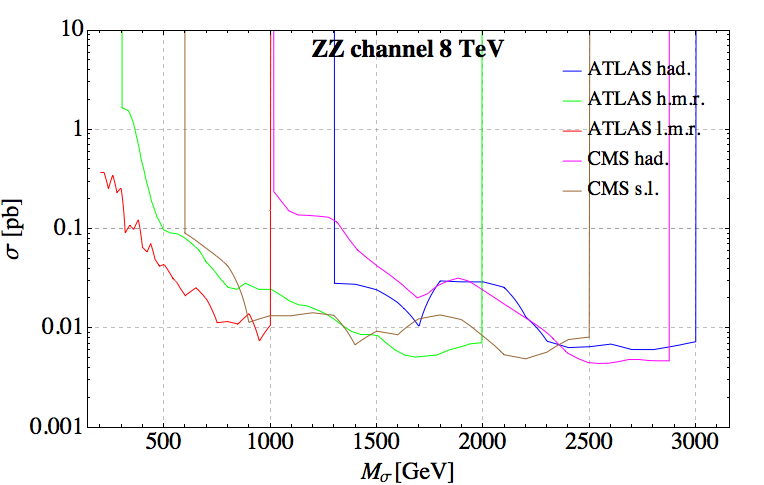} &
\includegraphics[width=.45\textwidth]{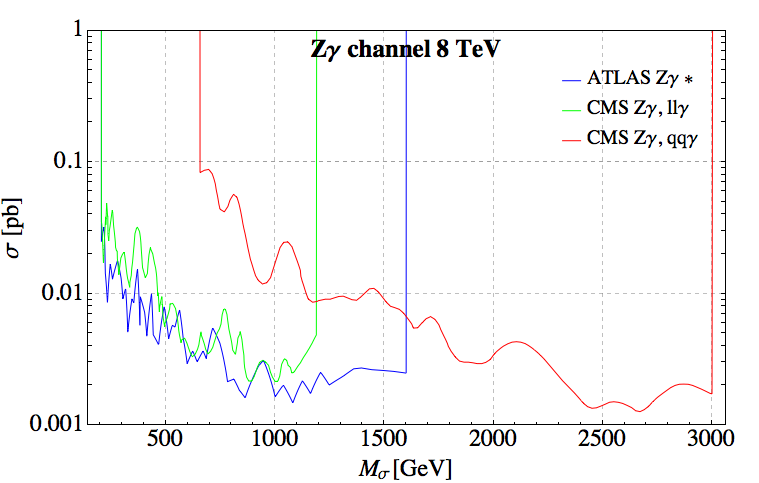} \\ 
\includegraphics[width=.45\textwidth]{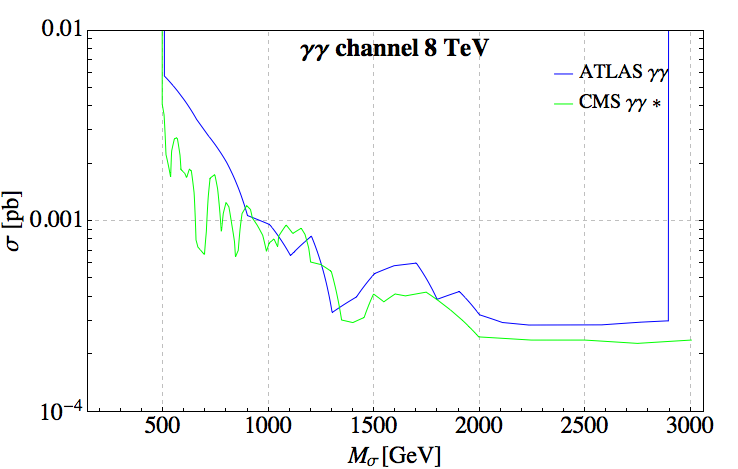} & 
\includegraphics[width=.45\textwidth]{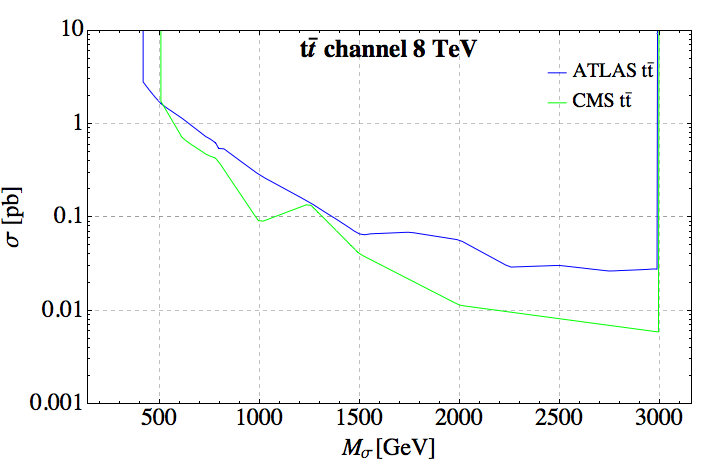} 
\end{tabular}
\caption{Bounds on the di-boson and $t\bar{t}$ channels from 8 TeV searches.}
\label{fig:bdsdet8}
\end{figure}

A summary of the di-boson bounds from run I are shown in Fig.~\ref{fig:bdsdet8}, where we used the following searches and assumptions:
\bigskip

\noindent $gg$-channel:\\
\noindent ATLAS: Ref.~\cite{Aad:2014aqa}, Fig. 9 ($gg$ fusion, lowest width). Bounds are given in terms of cross section $\times$ acceptance. We assume an acceptance of 50\%.\footnote{Acceptances for excited quark, scalar color octet, and quantum black hole searches are reported as 58\%, 61-63\% and 52-56\%  \cite{Aad:2014aqa}. In the absence of a full recast of the search for our pseudo-scalar resonance, we assume a slightly lower acceptance of 50\%, leading to a conservative estimate of the constraint of the folded cross section.}\\
\noindent CMS l.m.r.: Ref.~\cite{CMS:2015neg}, Fig. 3 ($gg$ fusion, low mass region). Bounds are given in terms of cross section $\times$ acceptance.We assume an acceptance of 50\%.\\
\noindent CMS h.m.r.: Ref.~\cite{Khachatryan:2015sja}, Fig. 4 ($gg$ fusion, high mass region). Bounds are given in terms of cross section $\times$ acceptance.We assume an acceptance of 50\%.
\bigskip

\noindent $WW$-channel:\\
ATLAS had.: Ref.~\cite{Aad:2015owa}, Fig. 6 (Randall-Sundrum Kaluza-Klein graviton).\\
ATLAS h.m.r.: Ref.~\cite{Aad:2015ufa}, Fig. 2 (Randall-Sundrum Kaluza-Klein graviton).\\
ATLAS l.m.r.: Ref.~\cite{Aad:2015agg}, Fig. 12 ($gg$ fusion).\\
CMS had..: Ref.~\cite{Khachatryan:2014hpa}, Fig. 8 (Randall-Sundrum Kaluza-Klein graviton).\\
CMS s.l..: Ref.~\cite{Khachatryan:2014gha}, Fig. 9 (Randall-Sundrum Kaluza-Klein graviton).
\bigskip

\noindent $ZZ$-channel:\\
ATLAS had.: Ref.~\cite{Aad:2015owa}, Fig. 6 (Higgs-like scalar produced in gluon-gluon-fusion).\\
ATLAS h.m.r.: Ref.~\cite{Aad:2014xka}, Fig. 2 (Randall-Sundrum Kaluza-Klein graviton).\\
ATLAS l.m.r.: Ref.~\cite{Aad:2015kna}, Fig. 12 ($gg$ fusion).\\
CMS had..: Ref.~\cite{Khachatryan:2014hpa}, Fig. 8 (Randall-Sundrum Kaluza-Klein graviton).\\
CMS s.l..: Ref.~\cite{Khachatryan:2014gha}, Fig. 9 (Randall-Sundrum Kaluza-Klein graviton).
\bigskip

\noindent $Z\gamma$-channel:\\
ATLAS $Z\gamma$*: Ref.~\cite{Aad:2014fha}, Fig. 3c (scalar). The article gives a bound on the fiducial cross section, only. Without a detailed recast, we are not able to interpret this bound in terms of the full cross section in order to compare it to other searches. Thus, we give the bound on the fiducial cross section of this study only for reference, and do not use it in our combined constraints.\\
CMS, $ll\gamma$: Ref.~\cite{CMS:2015lza}, Fig. 2 (Narrow signal model).\\
CMS, $qq\gamma$: ATLAS l.m.r.: Ref.~\cite{CMS:2016mvc}, Fig. 5 (results for narrowest width spin-0 resonance).
\bigskip

\noindent $\gamma\gamma$-channel:\\
ATLAS $\gamma\gamma$: Ref.~\cite{Aad:2015mna}, Fig. 4 (Randall-Sundrum Kaluza-Klein graviton).\\
CMS $\gamma\gamma$ *: Ref.~\cite{CMS:2015cwa}, Fig. 2 (Randall-Sundrum Kaluza-Klein graviton, narrowest available width). CMS  by now provides combined bounds of run I and run II searches in the $\gamma\gamma$ resonance. The bound shown here is the run-I bound, and is only given for reference. We do not use it in our analysis, but instead include the combined run-I and II bound from CMS.
\bigskip

\noindent $t\bar{t}$-channel:\\
ATLAS $\gamma\gamma$: Ref.~\cite{Aad:2015fna}, Fig. 11 (scalar resonance search).\\
CMS $\gamma\gamma$ *: Ref.~\cite{Khachatryan:2015sma}, Fig. 14 (narrow width $Z'$ search).
\bigskip

To combine the different bounds in each of the channels we take as a constraint the strongest bound in each channel at a given mass. In the model we discuss, the by far dominant single-production mechanism is gluon fusion, for which the ratio between the production cross section at 13 TeV and 8 TeV as a function of mass is determined by the cross sections given in Fig.~\ref{fig:prodX}. In order to compare the bounds from run I searches to those from run II searches, we rescale the bounds on the production cross section times branching ratios into the the di-boson and $t\bar{t}$ final states by $\sigma(gg\rightarrow \pi_0)_{13} / \sigma(gg\rightarrow \pi_0)_{8}$ and show the resulting bounds in Fig.~\ref{fig:bds}, labeled at ``8 TeV''.

\subsubsection*{13 TeV searches}

\begin{figure}[t]
\begin{tabular}{cc}
\includegraphics[width=.45\textwidth]{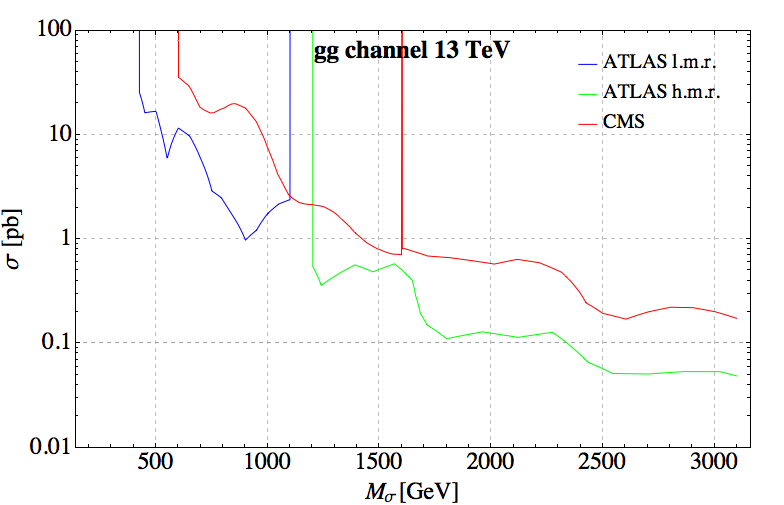} & 
\includegraphics[width=.45\textwidth]{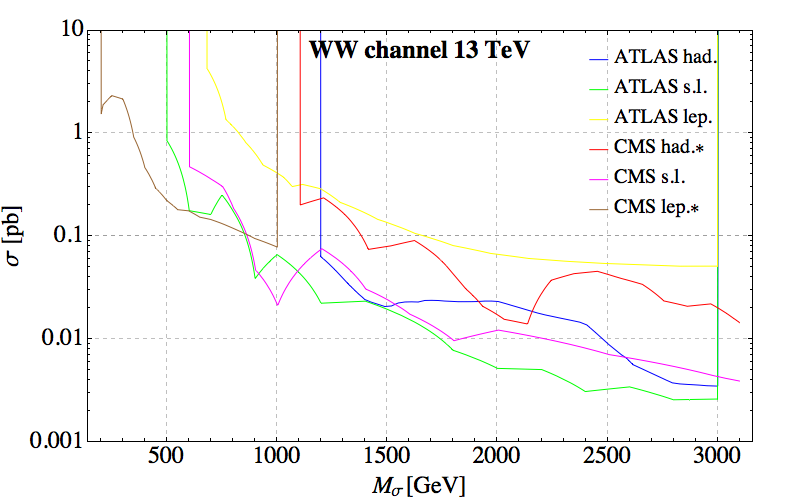} \\
\includegraphics[width=.45\textwidth]{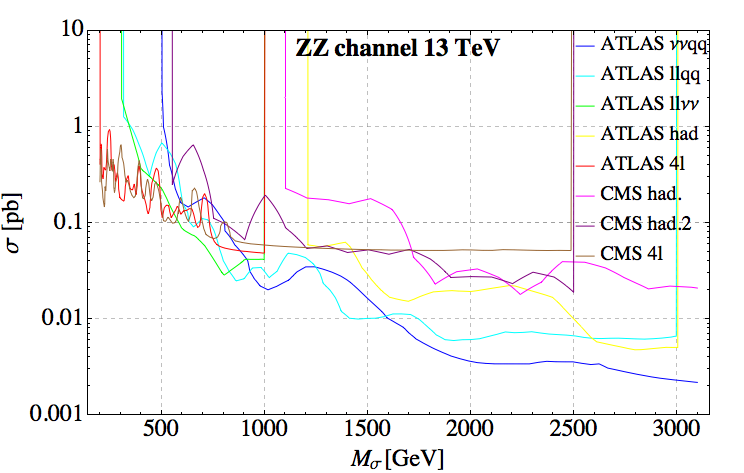} &
\includegraphics[width=.45\textwidth]{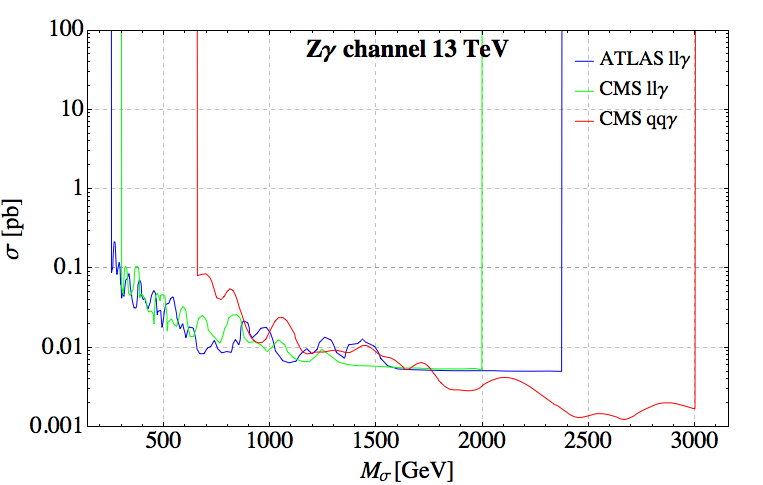} \\ 
\includegraphics[width=.45\textwidth]{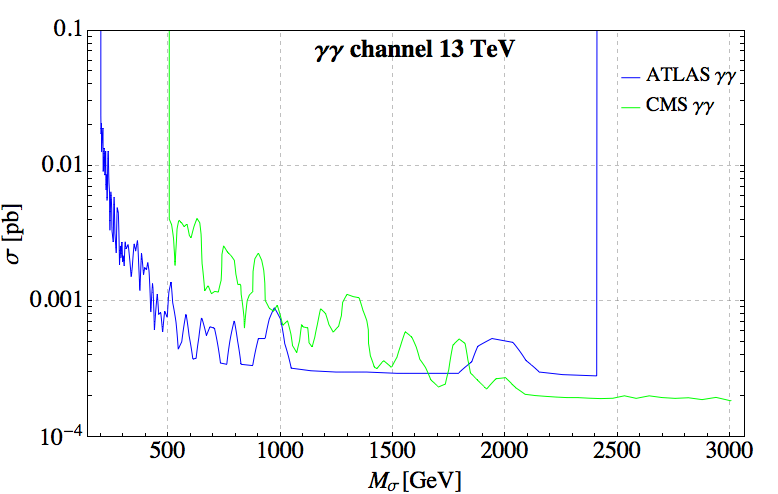} & 
\includegraphics[width=.45\textwidth]{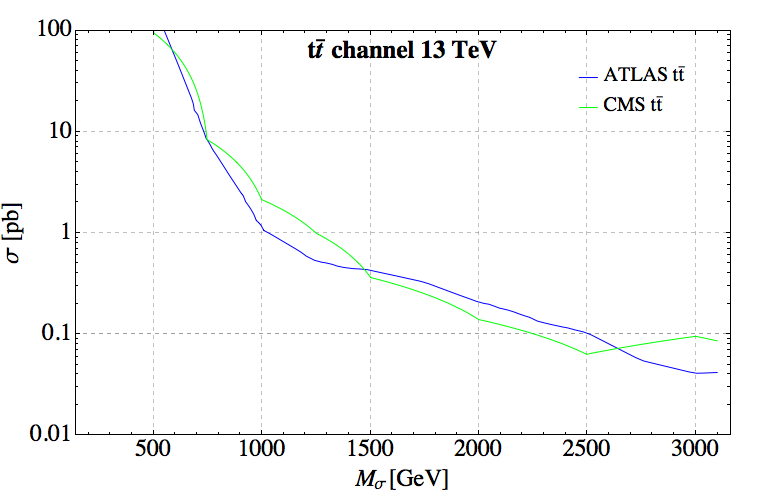} 
\end{tabular}
\caption{Bounds on the di-boson and $t\bar{t}$ channels from 13 TeV searches.}
\label{fig:bdsdet13}
\end{figure}

A summary of the di-boson bounds from run II are shown in Fig.~\ref{fig:bdsdet13}, where we used the following searches and assumptions:
\bigskip

\noindent $gg$-channel:\\
ATLAS l.m.r: Ref.~\cite{ATLAS:2016xiv}, Fig. 8 (Gaussian signal with detector resolution). Bounds are given in terms of cross section $\times$ BR $\times$ acceptance. For the acceptance, we use the acceptances provided in Tables 3 and 4 of Ref.~\cite{ATLAS:2016xiv}.\\
ATLAS h.m.r.: Ref.~\cite{ATLAS:2016lvi}, Fig. 5 (Gaussian signal with detector resolution). Bounds are given in terms of cross section $\times$ acceptance. We assume an acceptance of 50\%.\\
CMS: Ref.~\cite{CMS:2016wpz}, Fig. 6 ($gg$ fusion, low- and high mass region). Bounds are given in terms of cross section $\times$ acceptance. We assume an acceptance of 60\% (acceptance given in Ref.~\cite{CMS:2016wpz} for isotropic decays).
\bigskip

\noindent $WW$-channel:\\
ATLAS had.: Ref.~\cite{ATLAS:2016yqq}, Fig. 6 (Randall-Sundrum Kaluza-Klein graviton).\\
ATLAS s.l.: Ref.~\cite{ATLAS:2016cwq}, Fig. 2 (scalar produced in $gg$ fusion, narrow width approximation).\\
ATLAS lep: Ref.~\cite{ATLAS:2016kjy}, Fig. 5 (scalar produced in $gg$ fusion, narrow width approximation).\\
CMS had..: Ref.~\cite{CMS:2015nmz}, Fig. 10 (Randall-Sundrum Kaluza-Klein graviton).\\
CMS s.l..: Ref.~\cite{CMS:2016pfl}, Fig. 7 (Randall-Sundrum Kaluza-Klein graviton).\\
CMS lep: Ref.~\cite{CMS:2016jpd}, Fig. 5d (SM-like heavy Higgs, narrowest width available).\\

\bigskip

\noindent $ZZ$-channel:\\
ATLAS $\nu\nu qq$: Ref.~\cite{ATLAS:2016npe}, Fig. 12 (Higgs-like scalar produced in $gg$ fusion)\\
ATLAS $ll qq$: Ref.~\cite{ATLAS:2016npe}, Fig. 10 (Higgs-like scalar produced in $gg$ fusion)\\
ATLAS $ll \nu\nu$: Ref.~\cite{ATLAS:2016bza}, Fig. 7a (narrow width Higgs-like scalar produced in $gg$ fusion)\\
ATLAS had: Ref.~\cite{Aad:2015owa}, Fig. 6 (Higgs-like scalar produced in $gg$ fusion).\\
ATLAS $4l$: Ref.~\cite{ATLAS:2016oum}, Fig. 11a (scalar produced in $gg$ fusion, narrow width approximation).\\
CMS had.: Ref.~\cite{CMS:2015nmz}, Fig. 10 (Randall-Sundrum Kaluza-Klein graviton).\\
CMS had. 2:  Ref.~\cite{CMS:2016tio}, Fig. 12c (Randall-Sundrum Kaluza-Klein graviton).\\
CMS $4l$: Ref.~\cite{CMS:2016ilx}, Fig. 16 (SM-like heavy Higgs, narrow width approximation).\\
\bigskip

\noindent $Z\gamma$-channel:\\
ATLAS $ll\gamma$: Ref.~\cite{ATLAS:2016lri}, Fig. 6 (narrow width scalar).\\
CMS, $ll\gamma$: Ref.~\cite{CMS:2016pax}, Fig. 4.\\
CMS, $qq\gamma$: Ref.~\cite{CMS:2016cbb}, Fig. 5c (narrow width spin-0 resonance).
\bigskip

\noindent $\gamma\gamma$-channel:\\
ATLAS $\gamma\gamma$: Ref.~\cite{ATLAS:2016eeo}, Fig. 7a (spin-0 resonance, narrow width approximation). The study gives bounds on the fiducial cross section. To obtain bounds on the full cross section -- following the information on the fiducial volume given in  Ref.~\cite{ATLAS:2016eeo} -- we divide the bounds given by a fiducial volume function which is 54\% for a mass of 200 GeV, linearly extrapolated to 61\% at 700 GeV, and 61\% above. \\
CMS $\gamma\gamma$ *: Ref.~\cite{CMS:2016owr}, Fig. 8 (narrow width spin-0 resonance). CMS  provides combined bounds of run I and run II searches in the $\gamma\gamma$ resonance. The bound shown here is the combined bound.
\bigskip

\noindent $t\bar{t}$-channel:\\
ATLAS $\gamma\gamma$: Ref.~\cite{ATLAStt13}, Fig. 11 ($Z'$ search).\\
CMS $\gamma\gamma$ *: Ref.~\cite{CMS:2016zte}, Fig. 8a (narrowest width $Z'$ search).
\bigskip

To combine the different bounds in each of the channels,we take as a constraint the strongest bound in each channel at a given mass. The resulting bounds are shown in Fig.~\ref{fig:bds}, labeled at ``13 TeV''.

\subsection{Excited quark searches as constraints on the $\gamma g$ final state}

\begin{figure}[t]
\begin{tabular}{cc}
\includegraphics[width=.45\textwidth]{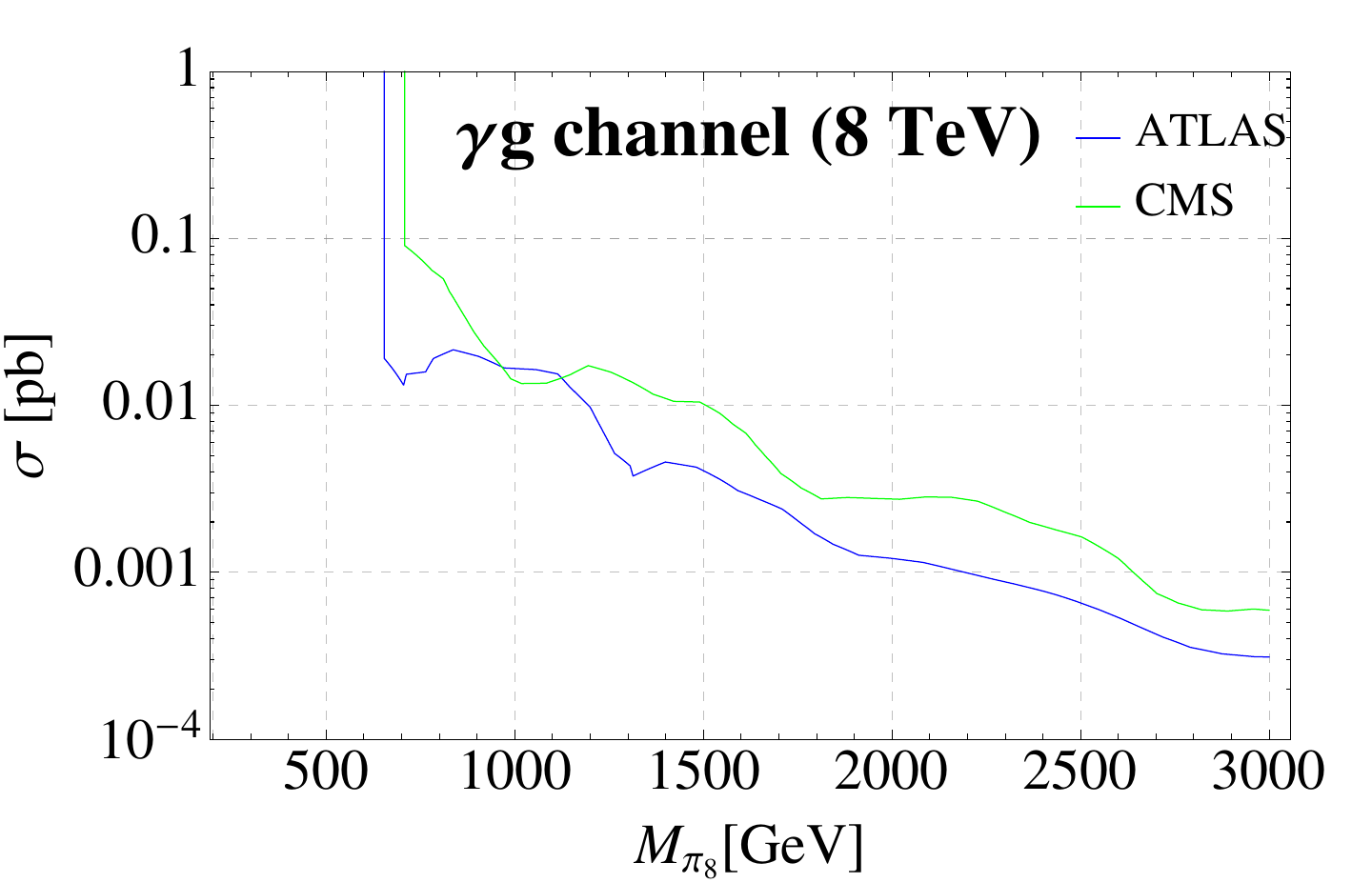} & 
\includegraphics[width=.45\textwidth]{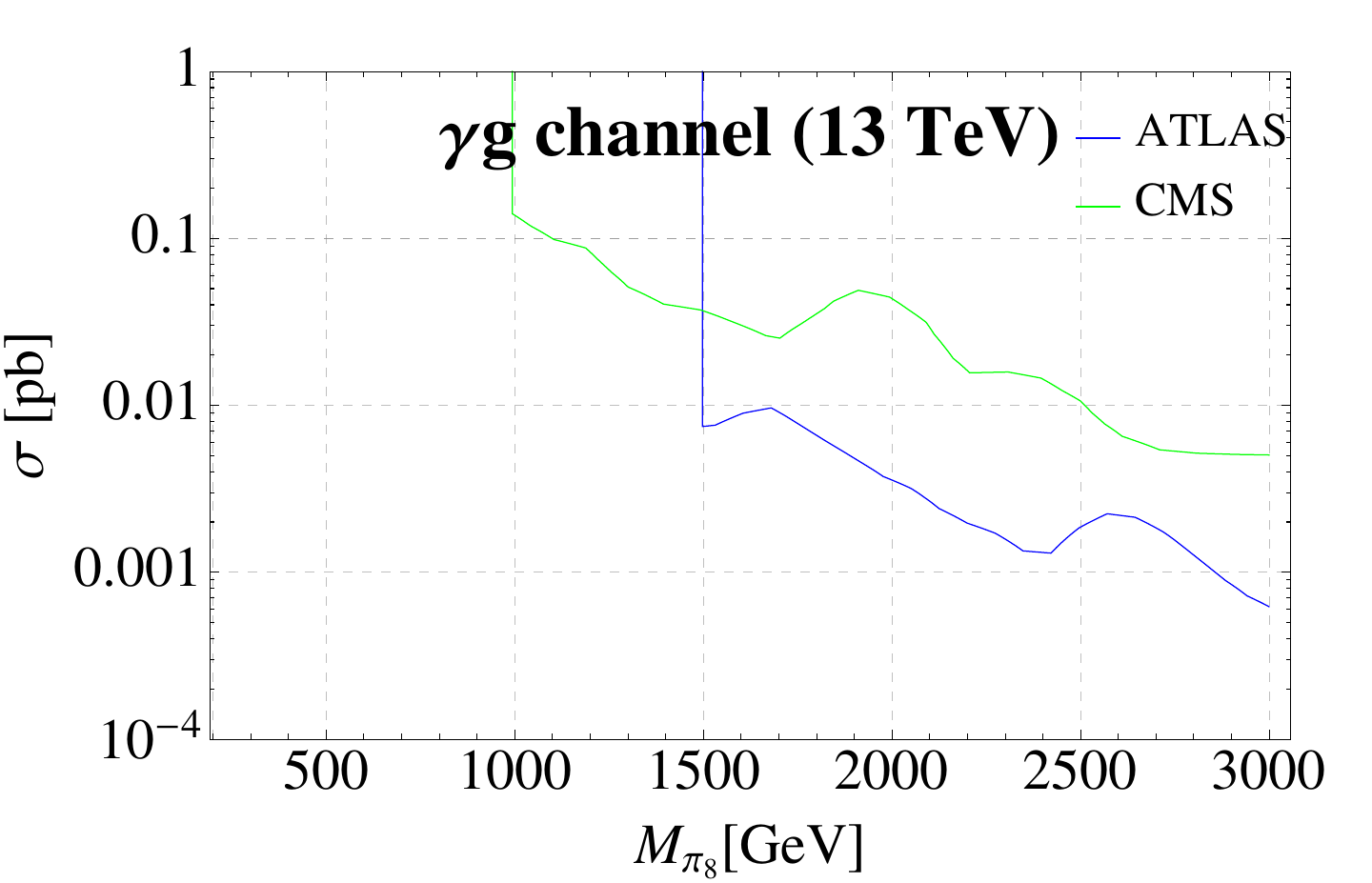}\end{tabular}
\caption{Bounds on $g\gamma$ channel  from 8 and 13 TeV searches.}
\label{fig:bdsdetggam}
\end{figure}

Excited quark searches for the final state $\gamma j$ can be used to constrain the $\gamma g$ channel relevant for the octet pseudo-scalar search. For the 8 TeV searches we used the following bounds shown in Fig.~\ref{fig:bdsdetggam} (left):

\noindent ATLAS: Ref.~\cite{Aad:2015zva}, Fig. 2 (excited quark, narrowest width). The bound is given on the cross section $\times$ acceptance $\times$ efficiency. To obtain the bound on the cross section we divide the bounds given by A =  60\% $\times$ $\epsilon= 60\%$, according to acceptances and efficiencies quoted in Ref.~\cite{Aad:2015zva} for excited quark searches.\\
CMS: Ref.\cite{Khachatryan:2014rra}, Fig. 2 (excited quark search).\\

The combined bound used in this article is obtained by taking the strongest bound at each resonance mass $m_{\pi_8}$. In order to compare the bounds from run I searches to those from run II searches, we rescale the bounds on the production cross section times branching ratios into $g\gamma$ by $\sigma(gg\rightarrow \pi_8)_{13} / \sigma(gg\rightarrow \pi_8)_{8}$ and show the resulting bounds in Fig.~\ref{fig:bdsoct}, labeled ``8 TeV''.

\bigskip

For the 13 TeV searches we used the following bounds shown in Fig.~\ref{fig:bdsdetggam} (right):

\noindent ATLAS: Ref.~\cite{Aad:2015ywd} Fig. 5a (excited quark search).\\
CMS: Ref.~\cite{CMS:2016qtb}, Fig. 5 (excited quark search)\\

The combined bound  is obtained by taking the strongest bound at each resonance mass $m_{\pi_8}$. The resulting bound is shown in Fig.~\ref{fig:bdsoct}, labeled ``13 TeV''.

\section{Additional mass mixing (and couplings) of the singlets} \label{app:coups}

In  this Appendix, we  briefly discuss the possible presence of mixing of the U(1) singlets $a_\psi$ and $a_\chi$ with pNGBs from the non-abelian flavor symmetries. 
These mixing terms can only arise from spurions explicitly breaking the flavor symmetries, and in particular from the mass terms of the fermions which also generate masses for the pNGBs.
In addition to mass mixing, couplings to two non-abelian pNGBs may also be generated thus opening the case for additional decay channels.
We will discuss each case individually, as the physics involved is very different.

\subsubsection*{Coset SU(4)/Sp(4)}

This coset, generated by $\psi$ in the pseudo-real representation, contains 5 pNGBs, which transform under the custodial symmetry as a singlet $\eta$ and a bi-doublet $H$ (which plays the role of the BEH field).

The underlying theory consists of 4 Weyl spinors: two transforming as a doublet of SU(2)$_L$, and the other 2 as a doublet of SU(2)$_R$, thus one can write down two independent mass terms, $m_L$ and $m_R$ respectively. We will parametrize the two masses as
\beq \label{eqApp:psimass}
m_L = \mu_\psi (1+\delta)\,, \quad m_R = \mu_\psi (1-\delta)\,,
\eeq
where $\mu_\psi$ is the common mass used in Section~\ref{sec:3_couplings} and $\delta$ is a parameter describing the deviation from Universality: $\delta$ is required to be small in order to preserve the stability of the vacuum.

Following the normalization adopted in this work, we find that:
\beq
\mbox{Im}\mbox{Tr} [\Sigma X^\dagger] = 2 B_\psi \mu_\psi \delta \left( - 4 \frac{\eta}{f_\psi} \frac{f_\psi \sin \sqrt{\eta^2 + |H|^2}/f_\psi}{\sqrt{\eta^2 + |H|^2}} \right)\,.
\eeq
Plugging this expression in Eq.(\ref{eq:Lmix}), we see that a linear mixing with the non-abelian singlet $\eta$ is generated, proportional to the universality breaking parameter $\delta$:
\beq
\mathcal{L}_m \supset 2 \frac{f_\psi}{f_{a_\psi}} m_{\pi_\psi}^2 \delta\ a_\psi \eta + \dots
\eeq
We also notice that no mixing nor coupling can be generated in the Universal case: this fact can be easily understood in terms of symmetries. In fact, there exists a symmetry acting on the pNGBs under which only $\eta$ is odd, provided that SU(2)$_L \leftrightarrow$ SU(2)$_R$ are exchanged~\cite{Arbey:2015exa}. Following the CP properties of the scalar fields, and invariance under Sp(4), the possible linear couplings of the $a_\psi$ singlet to the non abelian pNGBs need to have the following form:
\beq
a_\psi \eta^{2n-1} (H^\dagger H)^m\,, \quad \mbox{where $n$ and $m$ are integers.} 
\eeq
The odd power on $\eta$ derives from CP-invariance. This coupling, however, is odd under the $\eta$-parity described above, unless the coefficient is odd, i.e. proportional to $\delta \sim m_L - m_R$.

\subsubsection*{Coset SU(4)$\times$SU(4)/SU(4)}

This coset also arises in the EW sector when $\psi$ is complex. The 15 pNGBs transform as a singlet $\eta$, two bi-doublets $H_{1,2}$, an SU(2)$_L$ triplet $\Delta$ and an SU(2)$_R$ triplet $N$. 
Like in the previous case, the 4 $\psi$'s (and their conjugates) transform like doublets of the custodial symmetry, thus we can write down two mass terms $m_L$ and $m_R$. We can parametrize then as above, in Eq.(\ref{eqApp:psimass}).

The vacuum structure of this model is more complex that the previous one, and it has been discussed in detail in Ref. \cite{Ma:2015gra}. It is easier to describe the theory in the EW preserving vacuum, and think of the misalignment in terms of VEVs assigned to the pNGBs. The only pNGBs that can develop a VEV are:
\beq
\langle H_1^0 \rangle = \frac{v}{\sqrt{2}}\,, \quad \langle H_2^0 \rangle = i \frac{v_2}{\sqrt{2}}\,, \quad \langle \Delta_0 \rangle = \langle N_0 \rangle = v_3\,,
\eeq
where $v = v_{\rm SM}$.
It has been shown in \cite{Ma:2015gra} that the real VEV of the second doublet can always be rotated away without loss of generality, thus we do not consider it here further. Also, $v_2$ violates CP, and for simplicity we set it to zero here: if present, however, it will generate a tadpole for $a_\psi$ proportional to the product of the 3 VEVs. Note also that the triplet VEV is bound to be small as it does violate custodial invariance, thus we will neglect it in the following for simplicity.

We then study the mass mixing by expanding the Lagrangian terms in Eq.~(\ref{eq:Lmix}) up to the 3rd power in the pNGB matrix, thus capturing effects up to quadratic order in the VEVs. We find the following mass terms:
\beq
\mathcal{L}_m \supset - 2\frac{f_\psi}{f_{a_\psi}} m_{\pi_\psi}^2 \delta \left( 1 - \frac{v^2}{6 f_\psi^2} \right) a_\psi \eta +  \dots\,.
\eeq
We see then that a mixing to the singlet $\eta$ is generated in presence of Universality violation, as in the SU(4)/Sp(4) case. Additionally, a mixing to the second doublet is generated if the small triplet VEV is present. 

Couplings to two pNGBs can also be generated via the SM Higgs VEV $v$:
\begin{multline}
\mathcal{L}_m \supset \frac{m_{\pi_\psi}^2 v}{f_\psi f_{a_\psi}}\ a_\psi  \left[ \frac{2}{3} \delta\ \eta h + \sqrt{2} A \Delta_0 (1+\frac{\delta}{3} ) + \sqrt{2} A N_0 (1-\frac{\delta}{3} ) + \right.  \\
\left. + \sqrt{2} i H_2^+ \Delta^- (1+\frac{\delta}{3} )  + \sqrt{2} i H_2^+ N^- (1-\frac{\delta}{3} ) + h.c. \right] \,.
\end{multline}
The terms above can generate decays of the singlets to two pNGBs: note in particular the presence of couplings to the second doublet and triplets which are not suppressed by $\delta$. As these pNGBs may be odd under a conserved parity (see \cite{Ma:2015gra}), this opens the way to interesting invisible decay modes.
Nevertheless, such couplings are proportional to the mass $m_{\pi_\psi}$ which we assume being small, thus for the sake of simplicity we will neglect them in our study.

\subsubsection*{Coset SU(5)/SO(5)}

This coset arises when the $\psi$'s are in a real representation of Hypercolor. The 14 pNGBs transform like a singlet $\eta$, a bi-doublet $H$ playing the role of the BEH field, and a bitriplet, that can be decomposed into a real SU(2)$_L$ triplet $\phi_0$ and a complex one $\phi_+$ with hypercharge $+1$. The 5 fermions $\psi$ transform like a bidoublet of the custodial symmetry plus a singlet, thus one can assign 2 independent masses $m$ and $m_0 = m (1+\delta)$ respectively.

Besides the VEV of the Higgs field, the vacuum structure also allows the triplets to develop a VEV, so that in principle one can define 4 independent parameters:
\beq
\langle H^0 \rangle = \frac{v}{\sqrt{2}}\,, \quad \langle \phi_0^0 \rangle = v_0\,, \quad \langle \phi_+^- \rangle = \frac{v_1 + i v_2}{\sqrt{2}}\,.
\eeq
However, the triplets VEVs would violate custodial invariance unless a relation among them is imposed \cite{Georgi:1985nv}: $v_1 = 0$ and $v_2 = - v_0$. Plugging this vacuum structure in Eq.(\ref{eq:Lmix}), however, we see that a tadpole for the singlet $a_\psi$ is generated unless $v_0 = v_2 = 0$: this is easily understood, as the two VEVs correspond to CP-odd fields. Thus, in order to both preserve custodial invariance and avoid tadpoles for the singlets, the triplet VEVs must vanish: in the following, also for the sake of simplicity, we will impose this cancellation.

Expanding the mixing Lagrangian up to third order in the pNGB matrix, we find that the following mixing terms are generated:
\beq
\mathcal{L}_m \supset \frac{f_\psi}{f_{a_\psi}} m_{\pi_\psi}^2 \left[ \frac{4}{\sqrt{5}} \left( \delta - \frac{9+7\delta}{12} \frac{v^2}{f_\psi^2} \right) a_\psi \eta +  \left( 1+\frac{\delta}{3} \right) \frac{v^2}{f_\psi^2} a_\psi  (\phi_0^0 - 2 \mbox{Im} \phi_+^-) \right]\,.
\eeq
Like in the SU(4)$\times$SU(4)/SU(4) coset, we notice a mixing with the singlets proportional to the violation of the mass Universality, however at order $v^2/f_\psi^2$ mixings with the singlets and the two CP-odd neutral components of the triplets are generated also in the Universal limit.
Similarly, the following couplings to two pNGBs are generated at the leading order in $v$:
\beq
\mathcal{L}_m \supset 2 \frac{m_{\pi_\psi}^2 v}{f_{a_\psi} f_\psi} \left[ -\frac{9+7 \delta}{3 \sqrt{5}} a_\psi h \eta + \left( 1+\frac{\delta}{3} \right) a_\psi h (\phi_0^0 - 2 \mbox{Im} \phi_+^-) \right]\,.
\eeq

\section{Top loops} \label{app:toploops}

In the present framework of pNGB composite Higgs model top mass is induced via partial compositeness. This explicit breaking of the global flavor symmetry introduces direct couplings between the pNGBs and top. The octet coupling is model dependent but the coupling of the singlets is always present, as explained in Sec.~\ref{sec:PNGBtheory}. Such new interactions will induce loop corrections to the anomalous WZW terms, via a triangle fermionic loop. In this appendix we summarize the main results of the top loop contribution.

For a pseudo-scalar, the   amplitude is simply  proportional to an epsilon tensor and other gauge invariant tensor structures vanish due to the CP invariance\footnote {This argument  applies for the top loop contribution to  $H$-$Z$-$\gamma$ vertex in the SM,  where the Higgs is a scalar, thus the CP-Odd  epsilon-term  is exactly canceled because of  the opposite sign  in  the clockwise and anti-clockwise Feynman  diagrams.}.
For a generic coupling of the pNGBs to tops, i.e. $i\gamma_5 \,C_t m_t/f_\pi$ (were $\pi$ can be either $\pi_0$ or $\pi_8$), the amplitude of the process is given by:

\begin{equation} \label{eq:Mtoploop}
\mathcal{M}_{top}^{odd}= - \frac{\epsilon_{\mu\nu\rho\sigma}\epsilon^\mu(\vec{k})\epsilon^\nu(\vec{p})k^\rho p^\sigma}{4\pi^2}\frac{C_t}{f_\pi}\left[c_1\widetilde{C}_0(R_p,R_k,R_\pi;\xi)+c_2 \widetilde{C}_1(R_p,R_p,R_\pi;\xi)\right]\,.
\end{equation}
with $ R_i = \frac{p_i^2}{m_t^2} $, $ \xi = \frac{m_b}{m_t}$  for the $WW$ final state and $\xi=1$ otherwise. The second term on the right hand side will only be present for equal massive final states ($WW$ and $ZZ$), we have used the on-shell condition $R_k=R_p$ when writing it. The $c_{1,2}$ are combinations of SM couplings for the top (or top-bottom) to gauge fields, which already include the trace over the symmetry generators. For all neutral gauge bosons we have two loop diagrams contributing, corresponding to the fermions in the loop going clockwise or anti-clockwise. For the $W^+W^-$ final state only one fermion flow contributes, this multiplicity is also taking into account in the $c_{1,2}$ parameters. In Table~\ref{Tab: Apploop} we summarize these coupling combinations.    

\begin{table}[h]
\begin{tabular}{|c|cc|}
\hline
 channels&$c_1$&$c_2$\\
\hline
$\pi^0 gg$&$g_s^2 $&$0$\\
$\pi^0\gamma\gamma$&$\dfrac{8}{3} e^2$ &0\\
$\pi^0 Z\gamma$&$\dfrac{2ge}{c_W}\left(\dfrac{1}{2}-\dfrac{4}{3}s_W^2\right)$ &0\\
$\pi^0 ZZ$&$2g^2t_W^2\left(\dfrac{4}{3}s_W^2-1\right)$&$\dfrac{3 g_2^2}{2 c_W^2}$ \\
$\pi^0 WW$&0&$\dfrac{3 g_2^2}{2 }$\\
 \hline \hline
$\pi^8 gg$&$\dfrac{g_s^2}{2}d^{ABC}$&$0$\\ 
$\pi^8 g\gamma$&$\dfrac{2eg_s}{3}\delta^{AB}$&0\\
$\pi^8 gZ$&$\dfrac{g_sg}{2c_W}\left(\dfrac{1}{2}-\dfrac{4}{3}s_W^2\right)\delta^{AB}$&0\\
\hline
\end{tabular}
\caption{\label{Tab: Apploop}Coupling combinations $c_{1,2}$ induced by top loops for all di-boson channels. We use the shorthand notation $s_W\equiv \sin\theta_W$ and similar for the other trigonometric functions.}
\end{table}

The loop functions $\tilde{C}_{0,1}$ are related the usual Passarino-Veltman functions by a re-scaling, i.e. $C_0=-\widetilde{C}_0/m_t^2$ and $C_1=\widetilde{C}_1/m_t^2$~
\cite{Hahn:1998yk, Patel:2015tea}. Their integral form is given by
\begin{align}
\begin{split}
\widetilde{C}_0(R_p,R_k,R_\pi;\xi)=&\int_0^1\int_0^{1-x}\frac{dydx}{\Delta(R_p,R_k,R_\pi;\xi)}\\
\widetilde{C}_1(R_p,R_p,R_\pi;\xi)=&\frac{1}{2} \int_0^1\int_0^{1-x}\frac{x dydx}{\Delta(R_p,R_p,R_\pi;\xi)}
\end{split}
\end{align}
with
\begin{equation}
\Delta(R_p,R_k,R_\pi;\xi)=R_k(x^2-x)+R_p(y^2-y)-(R_\pi-R_p-R_k)xy+(x+y)(1-\xi^2)+\xi^2\,.
\end{equation}
These functions have in general very cumbersome analytic expressions. However, for scenarios with only massless gauge bosons or one massive they take a compact form
\begin{equation}
\widetilde{C}_0(0,0,R_\pi;1)=\frac{f(R_\pi)}{R_\pi}\,,\quad \widetilde{C}_0(R_Z,0,R_\pi;1)=\frac{1}{R_\pi-R_Z}\left(f(R_\pi)-f(R_Z)\right)
\end{equation}
with
\begin{equation} \label{eq:fx}
f(x)=\left\{
\begin{array}{ll}
2\,\text{arcsin}^2\sqrt{x/4}&0<x<4\\
-\dfrac{1}{2}\left[\ln\dfrac{1+\sqrt{1-4/x}}{1-\sqrt{1-4/x}}-i\pi\right]^2&x\geq 4
\end{array}\right.
\end{equation}
We can write the general $\widetilde{C}_1$ function in terms of  the scalar Passarino-Veltman functions\footnote {The definition of two-point function in D-dimension is  $ B_0 \left(R_i; \xi \right)= \dfrac{(2\pi \mu )^{4 - D}}{ i \pi ^2 }\int { \dfrac{d^D q}{[q^2 - m_t^2][(q + p)^2 - m_f^2]}}$, with  $R_i= p^2/m_t^2$ and $\xi = m_f/m_t$.}: 
\beq
\widetilde{C}_1(R_p,R_p,R_\pi;\xi) &=&  \frac{\left(R_p + \xi^2 -1 \right)}{ R_{\pi}-4 R_p} \widetilde{C}_0(R_p,R_p,R_\pi;\xi)  \nonumber \\ &+& \frac{1}{ R_{\pi}-4 R_p} \left( B_0 \left(R_{\pi}; 1\right) - B_0 \left(R_p; \xi \right) \right)
\eeq

From the above amplitude we can compute the corrections to the decays. We get for the singlet case
\begin{align}
\begin{split}
\dfrac{\Gamma_{WZW+top}(\pi_0\rightarrow gg)}{\Gamma_{WZW}(\pi_0\rightarrow gg)}=&
\left|1-\frac{C_t}{\kappa_{g}}\widetilde{C}_0(0,0,R_{\pi_0};1)\right|^2\,,\\
\dfrac{\Gamma_{WZW+top}(\pi_0\rightarrow \gamma\gamma)}{\Gamma_{WZW}(\pi_0\rightarrow \gamma\gamma)}=&
\left|1-\frac{8}{3}\frac{C_t}{\kappa_{B}+\kappa_W}\widetilde{C}_0(0,0,R_{\pi_0};1)\right|^2\,,\\
\dfrac{\Gamma_{WZW+top}(\pi_0\rightarrow W^+W^-)}{\Gamma_{WZW}(\pi_0\rightarrow W^+W^-)}=&\left|1-\frac{3}{2}\frac{C_t}{\kappa_{W}}\widetilde{C}_1(R_W,R_W,R_{\pi_0};\xi)\right|^2\,,\\
\dfrac{\Gamma_{WZW+top}(\pi_0\rightarrow Z\gamma)}{\Gamma_{WZW}(\pi_0\rightarrow Z\gamma)}=&\left|1-\frac{2}{c_W^2}\left(\frac{1}{2}-\frac{4s_W^2}{3}\right)\frac{C_t}{\kappa_{W}-t_W^2\kappa_B}\widetilde{C}_0(R_Z,0,R_{\pi_0};1)\right|^2\,,\\
\dfrac{\Gamma_{WZW+top}(\pi_0\rightarrow ZZ)}{\Gamma_{WZW}(\pi_0\rightarrow ZZ)}=&\left|1-\frac{C_t}{\kappa_{W}+t_W^4\kappa_B}\left(\frac{3}{2c_W^4}\widetilde{C}_1(R_Z,R_Z,R_{\pi_0};1)\right.\right.\\
&\left.\left.+2\frac{t_W^2}{c_W^2}\left(\frac{4s_W^2}{3}-1\right) \widetilde{C}_0(R_Z,R_Z,R_{\pi_0};1)\right)\right|^2
\end{split}
\end{align}
and for the octet
\begin{align}
\begin{split}
\dfrac{\Gamma_{WZW+top}(\pi_8\rightarrow gg)}{\Gamma_{WZW}(\pi_8\rightarrow gg)}=&
\left|1-\frac{C_t}{\kappa_{g}}\widetilde{C}_0(0,0,R_{\pi_8};1)\right|^2\,,\\
\dfrac{\Gamma_{WZW+top}(\pi_8\rightarrow g\gamma)}{\Gamma_{WZW}(\pi_8\rightarrow g\gamma)}=&
\left|1-\frac{4}{3}\frac{C_t}{\kappa_{gB}}\widetilde{C}_0(0,0,R_{\pi_8};1)\right|^2\,,\\
\dfrac{\Gamma_{WZW+top}(\pi_8\rightarrow gZ)}{\Gamma_{WZW}(\pi_8\rightarrow gZ)}=&
\left|1-\frac{1}{2s_W^2}\left(\frac{1}{2}-\frac{4s_W^2}{3}\right)\frac{C_t}{\kappa_{gB}}\widetilde{C}_0(R_Z,0,R_{\pi_8};1)\right|^2\,.
\end{split}
\end{align}

\section{Couplings and mixing in models M1 - M12} 
\label{sec:modeltables}

 \begin{table}[h]
{ \footnotesize
 \begin{tabular}{|c|c|c|c|c|c|c|c|c|c|}
 \hline
 \multirow{2}{*}{Model} & \multirow{2}{*}{} & \multirow{2}{*}{$\tan{\alpha}$} &\multirow{2}{*}{$\kappa_g$}&  \multirow{2}{*}{$\kappa_B$}&  \multirow{2}{*}{$\kappa_W$} &  \multicolumn{4}{c|}{$C_{t}\ (n_\psi\,,\,n_\chi)$}\\
\cline{7-10}
      &    & &    & & &$(2,0)$ & $(0,2)$ & $(4,2)$ & $(-4,2)$ \\ \hline
    \multirow{3}{*}{M3}    &$\pi_8$&    --                                  & 14.   & 18.7   &     --    &   0.  &  4. & 4. & 4. \\   \cline{2-10}      
                                      &$a$      &     \multirow{2}{*}{-.913\ (-.388)}  & -2.72\ (-1.46) & -3.53\ (.921) & 3.74\ (4.72)   & .934\ (1.18)   &   -.778\ (-.417)  & 1.09\ (1.94) & -2.65\ (-2.78)\\    
                                      &$\eta'$ &                                         & 2.98\ (3.77)   & 11.4\ (11.9)   & 3.41\ (1.83) &  .853\ (.457)    &    .853 (1.08)   & 2.56\ (1.99) & -.853\ (.162) \\                            
\hline                        
    \multirow{3}{*}{M4}    &$\pi_8$&    --                                  & 18.   & 24.  &     --    &   0.  &  4. & 4. & 4. \\   \cline{2-10}      
                                      &$a$      &     \multirow{2}{*}{-1.83\ (-.592)}  & -4.56\ (-2.65) & -7.29\ (1.64) & 4.86\ (8.71)   & .608\ (1.09)   &   -1.01\ (-.589)  & .203\ (1.59) & -2.23\ (-2.77)\\    
                                      &$\eta'$ &                                         & 2.50\ (4.47)   & 15.5\ (17.1)   & 8.88\ (5.16) &  1.11\ (.645)    &    .555 (.993)   & 2.77\ (2.28) & -1.66\ (-.296) \\                            
\hline                        
    \multirow{3}{*}{M8}    &$\pi_8$&    --                                  & 7.07   & 9.43  &     --    &   0.  &  2.83 & 2.83 & 2.83 \\   \cline{2-10}      
                                      &$a$      &     \multirow{2}{*}{-0.408\ (-.196)}  & -.771\ (-.393) & -1.13\ (-.067) & .926\ (.981)   & .926\ (.981)   &   -.309\ (-.157)  & 1.54\ (1.81) & -2.16\ (-2.12)\\    
                                      &$\eta'$ &                                         & 1.89\ (2.00)   & 5.42\ (5.53)   & .378\ (.193) &  .378\ (.193)    &    .756\ (.801)   & 1.51\ (1.19) & 0.\ (.416) \\                            
\hline                        
    \multirow{3}{*}{M9}    &$\pi_8$&    --                                  & 15.6   & 20.7   &   --    &  0.  &  2.83 & 2.83 & 2.83 \\   \cline{2-10}      
                                      &$a$      &     \multirow{2}{*}{-3.27\ (-.740)}  & -4.29\ (-2.67) & -9.11\ (-.689) & 2.34\ (6.43)   & .293\ (.804)   &  -.781\ (-.486)  & -.195\ (1.12) & -1.37\ (-2.09)\\    
                                      &$\eta'$ &                                         & 1.31\ (3.61)   & 11.2\ (14.4)   & 7.65\ (4.76) &  .956\ (.595)    &    .239\ (.656)   & 2.15\ (1.85) & -1.67\ (-.533) \\                            
\hline                        
    \multirow{3}{*}{M10}    &$\pi_8$&    --                                  & 14.1   & 18.9   &     --    &     0.  &   2.83 & 2.83 & 2.83 \\   \cline{2-10}      
                                      &$a$      &     \multirow{2}{*}{-3.27\ (-.740)}  & -3.90\ (-2.43) & -8.07\ (-.042) & 2.34\ (6.43)   & .293\ (.804)   &   -.781\ (-.486)  & -.195\ (1.12) & -1.37\ (-2.09)\\    
                                      &$\eta'$ &                                         & 1.20\ (3.28)   & 10.8\ (13.5)   & 7.65\ (4.76) &  .956\ (.595)    &    .239\ (.656)   & 2.15\ (1.85) & -1.67\ (-.533) \\                            
\hline                        
    \multirow{3}{*}{M11}    &$\pi_8$&    --                                  & 8.49   & 11.3   &     --    &     0.  &  2.83 & 2.83 & 2.83 \\   \cline{2-10}      
                                      &$a$      &     \multirow{2}{*}{-.816\ (-.356)}  & -1.55\ (-.822) & -2.58\ (-.309) & 1.55\ (1.88)   & .775\ (.942)   &   -.516\ (-.274)  & 1.03\ (1.61) & -2.07\ (-2.16)\\    
                                      &$\eta'$ &                                         & 1.90\ (2.31)   & 6.32\ (6.82)   & 1.26\ (.671) &  .632\ (.336)    &    .632\ (.769)   & 1.90\ (1.44) & -.632\ (.098) \\                            
\hline                        
    \multirow{3}{*}{M12}    &$\pi_8$&    --                                  & 14.1   & 18.9   &     --    &     0.  &  2.83 & 2.83 & 2.83 \\   \cline{2-10}      
                                      &$a$      &     \multirow{2}{*}{-.385\ (-.186)}  & -2.07\ (-1.05) & -3.20\ (-.355) & 2.33\ (2.46)   & .933\ (.983)   &   -.415\ (-.211)  & 1.45\ (1.76) & -2.28\ (-2.18)\\    
                                      &$\eta'$ &                                         & 5.39\ (5.68)   & 15.3\ (15.6)   & .898\ (.457) &  .359\ (.183)    &    1.08\ (1.14)   & 1.80\ (1.50) & .359\ (.770) \\                            
\hline                        
\end{tabular} }
\caption{Couplings for models with top partners of the form $\psi \psi \chi$, for $f_\chi = f_\psi$, normalised to $f_\psi$.}
\label{tab:models21}
\end{table}

 \begin{table}[h]
{ \footnotesize
 \begin{tabular}{|c|c|c|c|c|c|c|c|c|c|}
 \hline
 \multirow{2}{*}{Model} & \multirow{2}{*}{} & \multirow{2}{*}{$\tan{\alpha}$} &\multirow{2}{*}{$\kappa_g$}&  \multirow{2}{*}{$\kappa_B$}&  \multirow{2}{*}{$\kappa_W$} &  \multicolumn{4}{c|}{$C_{t}\ (n_\psi\,,\,n_\chi)$}\\
\cline{7-10}
      &    & &    & & &$(2,0)$ & $(0,2)$ & $(2,4)$ & $(2,-4)$ \\ \hline
    \multirow{3}{*}{M1}    &$\pi_8$&    --                                  & 16.   & 10.7   &     --    &   0.  &    4. & 8. & -8. \\   \cline{2-10}      
                                      &$a$      &     \multirow{2}{*}{-.913\ (-.388)}  & -3.11\ (-1.67) & 1.19\ (3.01) & 3.27\ (4.13)     &   .934\ (1.18)  & -.778\ (-.417)   & -.623\ (.344) & 2.49\ (2.01)\\    
                                      &$\eta'$ &                                         & 3.41\ (4.31)   & 5.26\ (4.47)   & 2.98\ (1.60)   &    .853\ (.457)  &  .853\ (1.08)    & 2.56\ (2.61) & -.853\ (-1.70) \\                            
\hline                        
    \multirow{3}{*}{M2}    &$\pi_8$&    --                                  & 32.  & 21.3   &     --    &  0.  &  4. & 8. & -8. \\   \cline{2-10}      
                                      &$a$      &     \multirow{2}{*}{-.456\ (-.217)}  & -3.84\ (-1.96) & 2.62\ (4.25) & 5.18\ (5.56)    &  1.15\ (1.24)  & -.479\ (-.245)  & .192\ (.745) & 2.11\ 1.73)\\    
                                      &$\eta'$ &                                         & 8.40\ (9.03)   & 7.97\ (7.23)   & 2.36\ (1.21)   &  .525\ (.269) &  1.05\ (1.13)    & 2.63\ (2.53) & -1.58\ (-1.99) \\                            
\hline                        
    \multirow{3}{*}{M5}    &$\pi_8$&    --                                  & 8.   & 5.33   &     --    &   0.  &  4. & 8. & -8. \\   \cline{2-10}      
                                      &$a$      &     \multirow{2}{*}{-1.83\ (-.592)}  & -2.03\ (-1.18) & .169\ (1.94) & 1.52\ (2.72)  &   .608\ (1.09)   & -1.01\ (-.589)   & -1.42\ (-.089) & 2.63\ (2.27)\\    
                                      &$\eta'$ &                                         & 1.11\ (1.99)   & 3.51\ (2.94)   & 2.77\ (1.61)    &    1.11\ (.645) &  .555\ (.993)   & 2.22\ (2.63) & 0.\ (-1.34) \\                            
\hline                        
    \multirow{3}{*}{M6}    &$\pi_8$&    --                                  & 8.   & 5.33   &     --    &     0.  &  4. & 8. & -8. \\   \cline{2-10}      
                                      &$a$      &     \multirow{2}{*}{-1.29\ (-0.490)}  & -2.58\ (-1.44) & .602\ (2.45) & 2.32\ (3.41)  &   .775\ (1.14)   & -1.29\ (-.719)   & -1.81\ (-.302) & 3.36\ (2.57)\\    
                                      &$\eta'$ &                                         & 2.\ (2.93)   & 4.33\ (3.63)   & 3.\ (1.67) &  1.\ (.557)    &   1.\ (1.47)   & 3.\ (4.50) & -1.\ (-2.38) \\                            
\hline                        
    \multirow{3}{*}{M7}    &$\pi_8$&    --                                  & 32.  & 21.3   &     --    &     0.  &   4. & 8. & -8. \\   \cline{2-10}      
                                      &$a$      &     \multirow{2}{*}{-.323\ (-.157)}  & -4.01\ (-2.03) & 3.34\ (4.89) & 6.02\ (6.25)    &   1.20\ (1.25)  & -.502\ (-.254)  & .201\ (.742) & 2.21\ (1.76)\\    
                                      &$\eta'$ &                                         & 12.4\ (12.9)   & 10.2\ (9.59)   & 1.94\ (.983)   &   .389\ (.197)   &  1.55\ (1.61)   & 3.50\ (3.42) & -2.72\ (-3.03) \\                            
\hline                        
\end{tabular} }
\caption{Couplings for models with top partners of the form $\psi \chi \chi$, for $f_\chi = f_\psi$, normalised to $f_\psi$.}
\label{tab:models12}
\end{table}

In this Appendix we present numerical values for all models, M1 to M12, assuming $f_\psi = f_\chi$ and normalising all couplings with $f_\psi$. The couplings of the singlets are shown for the two extreme values of the mixing angle $\alpha$: $\alpha = \zeta$ obtained when $m_{\eta'} \to \infty$, and $\alpha = \zeta/2$ obtained in the limit of minimal splitting. We checked that the couplings run approximately linearly with $\alpha$. In Table~\ref{tab:models21} we show models whose top partners are made of 2 $\psi$'s and one $\chi$, while in Table~\ref{tab:models12} the cases with 1 $\psi$ and 2 $\chi$'s.

\bibliography{diboson2}
\bibliographystyle{JHEP-2-2.bst}

\end{document}